\documentclass{pasa}%
\def \aj {AJ}
\def \mnras {MNRAS}
\def \apj {ApJ}
\def \apjl {ApJL}
\def \aap {A\&A}
\def \nat {Nature}
\def \araa {ARAA}
\def \aapr {A\&ARev}
\def \iauc {IAUC}
\def \pasp {PASP}

\def \actaa {Act.A.}
\def \physrep {Phys. Rep}
\def\lesssim{\mathrel{\hbox{\rlap{\hbox{\lower4pt\hbox{$\sim$}}}\hbox{$<$}}}}
\def\gtrsim{\mathrel{\hbox{\rlap{\hbox{\lower4pt\hbox{$\sim$}}}\hbox{$>$}}}}

\newcommand{\msun}{\mbox{M$_{\odot}$}}
\newcommand{\lsun}{\mbox{L$_{\odot}$}}

\newcommand{\rsun}{\mbox{R$_{\odot}$}}
\newcommand{\kms}{\mbox{$\rm{\,km\,s^{-1}}$}}
\newcommand{\logl}{\mbox{$\log L/{\rm L_{\odot}}$}}

\DeclareMathAlphabet{\mathsc}{OT1}{cmr}{m}{sc}
\def\testbx{bx}%
\DeclareRobustCommand{\ion}[2]{%
\relax\ifmmode
\ifx\testbx\f@series
{\mathbf{#1\,\mathsc{#2}}}\else
{\mathrm{#1\,\mathsc{#2}}}\fi
\else\textup{#1\,{\mdseries\textsc{#2}}}%
\fi}

\title[Progenitors of core-collapse supernovae]{Observational constraints on the progenitors of core-collapse supernovae : the case for missing high mass stars}
\author[S. J. Smartt]{S. J. Smartt\\
\affil{Astrophysics Research Centre,
School of Mathematics and Physics,    
Queen's University Belfast, 
Belfast, BT7 1NN, UK }%
}
\jid{PASA}
\doi{10.1017/pas.\the\year.xxx}
\jyear{\the\year}

\usepackage[authoryear]{natbib}
\bibpunct{(}{)}{;}{a}{}{,}
\begin{document}%
\begin{abstract}

Over the last 15 years, the supernova
community has endeavoured to directly identify progenitor stars of  core-collapse supernovae discovered in nearby galaxies. 
These precursors are often visible as resolved stars in high resolution 
images from space and ground based telescopes. The discovery rate of progenitor stars is limited
by the local supernova rate and the availability and depth of archive images of galaxies, 
with 18 detections of precursor objects and 27 upper limits. This review compiles these results (from 1999 - 2013) in a distance limited sample and discusses the implications of the findings.  The vast majority of the detections of progenitor stars are 
of type II-P, II-L or IIb with one type Ib progenitor system detected and 
many more upper limits for progenitors of Ibc supernovae (14 in all). 
The data for these 45 supernovae progenitors illustrate a remarkable
deficit of high luminosity stars above an apparent limit of 
\logl$\simeq5.1$\,dex. For a typical Salpeter initial mass function, one
would expect to have found 13 high luminosity and high mass progenitors by 
now.  There is, possibly, only one object in this time and volume limited sample that is unambiguously high mass (the progenitor of SN2009ip) although the nature of
that supernovae is still debated. The possible biases due to the influence of 
circumstellar dust, the luminosity analysis, and sample selection methods 
are reviewed.  It does not appear likely that these can explain the missing 
high mass progenitor stars. This review concludes that the community's work to 
date shows  that the observed populations
of supernovae in the local Universe are not, on the whole, produced by high mass ($M\gtrsim18$\msun) stars. Theoretical explosions of model stars also predict that 
black hole formation and failed supernovae tend to occur above an initial mass of 
$M\simeq18$\msun. The models also suggest there is no simple single mass division for neutron star or black-hole formation and that 
there are islands of explodability for stars in the $8-120$\msun\ range. The observational constraints are quite consistent with the bulk of stars above 
$M\gtrsim18$\msun\ collapsing to form black holes with no visible supernovae.

\end{abstract}
\maketitle%
\section{Introduction}
\label{sec:intro}

The link between massive progenitor star and the type of core-collapse event it produces at the 
end of its nuclear burning live is the fundamental piece of information that underpins our
understanding of the physical processes involved in stellar explosions. 
\citep[see][for a recent theoretical overview]{2012ARA&A..50..107L}. 
Since the end of the 1990's the supernova community has been extensively searching publicly available 
archives for high resolution and deep images of nearby galaxies which host supernovae (SN). The existence of 
these pre-explosion images has allowed direct identification of the progenitor stars of some 
of the nearest core-collapse supernovae 
\citep[e.g.][]{2003PASP..115.1289V,2004Sci...303..499S,2005MNRAS.364L..33M,2006ApJ...641.1060L}. 
Many of these, as individual events, are quite compatible with 
stellar evolutionary model predictions. The fact that numerous red supergiants are now observed as the stellar
end points that produce the most common class of type II SN is a validation of 
modern stellar structure theory.  In 2009, the number of detections of progenitor stars had reached a point that analysis of samples
was possible  \citep{2009MNRAS.395.1409S}  and a review of what these constraints meant for the field in the broader sense was 
warranted \citep{2009ARA&A..47...63S}.  Since then there has been further very significant progress and results which 
are more consequential than just an increase in sample size. This review will consider work done in the 
field since 2009, and review the results as a whole in the context of the SN population we observe in the Local 
Universe  as quantified by, for example, \cite{2011MNRAS.412.1419L},
 and \cite{2011MNRAS.412.1441L}.  Some of the highlights in the last 5 years are the significant increase in the sample size, nine 
further detections of progenitors, evidence for the disappearance of progenitor stars  many years after explosion and detections of possible binary companions. 
There has also been active discussion in the literature concerning the analysis methods, particularly
in the treatment of extinction toward progenitors  and how that may affect the luminosity and mass estimates. 
There are even some progenitor stars with multi-epoch data that allow pre-explosion variability 
to be probed, and as data sets grow then lightcurve monitoring of progenitors may become more 
common place \citep[e.g.][]{2009ApJ...706.1174E,2012ApJ...747...23S,2014MNRAS.439L..56F}.

In an attempt to set a well defined sample,  \cite{2009MNRAS.395.1409S}
\citep[and later][]{2013MNRAS.436..774E}
defined criteria for selecting a time and volume limited sample of SNe for which searches for 
progenitors were feasible :  all CCSNe in galaxies with recessional velocities 
$V_{\rm vir} \leq 2000$\kms (corrected for Virgo infall, which corresponds to a distance of 
$d \leq 28$\,Mpc for $H_{0}=72$\kms\,Mpc$^{-1}$).  This distance limit was imposed from 
practical experience of analysing HST archival image data and 
the difficulty in retrieving photometry of resolved individual massive stars beyond this 
distance. Additionally, as galaxy number count increases as the distance cubed, the relative number of galaxies
with high quality archival imaging rapidly decreases. Hence the probability of finding coincidences of
SN discoveries combined with existing pre-explosion images of the galaxies also falls off
quickly.  As we will see later in this review the SN which yield the most restrictive information on 
their progenitor stars have tended to be significantly closer than 28\,Mpc. Two groups have 
worked extensively in this area,  a California based team 
\citep[who started their concerted search with HST in][]{1999AJ....118.2331V}, and 
a UK based group  \citep[initiated with their first paper: ][]{2001ApJ...556L..29S}
 but others have also made significant contributions 
\citep[e.g.][]{2007ApJ...656..372G,2012ApJ...759...20K,2008ApJ...681L...9P}. 
The UK based group have focused their
attention on this volume and time limited survey to get high-resolution, deep imaging of 
all SN that have useful archival imaging so that the precise SN positions can be astrometrically 
placed on the pre-explosion images. The 
Van Dyk led team have also pursued a similar strategy and both teams have used combinations of 
HST imaging 
\citep[e.g.][]{2004Sci...303..499S,2011ApJ...742....6E,2014AJ....147...37V}
ground-based AO imaging 
\citep[e.g.][]{2011MNRAS.417.1417F,2012ApJ...759L..13F,2012ApJ...756..131V}
or good image quality natural seeing images
\citep[e.g.][]{2005PASP..117..121L,2014MNRAS.439L..56F}
to locate the SN positions to accuracies of $\sim$30-50 milliarcsec on the pre-explosion frames. 
\\

\section{Volume and time limited survey data}
\label{sec:data}

The definition of a time and volume limited survey has significant advantages in determining the overall 
properties of supernova progenitor systems as a function of initial mass. There have been 
some remarkable discoveries of individual stars before explosion, with tight 
physical constraints on their luminosity and temperature \citep{2004Sci...303..499S,2012AJ....143...19V}
some of which have been shown to have disappeared
\citep{2013ApJ...772L..32V,2014MNRAS.438..938M}. However the most interesting aspect of this area is now the statistical sample of more than a dozen unambiguous direct detections and many restrictive limits to produce a defined survey sample
of 44 objects.   Objects outside this distance limit have also contributed to our understanding, such as the massive progenitor of  SN2005gl at 60\,Mpc  \citep{2007ApJ...656..372G}
but the relative frequency of such an event needs to be put
in context with the fixed volume sample. 

This review  extends the time and volume limited sample introduced by \cite{2009MNRAS.395.1409S} and \cite{2013MNRAS.436..774E} for the type II and Ibc progenitors to the end of 2013. The discussions is extended to include all other progenitors within this survey sample including the IIb and IIn types and some beyond. The closest SNe provide the best opportunity for 
studying their progenitors, carrying out multi-wavelength 
monitoring until very late times. It is likely that we 
are not missing a large fraction of SNe in 
galaxies closer than $\sim$12\,Mpc  \citep{2012A&A...537A.132B}
as the starformation rates of these galaxies are in 
quite good agreement with the measured SN rate for a
a lower mass limit of 8\msun\ for core-collapse
\citep[although][suggests that at higher redshift there is a discrepancy]{2011ApJ...738..154H}. 
These closest SNe offer the best opportunities
for progenitor detection, but as we can't influence the
SN rate nor the discovery rate (assuming we are 70-80\% complete) the only way to increase the progenitor discovery
rate is through patience and time.  The following sections discuss the results published to date for
the different subtypes of progenitors and Section\,\ref{sec:discuss}
 presents a discussion of these empirical results
in the context of massive stars, the initial mass function, and the local SN population. 

\subsection{Type II supernovae}

\begin{figure}
\begin{center}
\includegraphics[width=6cm]{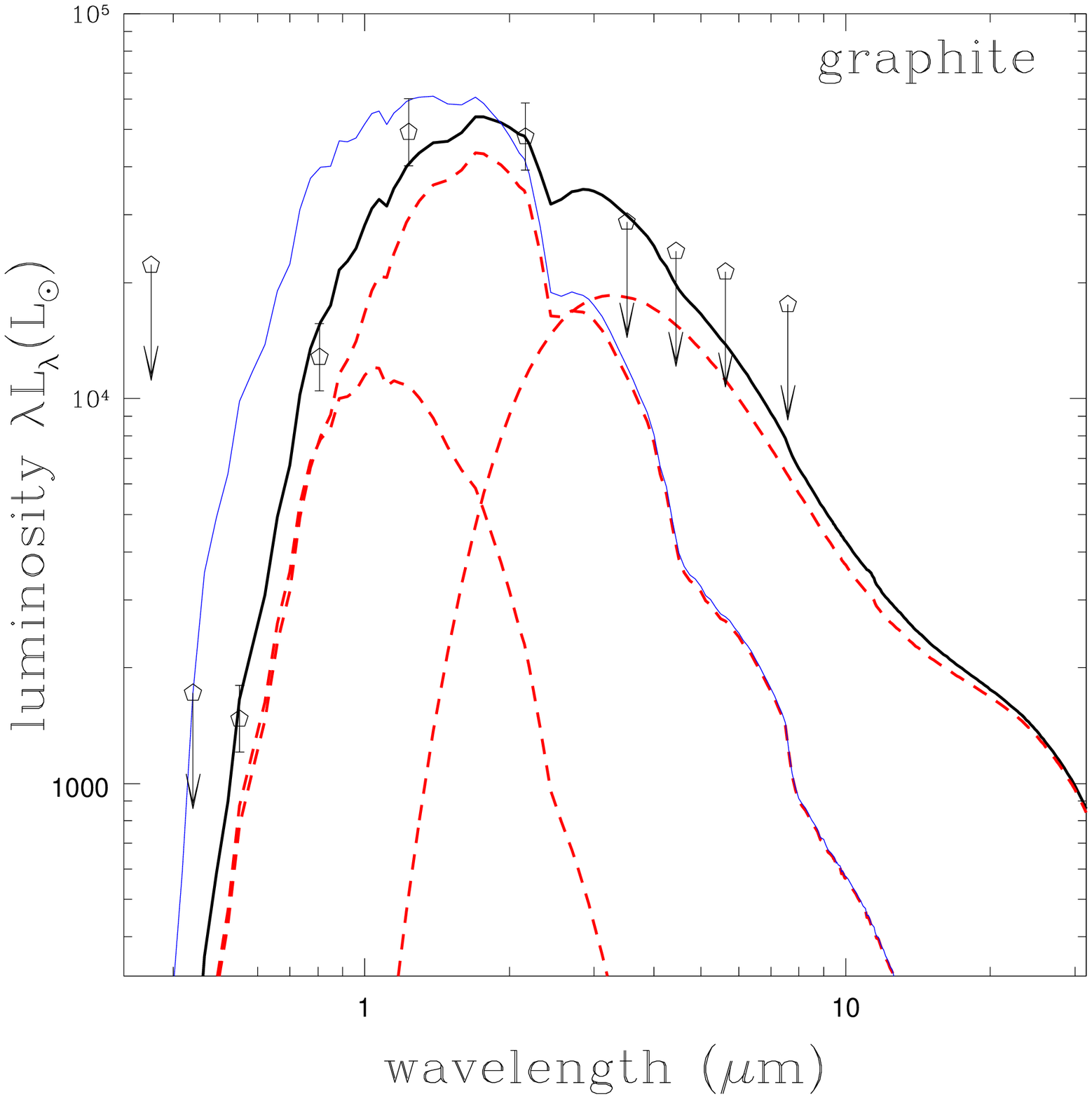}
\includegraphics[width=6cm]{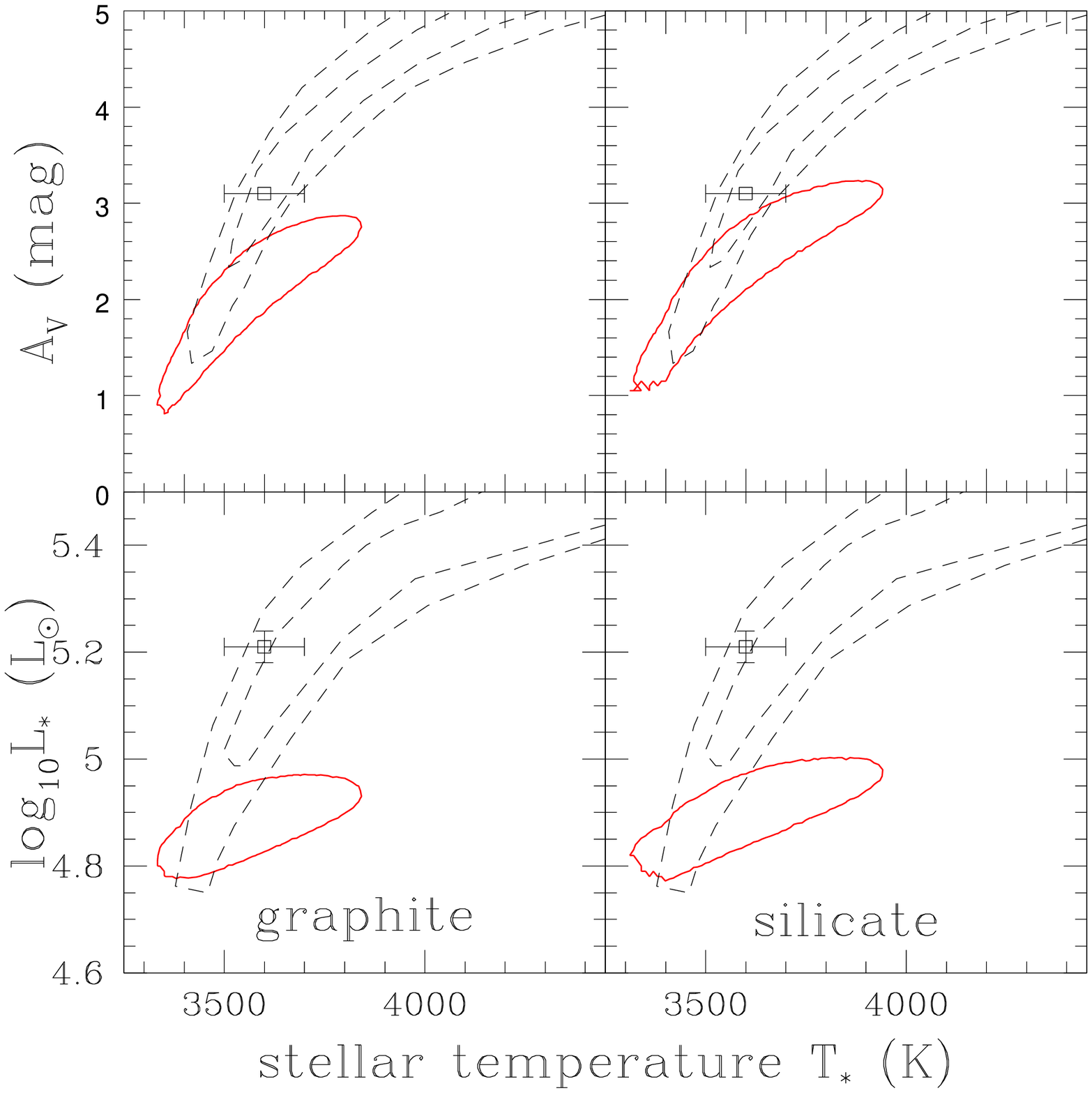}
\caption{Analysis of the progenitor of SN2012aw. {\bf Upper:}  The SED fit of a DUSTY model from \cite{2012ApJ...759...20K}
 to the detections in the optical and NIR filters presented in \cite{2012ApJ...759L..13F,2012ApJ...756..131V}
and the mid-IR Spitzer limits. The black curves show the SED model which is made up of the three components of 
scattered photons, direct photons and dust emission radiation (the red dashed curves from left to right respectively). The
blue line is the unobscured SED of the star itself. This is for a dust temperature of $T_d = 1000$\,K and a graphitic 
dust composition (see the original reference for a similar silicate fit). 
 {\bf Lower:} Comparison of the luminosity estimates of the three studies. The point with error bar
is  from 
\cite{2012ApJ...756..131V}, dashed contours from \cite{2012ApJ...759L..13F} and red locus from 
 \cite{2012ApJ...759...20K}, illustrating the  importance of considering the treatment of both dust absorption and emission.  As discussed in Section\,\ref{sec:12aw}, the proper inclusion of dust treatment lowers the luminosity estimates
substantially.  Reproduced from 
Figures 5 and 6 in {\em ``On Absorption by Circumstellar Dust, with the Progenitor of SN 2012aw as a Case Study”} by Kochanek et al. 2012, ApJ, 759, 20. }
\label{fig:koch}
\end{center}
\end{figure}

As illustrated in \cite{2009MNRAS.395.1409S}, the chance of having an image of a nearby galaxy with one of the 
three main imaging cameras onboard HST (WFPC2, ACS or WF3) for any nearby SNe (within $V_{\rm vir}  < 2000$\kms)
is about 25\%.  If multi-colour imaging is available then these colours can be fit with either observed supergiant 
colours or model atmospheres to determine approximate spectral types or effective temperatures, bolometric
corrections and hence bolometric luminosity. An alternative, but equivalent approach is to use stellar
evolutionary models that have associated model atmosphere spectral energy distributions (SEDs) for the surface characteristic $T_{\rm eff}$,  and $\log L$ (or $\log g$) and synthetic photometry calculated from these 
spectra.  There have been variations on these two methods from the early days of the 
first red supergiant detection of the progenitor of SN2003dg
\citep{2003PASP..115.1289V,2004Sci...303..499S}
through to the latest which have extended spectral colours in the  near infra-red
\citep{2012ApJ...759L..13F, 2012ApJ...759...20K,2012ApJ...756..131V}. 
Detections of red supergiants in $JHK$-bands are quite powerful indicators of the bolometric
flux, since extinction uncertainties are lower and the bolometric corrections are less uncertain
\citep{2006ApJ...645.1102L}. Deep enough near infra-red pre-explosions images of nearby SN are still the 
exception rather than common place \citep[although limits can set restrictions on super-AGB progenitors for example;][]{2007MNRAS.376L..52E}

Early results indicated that the detections of the progenitors of II-P SNe were red supergiants, with estimated
luminosities in the range $4.3 < \log L/{\rm L_{\odot}} < 5.0$ \citep[see review of results in][]{2009ARA&A..47...63S}, 
which would mean zero age main-sequence (ZAMS) masses in the region  7-15\msun.  Since then, 
improvement of the analysis methods have provided updated estimates of the stellar parameters both 
for previously published detections and new discoveries. Efforts have been made to systematically include the MARCs stellar models and consistently apply synthetic photometry to reduce comparative errors in 
colour 
corrections and homogenise the analysis \citep[for more in depth discussion see][]{2011MNRAS.417.1417F,2013ApJ...767....3D}. 
Observing the SN progenitor position 3-4 years after explosion to check that the identified stellar
progenitor has disappeared now allows difference imaging techniques to be applied 
in order to improve the precision of the pre-explosion photometry. 
\cite{2014MNRAS.438..938M} and \cite{2014MNRAS.438.1577M} 
discuss this in detail and  the case for the disappearance of the progenitor of SN2011dh is presented in 
\cite{2013ApJ...772L..32V}. 
The treatment of the extinction toward the progenitors has progressed and the paper of 
\cite{2012ApJ...759...20K}
highlights the need to consider interstellar medium (ISM)  and 
circumstellar medium (CSM) 
CSM extinction separately and consistently. 
\cite{2013A&A...558A.131G} have considered rotating and non-rotating models showing that
initial rotation of the progenitor introduces a further uncertainty on the progenitor mass estimate 
as the core luminosity changes. They suggest that any particular mass may be uncertain to 
within 4-5\msun\ but the overall mass range derived for type II-P progenitors is quite similar
to that originally derived in \cite{2009MNRAS.395.1409S}. All of this work has put the estimates of stellar parameters on a firmer footing with some 
changes in the stellar $T_{\rm eff}$ and $\log L$ for various progenitor stars. 
However it is fair to say that the 
improved estimates of stellar luminosity (and hence mass) are not significantly different to the
originally published work. 
The main result  from the data up to 2009  \citep{2009MNRAS.395.1409S}
still holds :  type II-P supernovae come from 
red supergiants for which the lowest mass estimated is around 8\msun and the 
highest mass progenitor has a luminosity of $L \simeq 5.0$L$_{\odot}$ (corresponding to a 
ZAMS of 15-18\msun). 

The most recent results, and in the opinion of the author, the most reliable estimates of the 
the physical parameters of detected progenitors of type II SNe are listed in  Table\,\ref{tab:SNII} with the 
references to the relevant papers. Where new and updated estimates are used, references
to the original detections are provided with comments. This review will not discuss all the 
detections individually, and the differences between the measurement and analysis methods. However two SNe are taken here as examples for a more in depth discussion as case studies. The issues either in updated stellar luminosities or the disagreement between 
analysis methods are discussed for SN2004A and SN2012aw. 

The two SNe which fell on compact, and likely coeval clusters (SN2004dj and SN2004am) in NGC2403 and M82 are not
considered here, although the turn-off mass estimated
for both is in the regime of the moderate to low masses that 
are derived for II-P progenitors. Hence their inclusion would
support the results discussed below. SN1999ev is also 
not included as \cite{2014MNRAS.438..938M}
showed that the progenitor object identified 
originally by \cite{2005MNRAS.360..288M} was likely
a stellar cluster. Finally SN2003ie, which was included
in \cite{2009MNRAS.395.1409S} is no longer considered as with the higher
extinction estimates used later in this analysis, the 
luminosity limit is not useful nor restrictive in any way.

\begin{figure}
\begin{center}
\includegraphics[width=6cm]{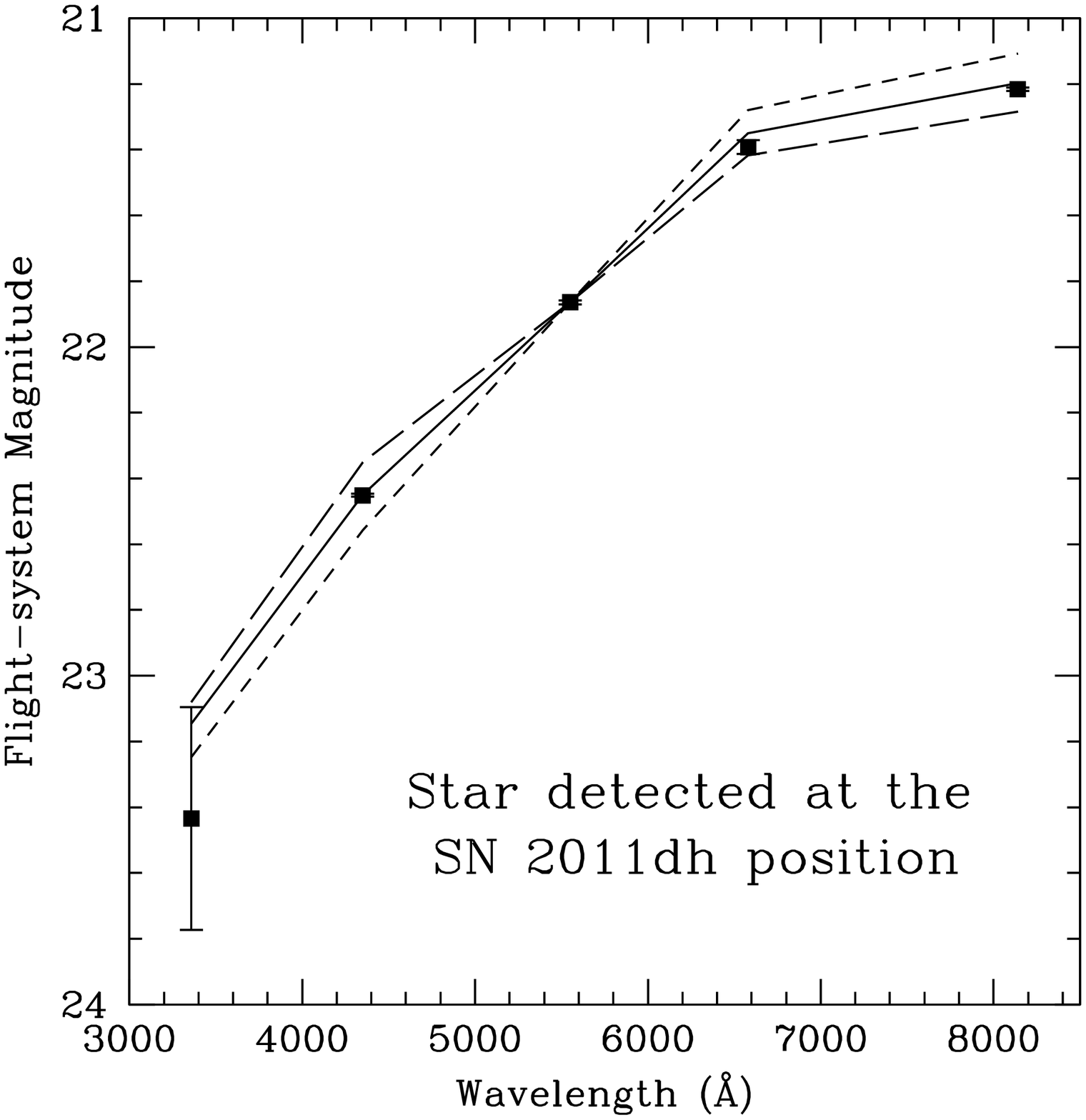}
\includegraphics[width=6cm]{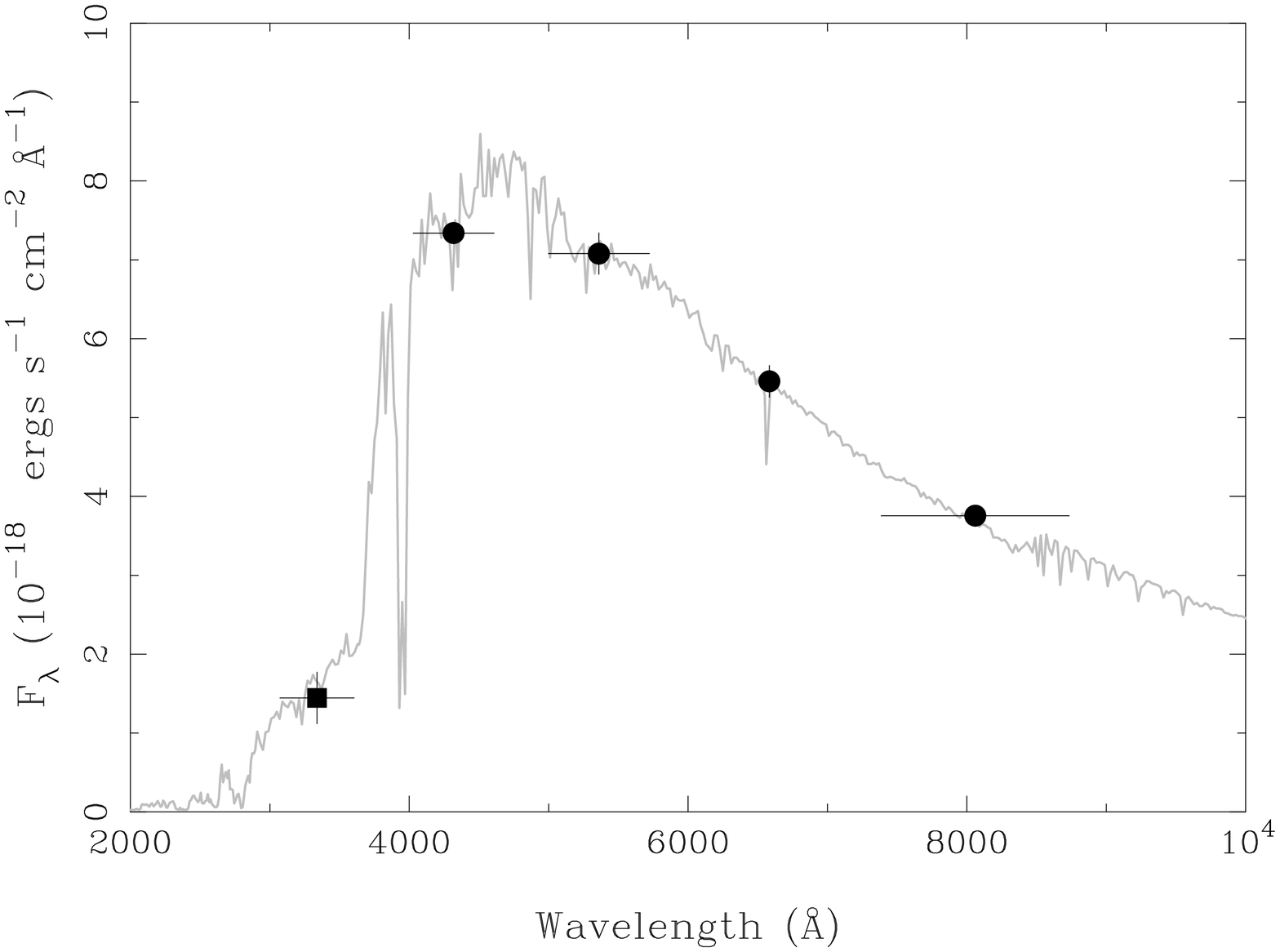}
\caption{The SED fit of the progenitor of the yellow supergiant progenitor of SN2011dh from two independent 
analyses of \cite{2011ApJ...741L..28V} (upper) and  \cite{2011ApJ...739L..37M} (lower). 
The pre-explosion HST data points (solid black with errors) are impressively fit with a single stellar source and an ATLAS model atmosphere for $T_{\rm eff}=6000$\,K and $\log g=1.0$ in both analyses. 
Reproduced from Figure  3 in {\em ``The Progenitor of Supernova 2011dh/PTF11eon in Messier 51”} by Van Dyk et al. 2011,ApJ, 741, L28  and Figure 2 in {\em ``The Yellow Supergiant Progenitor of the Type II Supernova 2011dh in M51” } by Maund et al. 2011, ApJ, 739, L37	. 
 }
\label{fig:11dh}
\end{center}
\end{figure}

\begin{table*}
\caption{List of detections of type II progenitors. The comments in the Reference column give guidance as to the 
source of the stellar progenitor data for the luminosities (\logl). The initial masses of model stars that 
end with these luminosities are given in the last two columns : $M$(S,G) are the masses (in \msun) from the STARS
and Geneva rotating models (which have similar final luminosities); $M$(K) are from the KEPLER models (see Section\,\ref{sec:lumfunc} for details). The values for the initial masses from KEPLER
for 2003gd, 2005cs and 2009md  
are uncertain since there are no  KEPLER models evolved to end points 
at this mass and luminosity. They are based on extrapolation, 
and quoted in parentheses. The errors for M(K) can be assumed to be the same as those quoted for 
$M(S,G)$.}  
\begin{center}
\begin{tabular}{lllllllll}
\hline\hline
SN & Type & $\log T_{\rm eff}$ & err & $\log L$ & err & Reference  & $M$(S,G) &  $M$(K) \\
\hline%
SN2003gd & II-P & 3.54 &  0.02 & 4.3 &  0.20  &  From (1),  similar to (2)              &      $7^{+4}_{-1}$ & (8)\\
SN2005cs & II-P & 3.55  & 0.05 & 4.4 & 0.20  &  From (3), similar to  (5) and (6)    &    $8^{+4}_{-1}$  & (9)\\
SN2009md & II-P & 3.55 & 0.01  & 4.5 & 0.20  & From (11)                                     &      $9^{+4}_{-2}$ &  (10) \\
SN2006my & II-P & 3.55 & 0.10   & 4.7 & 0.20   &  From (3), similar to (4) and (7)   &     $10^{+3}_{-2}$  & 12 \\
SN2012A  & II-P & 3.58 & 0.05   & 4.7 & 0.10 &  From (13)                                      &    $10^{+4}_{-2}$  & 12 \\
SN2013ej & II-P & 3.57 & 0.04   & 4.7 & 0.20  & From (16)                                       &     $10^{+4}_{-2}$  & 12 \\
SN2004et & II-P & 3.56   & 0.05  & 4.8 & 0.20   & From (4) and (17)                         &     $12^{+3}_{-3}$  & 13 \\
SN2008bk & II-P& 3.64   & 0.10    & 4.8 & 0.20    &  From (3), similar \logl=4.6 in (8) &    $12^{+3}_{-3}$ &  13 \\
SN2004A  & II-P &3.59    & 0.04  & 4.9 & 0.30   & From (3), see  Sect.\ref{sec:04A}  &   $13^{+6}_{-3}$   & 14 \\
SN2012aw & II-P& 3.56 & 0.04  & 4.9 & 0.10  & From (14), see Sect.\ref{sec:12aw} &     $13^{+2}_{-2}$ &  14 \\
SN2009hd & II-L& 3.72 & 0.15     &   5.0  & 0.20  & From (12)                                    &    $15^{+3}_{-3}$ &  16 \\
SN2009kr & II-L & 3.68  & 0.03  & 5.1  & 0.25  &  From (9), similar results in (10)      &   $16^{+5}_{-5}$ &18 \\
SN2012ec & II-P& 3.53 & 0.03  &  5.1 & 0.20 & From (15)                                          &    $16^{+5}_{-5}$ & 18 \\
\hline\hline
\end{tabular}
\end{center}
\label{tab:SNII}
(1)\cite{2009Sci...324..486M}; 
(2)\cite{2004Sci...303..499S};   
(3)\cite{2014MNRAS.438..938M};
(4)\cite{2011MNRAS.410.2767C};
(5)\cite{2009MNRAS.395.1409S};
(6)\cite{2005MNRAS.364L..33M};
(7)\cite{2007ApJ...661.1013L};
(8)\cite{2008ApJ...688L..91M};
(9)\cite{2010ApJ...714L.280F};
(10)\cite{2010ApJ...714L.254E};
(11)\cite{2011MNRAS.417.1417F};
(12)\cite{2011ApJ...742....6E} : note the $T_{eff}$  and \logl values are upper limits. 
(13)\cite{2013MNRAS.434.1636T} : note the $T_{eff}$ was an assumption, and not derived from multiple colours;
(14)\cite{2012ApJ...759...20K};
(15)\cite{2013MNRAS.431L.102M};
(16)\cite{2014MNRAS.439L..56F} : note the $T_{eff}$ was an assumption, and not derived from multiple colours;
(17) Fraser et al. (in prep)
\end{table*}

\subsubsection{SN2004A}
\label{sec:04A}
This SN progenitor is taken as an example were the original estimate of luminosity has a
 significant discrepancy with a new measurement. As the later measurement is based
on new data after explosion, with difference imaging applied it is worth discussing as a an 
example how significantly results could change.
 \cite{2009MNRAS.395.1409S} and \cite{2006MNRAS.369.1303H} originally 
estimated a \logl=4.5$\pm0.3$. With deep follow-up imaging 
\cite{2014MNRAS.438..938M}
 showed the progenitor object identified has disappeared and 
suggested a significantly higher value of \logl=$4.9\pm0.3$. The estimate of the magnitude of the progenitor 
in the HST F814W pre-explosion image is similar in  \cite{2006MNRAS.369.1303H} and 
\cite{2014MNRAS.438..938M}, illustrating the difference in final quoted luminosity  arises in the 
analysis methods. The methods differ in that a higher extinction of $E(B-V)=0.16$ was employed by 
\cite{2014MNRAS.438..938M}, together with application of the MARCS model atmospheres. 
In his PhD thesis work,  \cite{2011PhD_MFT} undertook a re-analysis of the type II-P progenitor
sample. As an experiment, he suggested that 
adding in extinction values that are seen toward the red supergiant population 
of local group galaxies M31, the 
Large Magellanic Cloud (LMC) and
the
Small Magellanic Cloud (SMC) would be appropriate to test if the typical 
extinctions estimated for SN progenitors are systematically lower than those observed toward 
existing red supergiants.  \cite{2009MNRAS.395.1409S} had also discussed this method 
for adding extra ad hoc extinction to those progenitors for which foreground estimates only
where available.  
While \cite{2009MNRAS.395.1409S} had adopted an additional  $A_V \simeq 0.3$, 
 \cite{2011PhD_MFT} adopted an additional $A_V \simeq 0.53$. Hence there
are three estimates of \logl=4.5, 4.8 and 4.9  from 
\cite{2009MNRAS.395.1409S,2011PhD_MFT,2014MNRAS.438..938M}, each with an
uncertainty of  $\pm0.3$\,dex.  This is probably the most extreme case of a 
discrepant result due to different  methods, although formally, the uncertainties 
overlap.  The value of  \logl=$4.9\pm0.3$ is reported in Table\,\ref{tab:SNII}, and it 
serves to illustrate the point made earlier that results from improved analysis methods 
should be adopted for individual objects were possible. However
it will be shown that the luminosity shifts do not affect the important overall 
trend of a lack of high mass stars.

\subsubsection{SN2012aw}
\label{sec:12aw}
The second specific case study example is the progenitor of SN2012aw. 
This was a very well studied type II-P SN in M95 ($d=10$\,Mpc) discovered within 1-2 days of explosion
and followed well into the nebular phase  \citep{2013MNRAS.433.1871B,2014ApJ...787..139D}. 
\cite{2012ApJ...756..131V} and 
\cite{2012ApJ...759L..13F} both identified the same progenitor object as the culprit star with 
detections in HST filters $F555W$, $F814W$ and in  near infra-red ground based images
through filters $J$ and $K$. The galaxy was also imaged by the SINGS
\citep{2003PASP..115..928K}
Spitzer survey but no detection of a single point source was possible, as discussed in detail in 
\cite{2012ApJ...759...20K}. Both \cite{2012ApJ...759L..13F} 
and 
\cite{2012ApJ...756..131V} carried out a similar analysis and suggested the progenitor
star was quite luminous and relatively high mass in the range 14-26\msun\ and 15-20\msun\
respectively. They also suggested that the progenitor suffered significant reddening. 
Since the SN itself suffered low line of sight extinction, this was taken as evidence that 
the progenitor reddening was due to CSM dust which was destroyed in the explosion. 
Up to this point in the history of progenitor analysis, all treatments of reddening towards
progenitors had assumed that applying interstellar laws and relations were applicable 
also to circumstellar dust. However \cite{2012ApJ...759...20K} presented an alternative 
analysis in which they discussed the three consequences of such assumptions. 
Attempting to model CSM dust with ISM laws ignores dust emission in the near-IR, 
assumes (wrongly) that all scattered photons are lost to the observer and do not treat the (probable) dust composition correctly.  In the case of SN2012aw, 
\cite{2012ApJ...759...20K} argued that  \cite{2012ApJ...759L..13F} 
and 
\cite{2012ApJ...756..131V} had significantly overestimated the luminosity of the progenitor. 
Using the DUSTY code of \cite{1997MNRAS.287..799I} to compute how photons are scattered, absorbed and reemitted by 
a dusty region enclosing a stellar source, 
\cite{2012ApJ...759...20K} calculated a model to fit the detected magnitudes and the upper limits from the mid-IR
(see Fig\,\ref{fig:koch}). They suggested a significantly lower luminosity progenitor, 
a stellar temperature of $T_{\rm eff}\simeq 3600^{+300}_{-200}$\,K and 
an upper  mass limit of $M_{\rm ZAMS}<15$\msun. 
The amount of CSM gas, due to the progenitor's stellar wind was also constrained by  the x-ray and radio detections. The mass loss estimate of 
$\dot{M} \lesssim 10^{-5}$\,\msun\,yr$^{-1}$
is  was roughly consistent with a star of this temperature and luminosity. 

This progenitor is important as the discussion over the CSM reddening has implications for estimates 
of the most massive stars that can explode as bright supernovae, as we shall see in Section\,\ref{sec:discuss}.



\subsubsection{Upper luminosity limits }
\label{sec:IIsum}

\begin{table*}
\caption{Upper limits for the luminosity of type II and type IIP SNe. Most of the limits quoted here were re-calculated by Fraser (2011), based on the original published photometry limits  and with an additional, but adhoc extinction of $A_{V} \simeq 0.5^{\rm mag}$ applied as a conservative estimate. The values for the initial mass of star which ends it life with 
that luminosity is given in the last two columns. As in Table\,\ref{tab:SNII} these are for the STARS and Geneva models $M$(S,G) and the KEPLER models $M$(K), as discussed in Section\,\ref{sec:lumfunc}.}
\begin{center}
\begin{tabular}{lllll}
\hline\hline
SN & $\log L$  & Reference & $M$(S,G) & $M$(K) \\
\hline%
SN2006ov & $<4.7$  & From (1) ; $<$4.4 from (2)   &   $<10$  &  $<12$  \\
SN2004dg  & $<4.7$   & From (1) ; $<$4.5 from (2) &  $<10$    & $<12$   \\
SN2001du &  $<4.7$    & From (1) ; $<$4.7 from (2)  &   $<10$   & $<12$  \\
SN2006bc & $<4.9$     & From (1); $<$4.4 from (2)    &   $<13$  &  $<14$  \\
SN2007aa &  $<4.9$    & From (1) ; $<$4.6 from (2)   &  $<13$  &  $<14$  \\
SN1999gi &   $<4.9$    & From (1); $<$4.6 from (2)   &  $<13$    & $<14$  \\
SN2009N  &  $<4.9$      & From (1)                             &  $<13$   & $<14$  \\
SN1999br  &  $<5.0$    & From (1)  $<$4.9 from (2)     &  $<15$   & $<16$  \\
SN1999em &  $<5.0$    & From (2)                                &  $<15$   & $<16$  \\
SN2009ib  &  $<5.0$   &  From (3); same result in (1)   &  $<15$  & $<16$  \\
SN2002hh &  $<5.0$    & From (1); $<$5.0 from (2)     &  $<15$   & $<16$  \\
SN2009H  & $<5.1$    & From (1)                                 &  $<16$  & $<18$  \\
SN1999an &  $<5.2$  & From (1);  $<$5.2 from (2)   &     $<17$  &  $<20$  \\
\hline\hline
\end{tabular}
\end{center}
\label{tab:II-lims}
(1) \cite{2011PhD_MFT} ; (2)\cite{2009MNRAS.395.1409S};
(3)\cite{2015arXiv150402404T}
\end{table*}

While 
Table\ref{tab:SNII} lists the thirteen type II SN that have secure detections of their progenitors in the period
from 1999-2013 (inclusive), there are a further 13 type II SNe which have pre-explosion HST data (or 
alternative high quality ground based images) in which no progenitor is detected. In these
cases the progenitor falls below the detection limit of the images and no information on the 
colour and effective temperatures of the stars can be gained a priori. Any individual event
does not therefore add a lot of new information or insights compared to the large number of 
detections now available. However the statistical significance of the number of non-detections 
does provide a constraint on the mass and luminosity range of the progenitor population
as discussed in \cite{2009MNRAS.395.1409S} and in \cite{2008ApJ...684.1336K}. Since the 
Smartt et al. (2009) study, there have been 3 
more non-detections that can boost the statistical significance of the sample further. 
These are  SN2009H \citep{2009CBET.1656....1L},
 SN2009N \citep{2014MNRAS.438..368T} 
and SN2009ib 
\citep{2015arXiv150402404T}.
As discussed in  \cite{2009MNRAS.395.1409S} and previous papers
such as 
\cite{2003MNRAS.343..735S}
and
\cite{2003PASP..115....1V},
if one assumes that the progenitors of these type II (mostly II-P) SNe are red supergiants then a luminosity limit can be calculated for this 
effective temperature regime. An $I$-band like filter sampling the 
wavelength region between 7000-9000\AA\ is particularly powerful since the stellar spectral energy distribution peaks in this region and the bolometric correction change means the luminosity limit is roughly 
constant across the K to M-type supergiant colours and temperatures. 
The major uncertainty is the extinction to adopt toward the progenitor
and different methods have been proposed.

\begin{table*}
\caption{Type IIb detections}
\begin{center}
\begin{tabular}{llllll}
\hline\hline
SN & $T_{\rm eff}$ & err & $\log L$ & err & Comments  \\
\hline%
SN2008ax & 3.95 & 0.2 & 5.1 & 0.2 &   (1) ; See discussion in Sect.\ref{sec:IIb}  \\
SN2011dh & 3.78 & 0.01 & 4.9 & 0.2 &  From (2). Similar results in  (3) \\
SN2013df & 3.62 & 0.01 & 4.94 & 0.1 & From (4)\\
\hline\hline
\end{tabular}
\end{center}
\label{tab:IIb}
(1)\cite{2008MNRAS.391L...5C}; 
(2)\cite{2011ApJ...739L..37M}; 
(3)\cite{2011ApJ...741L..28V}; 
(4)\cite{2014AJ....147...37V};
\end{table*}

Initial work in this area simply adopted the estimated line of sight 
extinction toward the SN as appropriate to apply to the progenitor. 
The extinction could be estimated from the spectral slope 
of the SN,
\citep{2003MNRAS.343..735S}
 the absorption of Na\,{\sc i} lines
\citep{2011MNRAS.415L..81P,2012MNRAS.426.1465P}
or from three colour photometry of the surrounding stellar population
\citep{2014MNRAS.438..938M}.
There are problems with all of these estimates, but even more 
important is that this line of sight extinction may either not be directly
 applicable to the progenitor (as in the case of the surrounding
stellar population) or may be simply a lower limit since it does
not take into account circumstellar medium extinction in the locality
of the progenitor star (see the discussion in Section\,\ref{sec:12aw}). 
It is quite possible, and indeed probable
that CSM dust (within a radius of approximately 10-100$R_{\star}$, where $R_{\star}$ is a typical 
red supergiant
(RSG)  
radius of a 500-1000\rsun) is destroyed 
during the UV flash of the SN. The discovery of a dust enshrouded
progenitor, and a relatively unobscured transient for SN2008S
\cite{2008ApJ...681L...9P,2009MNRAS.398.1041B}
illustrated this possibility, followed by the case of SN2012aw (Section\,\ref{sec:12aw}).
The upper limits suffer much more uncertainty than the 
direct detections as in the latter cases the extinction can be constrained
by SED fits, at least for the cases with the largest wavelength coverage
in the broad band colours. Nevertheless, under reasonable assumptions, 
the limits are a useful constraint. Approaches have been to 
adopt the best estimate of reddening toward the SN, 
 to adopt  an arbitrary extinction to 
represent the CSM, or to model it from theoretical assumptions. 
\cite{2009MNRAS.395.1409S} 
added an additional $A_V \simeq 0.3$ for those progenitors
which had only Milky Way foreground estimates (the value 
of 0.3 coming from comparisons between the SN sample and
the red supergiant population of the LMC). 
\cite{2011PhD_MFT} reanalysed this sample, using a better method of 
model star SEDs (MARCs models) and adopting extra 
extinction of $A_V \simeq 0.5$ for all (again from comparing the 
M31, LMC and SMC populations of red supergiants). To be 
conservative the values for the luminosity limits in 
Table\,\ref{tab:II-lims} are quoted from 
\cite{2011PhD_MFT}
with this higher, ad hoc (but physically plausible) extra extinction of 
$A_V \simeq 0.5$.  The results of the theoretical modelling of 
stellar mass-loss rates and dust production of \cite{2012MNRAS.419.2054W} will be discussed
further in Section\,\ref{sec:explain}.

\begin{figure*}
\begin{center}
\includegraphics{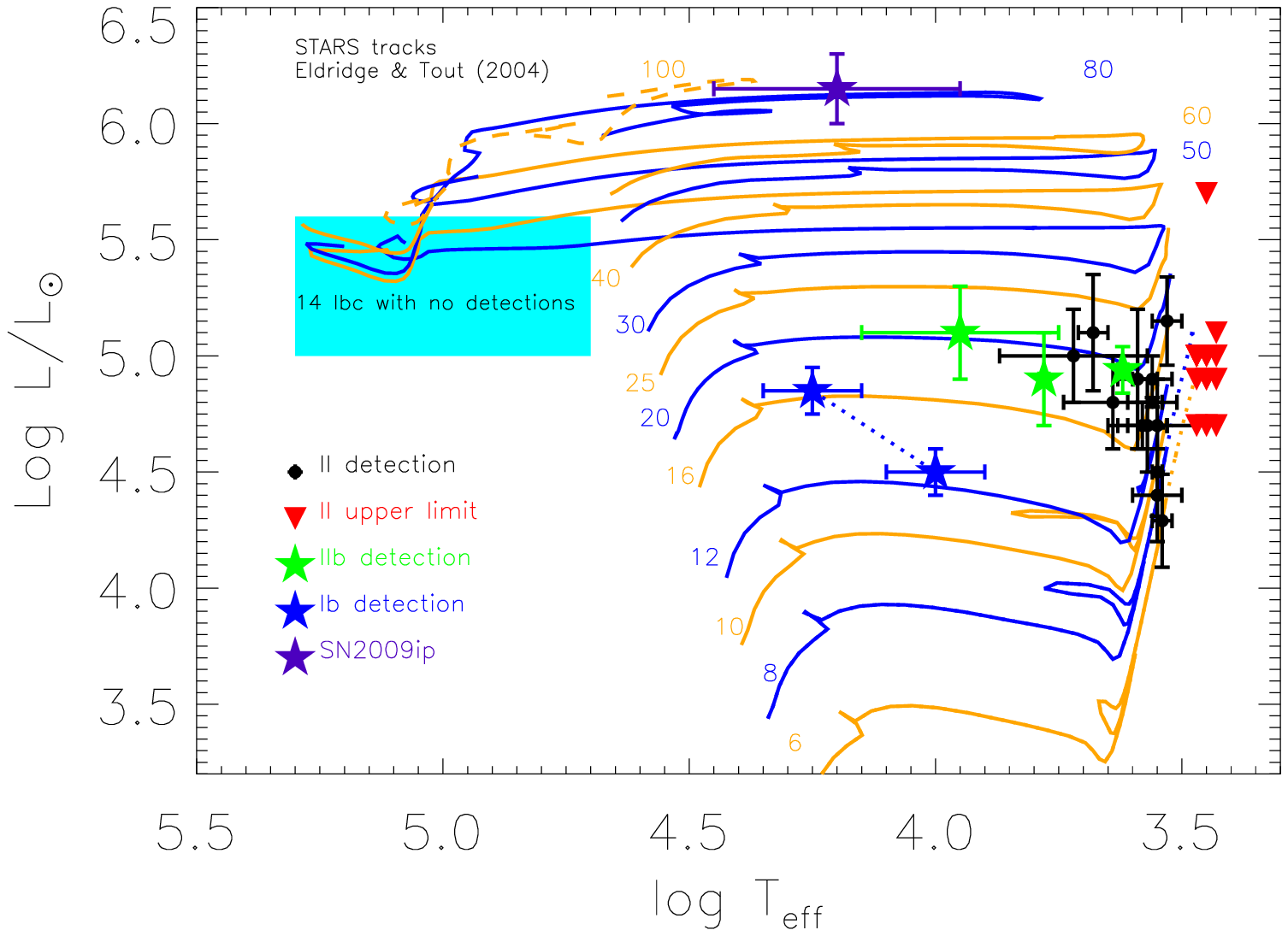}
\caption{The positions of the detected progenitors and upper limits to the type II SNe as discussed in 
Section\,\ref{sec:data}. The stellar evolutionary tracks are from 
\cite{2004MNRAS.353...87E}. 
The possible positions of the progenitor star of PTF13bvn is marked with two symbols, 
joined by the dotted line. These show the two positions of the progenitors proposed by 
\cite{2014AJ....148...68B}
and 
\cite{2015MNRAS.446.2689E}  in their binary models. The position of the progenitor of 
SN2009ip is shown as the magenta symbol, as estimated from the faintest magnitude the 
LBV star was found at (see Section\,2.4 for more details).  The 
14 Ibc progenitors with no detections are not quantitatively marked here. If they were
WR stars then one would expect to find them around the blue shaded area
\citep[although the box position is illustrative as some models predict progenitors outside this locus, e.g.][]{2013A&A...558L...1G,2013A&A...558A.131G}
There are 30 progenitors below $\log L = 5.1$, and only one (SN2009ip) above, if
indeed SN2009ip is a genuine core-collapse supernova. }
\label{fig:hrd}
\end{center}
\end{figure*}

\subsection{Type IIb supernovae}
\label{sec:IIb}

As discussed in the previous two sections, the pre-discovery images of type II-P SNe are often not deep 
enough to detect the progenitor stars. This is not surprising since for distance moduli of $\mu=30 - 32$  (10-25\,Mpc)
and the typical depth of HST images ($m_{AB}\sim 25^{m}$), the lowest luminosity red supergiants would 
go undetected. If the broad picture of core-collapse, neutron star formation and successful SNe occurring in stars above an initial mass of $\sim$8\msun is true, then the initial mass function {\em must} dictate that the bulk of CCSNe we 
detect are from the lower mass region. The IIb SNe are those which begin by resembling type II SNe with unmistakable
H\,{\sc i} Balmer lines and evolve to have H-weak and He-strong spectra 
\citep{1997ARA&A..35..309F}, 
the prototype of which is SN1993J
\citep{1993ApJ...415L.103F,2000AJ....120.1499M}. 
It's  remarkable that the three IIb SNe in the 1999-2013 distance and time limited survey which have high quality imaging {\em all} have relatively  bright progenitors
(or progenitor systems) detected at the SN position. 
The detections of the progenitors of SN2008ax \cite{2008MNRAS.391L...5C}; 
SN2011dh \citep{2011ApJ...739L..37M,2011ApJ...741L..28V}; and 
SN2013df  \citep{2014AJ....147...37V}
are summarised in Table\,\ref{tab:IIb}.  The spectral energy distribution of the progenitor of SN2011dh is 
very well fit with a spectrum of a single star with $T_{\rm eff}=6000$\,K and  $\log L=4.9$\lsun\ 
as shown in Figure\,\ref{fig:11dh}. The progenitor of SN2013df is cooler, with 
 \cite{2014AJ....147...37V} finding $T_{\rm eff} = 4250$\,K and remarking that the detected progenitor appears
quite similar to that of SN1993J
\citep{1994AJ....107..662A}. 
The progenitor of SN1993J was postulated, and shown to be, a binary system of a K-type supergiant and 
hotter companion. It too had a luminosity of $\log L=5.2\pm0.3$\lsun\  
\citep{2002PASP..114.1322V,2004Natur.427..129M,2014ApJ...790...17F}. 

The colours of the progenitor of SN2008ax could not be satisfactorily fit with a single stellar
SED 
\citep{2008MNRAS.391L...5C}
or with a plausible binary which led the authors to suggest it might be a single WR-type star
or a system which has contaminated flux from a physically associated cluster or nearby stars. 
Crockett et al. found the absolute magnitude of the star coincident with the SN position to be 
$M_{\rm F606W}=-7.4\pm0.3$, and if we assume a bolometric correction of $-0.6$ (to cover possible 
supergiant stars from late B-type to early K-type) then this implies a stellar  $\log L\sim5.1$\lsun\
but with quite large uncertainties due to the difficulty in fitting a single SED. 
\cite{2013A&A...550L...7G} suggest that the progenitor could have been a rotating star 
of around 20\msun\ that ends its nuclear burning life as a ``compact LBV" with a luminosity of 
around $\log L=5.2 - 5.3$\,dex. 
Nevertheless, taken together the IIb progenitors tend to be brighter, more often detected, 
and of systematically higher luminosity than the bulk of the type II-P SNe. 

The model of IIb SNe coming from extended supergiants which have had mass stripped
due to a binary companion was supported initially by the detection of a progenitor companion
in SN1993J
\citep{1993Natur.364..509P,1993Natur.364..507N,1994ApJ...429..300W,2004Natur.427..129M,2014ApJ...790...17F}
and recently also for SN2011dh 
At the time of discovery of SN2011dh, a yellow star  
dominated the SED of the progenitor system and the community had 
split views on whether the yellow supergiant  or its putative compact 
companion had exploded
\citep[e.g.][]{2012ApJ...752...78S}.
The disappearance of the bright yellow star \citep{2013ApJ...772L..32V}
and the detection of 
a UV source remaining at the SN position by 
\cite{2014ApJ...793L..22F}
is strong evidence that it was indeed the 
yellow supergiant that exploded and that it had a more compact companion
\cite[see the model in][]{2013ApJ...762...74B}. 

It remains to be seen if the existing, and follow-up data for SN2013df and SN2008ax also can be explained 
with binary models. However it is certain that the progenitors all have luminosities around 
 $\log L\simeq5.0$\lsun, albeit with the assumption that the bolometric correction for the SN2008ax progenitor is 
not unusually high. The stellar evolutionary models employed in the studies discussed in this section to explain the progenitor luminosities have  ZAMS masses in the range 13-17\msun. It would appear that IIb progenitors 
are consistently among the highest luminosity (and hence highest mass) progenitors
so far detected.  There are 
few other upper limits, and with four cases of nearby IIb SNe all with bright, detected progenitors it seems  reasonable
to conclude that they are not produced by lower mass systems in the 8-12\msun\ range. 
 The physical reason for this is not determined and it will be interesting in the future to test
quantitative models.  One may speculate that it could be due to higher mass stars forming closer binaries more easily or 
brighter progenitors being more extended and hence their envelopes more efficiently undergoing 
Roche lobe overflow when they reach the L1 point. 

 A spectrum of the light echo of Cas A showed this Milky Way  supernova to be of type IIb
\citep{2008Natur.456..617K,2008ApJ...681L..81R}.
There is no sign of an obvious companion star in the centre of the Cas A remnant, leading
to the speculation that this was a high mass single star that lost most of its hydrogen envelope
through winds. However the total amount of gas either ejected or residing in the CSM, and therefore
recently ejected by the progenitor is not particularly high. The mass of  the shocked and unshocked ejecta is estimated at $3-4$\msun\
\citep{2012ApJ...746..130H}
and the mass in the CSM which the blastwave has swept up is around 8\msun\
\citep{2009ApJ...703..883H,2009ApJ...697..535P}. 
Adding in $\sim2$\msun\ for a neutron  star remnant, results in a total progenitor mass
estimate of about 15\msun.  The internal structure seen in the ejecta  originate 
from turbulent mixing processes and plumes of $^{56}$Ni-rich material
\citep{2015Sci...347..526M}.
These detailed constraints are in reasonable agreement with the the results from the 
three IIb SNe with progenitors discussed above - originating from stars with initial masses
in the range $\sim15\pm3$\msun. However the puzzle remains for Cas A - if the star had no companion,
and the initial mass was only $\sim$15\msun, then how did it lose its envelope to become a IIb SN? One might speculate it was a high mass, obscured star, although it seems the 
extinction toward Cas A is no more than $A_{V}\simeq6-7^{\rm mag}$ \citep{2009ApJ...697...29E} and this is integrated along the Galactic line of sight rather than being local to the SN and the progenitor.

\subsection{Type Ibc supernovae}
\label{sec:Ibc}

There are two possible models for the progenitors of 
type Ibc supernovae. The first are Wolf-Rayet stars 
which are evolved, single massive stars that have lost their
hydrogen envelopes (in the case of WN stars) and 
also helium layers (in the case of WC and WO stars) 
primarily through radiatively driven winds, 
\citep{2000ARA&A..38..613K,2005A&A...442..587V}
or episodic mass-loss
\citep{2006ApJ...645L..45S} or 
through rapid rotation and chemically homogenous evolution 
\citep{2005A&A...443..643Y}. 
The observed  Wolf-Rayet stars in our galaxy and the Magellanic clouds 
are observationally constrained to come from stars (either single or binary) above 
about 25-30\msun\ from considerations of the turn off masses in stellar clusters
\citep[e.g. see review of][]{2007ARA&A..45..177C}. 
From a theoretical standpoint, the stellar winds of single stars below this mass 
are not strong enough alone to cause sufficient mass loss
to expose the He and CO cores. These stars are predominantly 
found in young stellar clusters and OB associations within the 
Local Group and turn-off masses for coeval stellar populations
have provided estimates of mass and age
\citep[e.g.][]{1995ApJ...438..188M}. 
The second alternative is lower mass stars in binary systems
with initial masses lower than that required for single stars
\citep{1967AcA....17..355P,1998A&ARv...9...63V,1995PhR...256..173N,1992ApJ...391..246P,2008MNRAS.384.1109E,2010ApJ...725..940Y}
Theoretical stellar populations including binaries 
with reasonable distributions of stellar masses and separations
can produce large number fractions of He stars with core 
masses large enough to undergo core-collapse
\cite[e.g.][]{2013MNRAS.436..774E}. However one problem with 
these models is that the systems they predict are not observationally identified in the Milky Way.

\begin{figure}
\begin{center}
\includegraphics[width=\columnwidth]{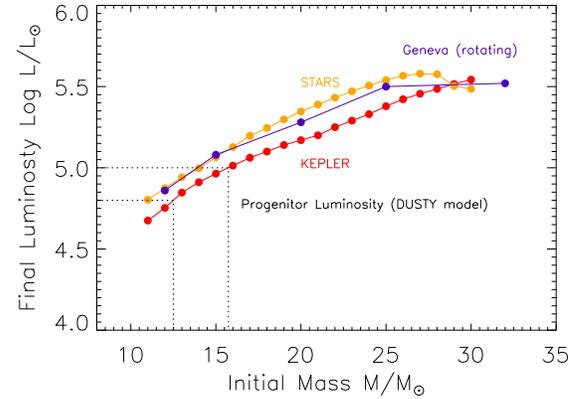}
\caption{
A comparison of the final pre-SN luminosity as a function of stellar mass, from three sets of stellar evolution models, discussed in the text
 \citep{2014MNRAS.439.3694J}. The STARS and rotating Geneva models have 
very similar core masses throughout the mass range, whereas the KEPLER models are $0.1-0.2$\,dex 
less luminous at the same  stellar mass. This would corresponds to difference in estimated progenitor mass of 2-3\msun\
if the final luminosity is used as an initial mass tracer. The progenitor limis for the "DUSTY" model are those 
for SN2012aw, from \cite{} and are discussed further in Section\,\ref{sec:nucleo}.
Reproduced from Figure 5, in
{\em ``The nebular spectra of SN 2012aw and constraints on stellar nucleosynthesis from oxygen emission lines''}, A. Jerkstrand et al, 2014, Monthly Notices of the Royal Astronomical Society, 439, 3649. 
}
\label{fig:endpoints}
\end{center}
\end{figure}

As discussed in Section\,\ref{sec:data}, the volume and time limited survey of 
\cite{2013MNRAS.436..774E} presented an extensive search for the progenitors of all type Ibc SNe in pre-discovery
images (from 1998 to 2012.25). This paper described the compilation of literature data, and new analysis of images
of the progenitors of  a total of 12 type Ibc SNe. There are no detections of progenitors, or progenitor systems in 
any of these. The deepest limits in the typical $BVR$-bands were between $-4$ and $-5$ and, 
the authors compared these limits with the observed magnitudes of WR stars in the Large Magellanic Cloud. 
As WR stars, or stripped He stars in binary systems, have diverse temperatures and radii (and hence 
diverse colours, SEDs and bolometric corrections) the quantitative fitting methods which have been 
successful for type II progenitors cannot be applied to the Ibc limits without major uncertainties in the results. 
Hence a more empirical approach was applied which simply postulated that the observed WR star population of 
the LMC are plausible progenitor systems of Ibc SNe and then determined what was the statistical
uncertainty that no detections were made. Since then, there have been two further Ibc SNe within the 
distance limit, also with no detections of progenitors: SN2013dk 
\citep{2013MNRAS.436L.109E}
and  SN2012cw 
\citep{2012ATel.4535....1G}. 
\cite{2013MNRAS.436..774E} originally estimated the 
probability of not finding a progenitor system detection, 
{\em if the LMC WR population are direct progenitors of Ibc SNe} to be 16\%, and with the inclusion of the limits of SN2013dk and SN2012cw this drops to $\sim$12\%. This is an interesting constraint but not in itself strong enough to rule out WR stars
being the progenitors of the normal Ibc population we see
in the local Universe. Several other authors have argued
that the temperature and luminosity (which of course dictate
optical and NIR fluxes) of the observed WR population are not
applicable to WR stars at the point of collapse and that 
significant evolution to higher temperatures and 
fainter optical magnitudes occur in their models
\citep{2012A&A...544L..11Y,2013A&A...558A.131G}. 
These
of course are dependent on the mass loss rates and 
the time to core-collapse from where we observe them now. 
It is also possible that the extinction toward these progenitors, 
or CSM extinction, is systematically underestimated
as discussed above for the type II-P progenitors and limits
(in Section\,\ref{sec:IIsum}). 

There are two other strong constraints on the progenitors of 
type Ibc SNe. The first is their relatively high rate which has been 
explained with binary population models  \citep[for example as discussed in][]{1992ApJ...391..246P,2008MNRAS.384.1109E}.
More recently \cite{2011MNRAS.412.1522S} pointed out that the 
the SN rates from the volume limited 
Lick Observatory Supernova Search 
(LOSS)  imply a rate (26\% of all CC) that is a factor 
of $\sim$2 too high for massive, single WR stars to be the sole 
producers just from simple IMF arguments. 
An IMF of single stars would produce a WR fraction
(and hence a Ibc fraction) of 
14-18\% of all massive stars between 8-100\msun\ if
the minimum mass to produce a WR is 25-30\msun. 
\cite{2013MNRAS.436..774E} built on this with 
the Binary Population and Spectral Synthesis (BPASS) code
to calculate rates of progenitors in stellar population models
including binaries. The rate of Ibc SNe  within 
28\,Mpc (during 1998-2012.25) is also 26\%, and 
\cite{2013MNRAS.436..774E} find that the rates can 
only be produced if a  mixed population of single stars
and binaries are used with one single star for every binary system and the mass ratio and initial separations are set such that approximately two thirds of the binaries 
interact. This value is compatible with the observed result of
\cite{2012Sci...337..444S} 
who showed that this fraction of massive binaries are likely to interact. These multiple and consistent
observational
constraints from relative SN rates and the binary fraction
of massive stars are rather compelling arguments that 
binary systems are likely to produce the bulk, if not all
of the Ibc SN progenitors. 
\cite{2013MNRAS.436..774E}  points out that the reason that
Galactic analogues may be difficult to identify  is that stars in binaries with initial masses below 20\msun\
tend to retain a low-mass hydrogen envelope until 
quite close to core-collapse, $\simeq 10^{4}$\,yrs. Hence 
they are not easily detectable as H-free compact helium 
stars. 

The second additional argument for Ibc SNe coming from
interacting binaries is that the ejecta masses calculated 
from lightcurve modelling are typically in the 
range 1-4\msun\ 
\citep[see][the results of which are summarised in Eldridge et al. 2013]{2008MNRAS.383.1485V,2011ApJ...741...97D}. 
These final masses are somewhat low 
compared to the estimated masses of WR stars 
in the Local Group, which are typically in the range
8-20\msun\ 
\citep[from binary orbits, Fig. 4 of][]{2007ARA&A..45..177C}. 
The binary models of Eldridge et al. (2013)
produce helium stars that would have typical 
ejecta masses of $4.2\pm2.4$\msun, in agreement with those
measured \citep[also see][]{2014arXiv1406.3667L}. 

\cite{2012MNRAS.424.1372A} and 
\cite{2008MNRAS.390.1527A} have matched  the positions of 
nearby SNe to the H$\alpha$ emission produced in starforming regions. 
The SNe Ibc in their sample are more closely associated with the
H$\alpha$ emission than type II SNe in general. They interpret this as 
the SNe Ibc (and SN Ic in particular) arising from younger regions and 
hence coming from higher mass progenitors. 
\cite{2013MNRAS.428.1927C} compared a much closer sample 
of CCSNe ($\leq$15\,Mpc) with higher resolution images
and also concluded that the association of Ibc with H$\alpha$ regions
was stronger than for type II SNe. However the diagnostic power of 
this association in quantitatively determining the mass ranges  was shown
to be difficult to interpret. The Ibc progenitors do appear to come from 
younger populations than the bulk of the type II progenitors. However 
the latter are dominated by stars in the 8-12\msun\ regime, and a 
quantitative age and mass estimate for the Ibc progenitors has not yet
been possible. Although they may on the whole be from more massive 
progenitors, the data can't at present distinguish between WR stars
and lower mass (say 12-20\msun) binaries.

\subsubsection{PTF13bvn}
\label{sec:13bvn}
Just after the publication of the Eldridge et al. (2013) summary,
the Palomar Transient Factory discovered a nearby Ib SN in 
NGC5806 ($d=22$\,Mpc) and set a very restrictive time on the explosion  epoch
\citep{2013ApJ...775L...7C}.
The non-detections indicate that the first detection
was within 24hrs of the shock breakout. 
The HST pre-explosion images of the site indicate 
a blue star at the position of the SN. 
\cite{2013ApJ...775L...7C} 
noted that their alignment of a high resolution ground-based
image of the SN was close to, but not formally within the 
1$\sigma$ error circle of the alignment calculation. 
However one cannot confidently reject  a physical association
unless there is a 3$\sigma$ difference and the $\sigma$
is correctly defined and measured. Subsequent alignment
in 
\cite{2015MNRAS.446.2689E}
with HST imaging again found a similar result but supported
the suggestion by 
\cite{2013ApJ...775L...7C} 
that this was likely the progenitor system and the 
first detection of a stellar progenitor of a Ibc SN. 
\cite{2013ApJ...775L...7C} 
and 
\cite{2013A&A...558L...1G}
proposed that the broad band magnitudes of the 
source in the HST images were similar to massive WR stars
and a stellar model of initially $M_{\rm ZAMS}\simeq30$\msun\
evolved into a WN star with model parameters 
$\log L/{\rm L_{odot}}=5.6$ and $T_{\rm eff} = 46$kK (producing  
a final CO star mass of 10\msun). A reassessment of the HST 
magnitudes of the progenitor led 
\cite{2015MNRAS.446.2689E}
to argue that a binary system could explain the updated 
stellar magnitudes better. They found brighter magnitudes  by about 0.5$^{m}$ ; 
$M_{\rm F435W}=-6.0\pm0.4$, 
$M_{\rm F555W} =-6.0\pm0.4$, 
$M_{\rm F814W} = -5.9\pm0.4$. These updated magnitudes do not
rule out the single WR models of \cite{2013A&A...558L...1G}, but the
model binary systems reproduce the measurements quite well.

Both 
\cite{2014AJ....148...68B}
and 
\cite{2015MNRAS.446.2689E}
presented binary models as alternatives which have initial masses of 
either $20+19$\msun\ or $10+8$\msun\ components. 
These produce a stripped He-core with a final mass of 
$M_{final}=3-4$\msun\ and the combined flux from
the binary systems can reproduce the progenitor magnitudes. 
The ejecta mass would then be of order 2\msun\ in
agreement with the estimate of  
\cite{2014A&A...565A.114F}
from lightcurve modelling. 
While a single, massive WR star cannot be definitively ruled
out, the progenitor magnitudes and ejecta mass estimates
are quite well reproduced with a binary system (see Section\,\ref{sec:lumfunc} for a position on the Hertzsprung Russell Diagram). 

In a recent development in this area, 
\cite{2014Natur.509..471G} took a spectrum of 
SN2013cu within the first day after explosion showing that it has 
spectral features similar to those seen in WR stars (specifically 
H\,{\sc i}, He\,{\sc i}, He\,{\sc ii}, C\,{\sc iv} and N\,{\sc iv} seen in WN type stars). 
While this could be interpreted as evidence for a direct WR progenitor star, 
\cite{2014Natur.509..471G} note that the inferred wind density and mass-loss rate 
are rather extreme when compared to typical WR stars. 
\cite{2014A&A...572L..11G} applied a more detailed analysis code to suggest that
a WR progenitor star, with a classical radiatively driven wind, is not likely and the precursor
was possibly an LBV-type or yellow hypergiant with a recent eruptive
phase. SN2013cu was a type IIb and the mass-loss event that produced
 the circumstellar
mass illuminated in the wind could have plausibly occurred during mass-transfer in a binary  or in an eruptive phase. The remarkable spectrum is  similar to Galactic
WR star spectra, although the emitting surface is likely a factor $\sim$10 larger than 
the typical WR stellar radii and the the narrow cores of the emission lines seen in SN2013cu are not typical of
fast WR wind lines.   
As \cite{2015MNRAS.449.1876S} and \cite{2014A&A...572L..11G}
point out, the spectrum of the early SN is not a direct indicator of the spectral type of the progenitor
star before explosion, but it is a powerful measure of the recent mass-loss history of the progenitor. 
It shows that the CSM material within $\sim10$\,AU  is similar
in H and He composition to WN star winds, rather than the star definitely being a WR-type progenitor \citep{2014A&A...572L..11G,2014arXiv1408.1404S,
2015MNRAS.449.1876S}.

\subsection{Type IIn progenitors : LBVs  and SN2009ip}
\label{sec:IIn}

There are three SNe which have been proposed as core-collapse SNe of
type IIn with either identified high luminosity stellar progenitors
detected or reasonably strong arguments in favour of high-mass
progenitors \citep[other work has suggested LBV-like progenitors from
modulations in radio lightcurves or in optical spectral
features][]{2006A&A...460L...5K, 2008A&A...483L..47T}.  The first
discovered was the progenitor of SN2005gl which was a bright $M_{V}
\simeq -10$ progenitor detected at the surprisingly large distance of
60\,Mpc \citep{2007ApJ...656..372G,2009Natur.458..865G}

The type IIn SN2010jl has  HST pre-explosion imaging
of its host galaxy (at 50\,Mpc) and 
\cite{2011ApJ...732...63S}
identified the site of the explosion. At this distance the relatively
crowded region of the explosion site prevents a confident 
detection of a resolved point source, but the blue flux 
coincident with the SN is likely to be either a very young 
cluster or a single massive star. 
\cite{2011ApJ...732...63S}
argued that either of these two cases suggest that the 
progenitor had an initial mass higher than $M_{\rm ZAMS}>30$\msun. Both of these SNe occurred within the
time limit of the survey defined by 
\cite{2009MNRAS.395.1409S}
and 
\cite{2013MNRAS.436..774E} but are much more distant than 
the 28\,Mpc limit. 

The now infamous SN2009ip is a type IIn SN 
\citep{2013MNRAS.430.1801M}
with a 
progenitor detected in outburst over many years. 
After the initial discovery of a
non-terminal, but luminous ($M\sim -14$) outburst
in 2009 \citep{2009CBET.1928....1M}
at which point it was named SN2009ip, 
a luminous progenitor star ($M_{V}= -10$) was detected in HST 
archival imaging taken in 1999 
\citep{2010AJ....139.1451S,2011ApJ...732...32F}. 
Subsequent monitoring over the next three years, 
mostly by 
\cite{2013ApJ...767....1P} illustrated that the 
explosion in SN2009ip was part of a regular, but 
temporally sporadic, series of outbursts of a massive star
which photometrically shared similarities with the 
Luminous Blue Variables (LBVs) of the Local Group. 
Indeed it is not even clear if the detection in the
HST imaging is the star in outburst or quiescence
and hence attaching a bolometric luminosity to any of the 
data points is not easy to interpret with hydrostatic stellar
evolutionary models.  It may be that the luminosity is powered at all 
phases, including that in the archival HST images, by interaction 
(and the same could be true for SN2005gl). 

In 2012, the star had a 30 day outburst that 
peaked at a luminosity similar to the brightest 
known LBV giant outbursts 
(the``2012a" event ; $M_V \simeq -15$, $\log L =  41.5$\,erg\,s$^{-1}$), 
which was followed by a 10 day rise to SN luminosities 
(the``2012b" event; $M_V \simeq -18$, $\log L =  43$\,erg\,s$^{-1}$)
\cite[monitored by][]{2013ApJ...767....1P,2013MNRAS.430.1801M,2013ApJ...763L..27P}. 
The cause and nature of the two are still debated. 
\cite{2014MNRAS.438.1191S}
fervently argue that core-collapse must have occurred
to cause the bright 2012b peak, and indeed many observational 
characteristics of this explosion are similar or identical 
to type IIn supernovae. 
\cite{2013MNRAS.430.1801M} and 
\cite{2014MNRAS.438.1191S}
propose that the fainter ``2012a'' event was a non-terminal 
giant eruption, that core-collapse caused gas to be accelerated 
to velocities of $>10,000$\kms\ and 
the kinetic energy of this gas is efficiently converted to 
radiative energy in colliding with a dense, 
but possibly asymmetric CSM to produce a bright SN. 
Both \cite{2013ApJ...767....1P} and 
\cite{2013MNRAS.433.1312F}
argued that it is still possible that the last 
SN-like event in the history of SN2009ip was not due to 
core-collapse, but caused by the collisions of massive shells
ejected from the star with dense a CSM. The energetics
are plausible, although are disputed by 
\cite{2014MNRAS.438.1191S}. Other papers on the topic
\cite{2014ApJ...787..163G,2014ApJ...780...21M}
have also presented extensive data and analysis 
 but none as yet have 
definitively proven the true nature. 
\cite{2014ApJ...789..104O} 
have suggested that 
at least 50\% of type IIn SNe have 
outbursts within the last few years 
of their explosion. 
 This seems plausible 
given the Palomar Transient Factory (PTF) discovery
rate, the other discoveries of 
pre-explosion outbursts 
\citep{2007Natur.447..829P,2013ApJ...779L...8F} and the theoretical 
ideas of hydrodynamic instabilities discussed in  \cite{2014ApJ...785...82S}. 
Whatever 
mechanism powers the bulk of IIn SNe 
it appears clear that eruptions in the few years before explosion are much 
more frequent that previously thought. 

For the purposes of this review, the interesting question
concerns the possibility that very massive stars 
produce successful SNe through the core-collapse 
mechanism. The faintest magnitude that the progenitor
star of SN2009ip was observed at was $M_{V} = -10$, 
and with a bolometric correction appropriate for a 
blue supergiant ($-1.0 \lesssim BC \lesssim -0.5$), this equates to 
$\log L/{\rm L_{\odot}}$ = 6.0 - 6.3\,dex, or an initial mass of 
60-80\msun.  The discussion that follows,  will consider 
the possibility that SN2009ip has undergone a core-collapse and
successful  explosion. 

Although it has been reasonable to link supernovae of types IIn to LBVs or very 
massive stars that have had violent recent mass-loss history, an alternative 
explanation was recently put forward by \cite{2014Natur.512..282M}. In this 
model, a red supergiant with a fairly standard mass-loss rate (such as Betelgeuse with
$\dot{M}=1.2\times10^{-6}$\,\msun\,yr$^{-1}$) produces a static shell of circumstellar 
material through pressure from photoionization by radiation from external sources. 
The steady wind gets confined into a static  photoionization-confined shell
which could contain up to 35 per cent of all mass lost during the red supergiant phase. 
This  gas shell is confined close to the star, giving a possibility of a type IIn supernova
resulting from a moderate mass red supergiant with standard mass-loss.

\section{DISCUSSION}
\label{sec:discuss}

\subsection{The luminosity distribution of progenitors}
\label{sec:lumfunc}

A summary of the detected progenitors and their limits are illustrated in a Hertzsprung Russell Diagram in 
Figure\,\ref{fig:hrd}, along with a set of theoretical stellar evolutionary models of \cite{2004MNRAS.353...87E}. 
This represents a complete summary of the progenitors detected (and limits set) within the time and volume limit 
of \cite{2009MNRAS.395.1409S} and \cite{2013MNRAS.436..774E} (extended to all data published up to end of 2013) and hence the relative numbers 
and distribution of the stars in this plot are meaningful. The striking result is that there is a 
deficit of progenitors above an estimated luminosity of \logl$\simeq5.1$. 
There are 30 progenitors (type II detections, II limits, IIb and Ib detections) below this luminosity limit. 
The final luminosity of a supernova progenitor is determined by the He core mass and luminosity, 
and this luminosity  equates to a slightly different mass depending on the stellar 
evolutionary models used. 
This \logl$\simeq5.1$ equates to 
16\msun for the STARS models of  
\cite{2004MNRAS.353...87E} ; 
16\msun\ for the rotating Geneva models of  \cite{2004A&A...425..649H} ; 
18\msun\ for the \cite{2007PhR...442..269W} (see Figure\,\ref{fig:endpoints})
and other stellar evolutionary codes produce model progenitors that are 
broadly in this range.

For a Salpeter IMF with $\alpha =-2.35$, 70\% of 
stars between 8-100\msun\ are in the 8-18\msun\
range. Hence with this number of stars  (30) lying below 
the mass limit of 18\msun\ one would expect to 
have found 13 higher mass progenitors with 
\logl$>5.1$\,dex and $M_{\rm ZAMS} > 18$\msun
\footnote{Or alternatively one could propose that these 30 objects
are the total sample and actually do contain 
the full range of masses. This would mean 
around 9 stars of this sample would need to have 
had their luminosities significantly underestimated. }
There is possibly one (SN2009ip; Section\,\ref{sec:IIn}), 
but the Poissonian probability of finding between 0 and 1, if the expectation value is 13 would be $p = 3\times10^{-5}$. 

The number 13 is a very close match to the number
of Ibc progenitors for which we have upper limits and one 
detection. One could argue that the bulk of stars above 
18\msun\ evolve into  WR stars and evade detection due
to them being too hot and faint at the point of core-collapse. 
And the numbers would match (almost too well).
This is not strictly ruled out by any of the 
observational constraints to date but it appears to be unlikely. 
The three constraints of having no detected progenitors, 
the high relative rate of Ibc SNe and the low ejecta masses
estimated from lightcurve fitting (presented in Section\,\ref{sec:Ibc})
all suggest that the binary channel dominates the production of 
type Ibc SNe. 

The missing high mass stars clearly exist in substantial numbers
in starforming regions and we observe them following a Salpeter 
(or similar) initial mass function. Therefore the question is, why are they not detected as SN progenitors ? It does not appear likely that  they are all WR stars that have evaded detection, 
given the arguments above. One argument put forward
is a bias in detecting high mass stellar progenitors as they may exist in 
denser, younger starforming regions - dense clusters or associations. Thus the 
surface brightness of these regions could 
make resolving individual stars more 
difficult. 
However there are only three known SNe 
within the distance and time limit that fall on compact
clusters. Two of them SN2004am and SN2004dj are relatively old clusters and 
very unlikely to host stars of $M_{\rm ZAMS} > 20$\msun, 
\citep{2013MNRAS.431.2050M,2004ApJ...615L.113M}
and the other is 
SN1999ev which is also unlikely to be a very young cluster \citep{2014MNRAS.438..938M}. 
It is also unlikely that these missing high mass progenitors stars produce normal luminosity SNe that 
are systematically missed by nearby surveys because of surface brightness
effects - since SN2004am was a relatively faint and reddened 
SN, recovered on top of a 
bright super star cluster in M82  
\citep{2004IAUC.8297....2S,2013MNRAS.431.2050M}. And 
SN2004dj was also easily detected in amateur surveys
\citep{2004CBET...74....1N}. 
It would appear that the SN population we observe in the local Universe are not being produced, 
in large proportions, by stars with \logl$>5.1$ ($M_{\rm ZAMS} \gtrsim 18$\msun).

\subsection{The mass function of progenitors}
\label{sec:massfunc}
\cite{2009MNRAS.395.1409S} calculated the mass function of progenitors of type II SNe, 
assuming that there are no major biases in selecting the SNe with progenitor 
information from the broader local SN population. This paper 
used a consistent approach, employing the STARS models to attach an initial stellar mass to 
the luminosities and luminosity limits determined.  They defined the ``red supergiant problem" which articulates the fact that while red supergiants are commonly found as direct progenitors of II-P SNe,  there were no progenitors found above an estimated initial mass of 
around 17\msun, or \logl$\simeq5.1$. 
As discussed in 
\cite{2009AJ....137.4744L}, 
red supergiants are found up to 
luminosities of  \logl$\simeq5.5$.

\begin{figure}
\begin{center}
\includegraphics[width=\columnwidth]{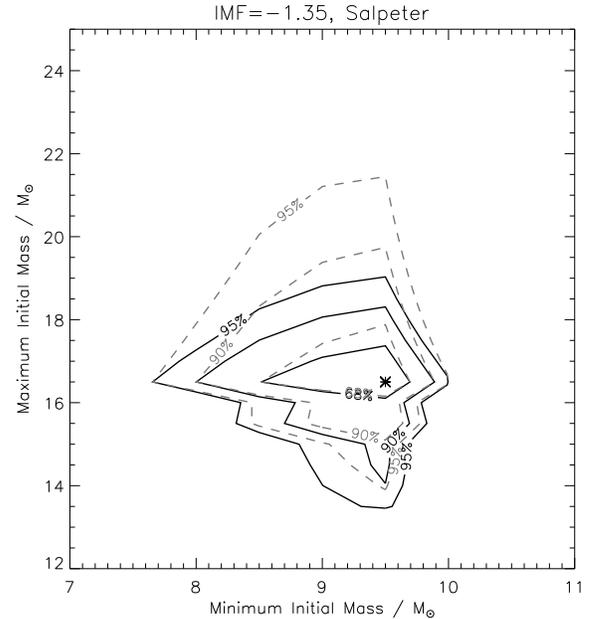}
\caption{The maximum likelihood of the minimum and maximum initial masses of the type II progenitor distribution, assuming the stars follow a Salpeter IMF mass function. Originally calculated in in Smartt et al. (2009), and reproduced here with the updated and extended masses in this review. The dashed lines show the confidence contours (68, 90 and 95\%) for the detections only and the solid lines show the confidence contours for the detections and upper limits
combined.  The star symbol marks the best fit, as described in 
Section\,\ref{sec:massfunc}, of $m_{\rm min} = 9.5^{+0.5}_{-2}$ and $m_{\rm max} = 16.5^{+2.5}_{-2.5}$. This is for the masses from the STARS and Geneva rotating models, the values for the 
KEPLER masses are given in the text. }
\label{fig:maxlike}
\end{center}
\end{figure}

\begin{figure}
\begin{center}
\includegraphics[width=\columnwidth]{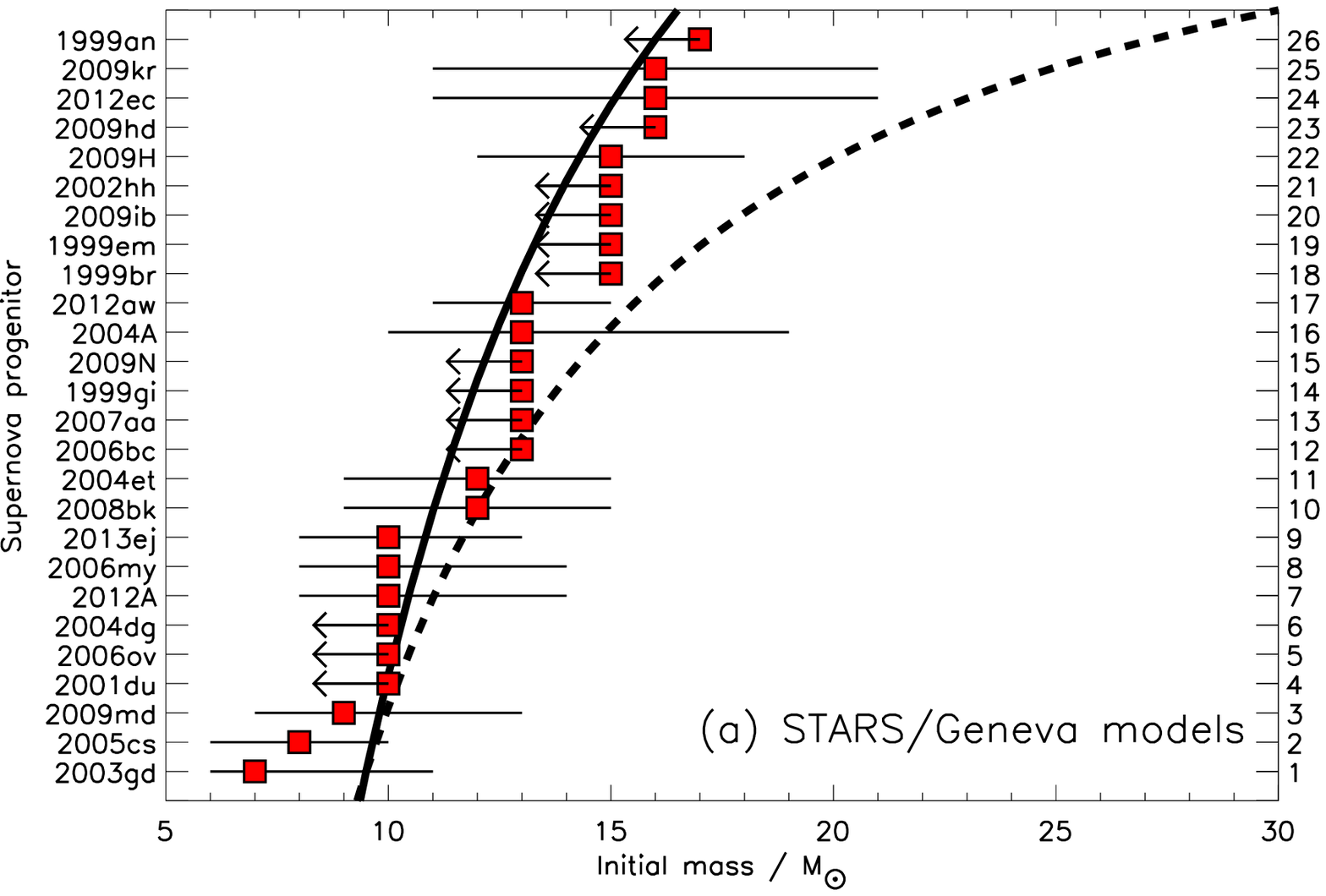}
\includegraphics[width=\columnwidth]{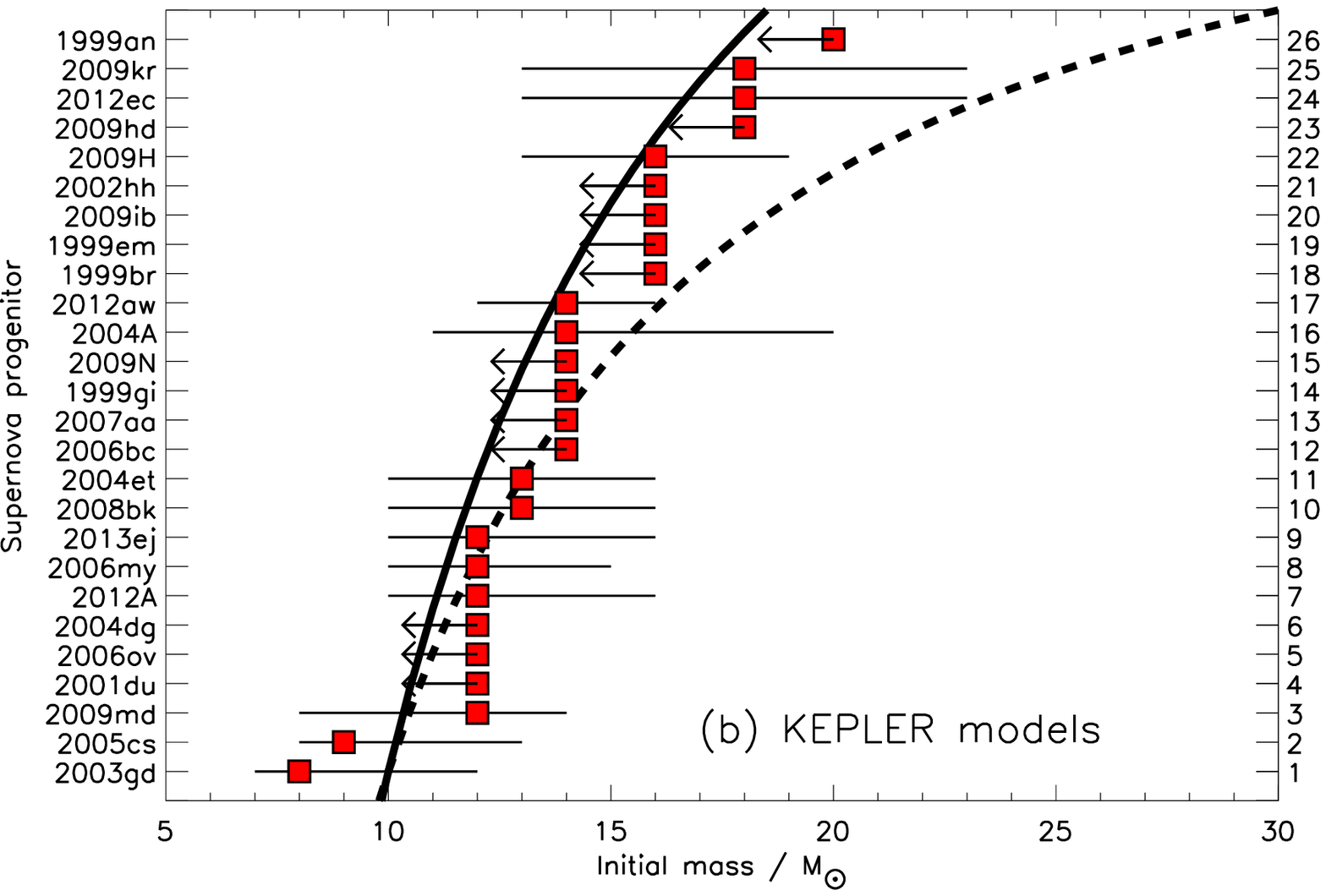}
\caption{The progenitor detections are marked with error bars (data from Table\,\ref{tab:SNII} and the  limits
are marked with arrows (data from Table\,\ref{tab:II-lims}). The  lines are cumulative IMFs with different 
minimum and maximum masses.}
\label{fig:imf}
\end{center}
\end{figure}

It is timely to review this in light
of the discoveries made since 2009 and the increase in statistics now available
and updated values using different methodologies. 
The luminosities  compiled in this paper for detections and non-detections are
used with three sets of models to estimate initial masses of the progenitor stars. 
The end points of three sets of models are plotted in Figure\,\ref{fig:endpoints} showing the initial 
mass - final luminosity relation for the 
STARS models  \cite{2004MNRAS.353...87E},  the rotating Geneva models of  \cite{2004A&A...425..649H} 
and the KEPLER models of \cite{2007PhR...442..269W}. The STARS and rotating Geneva models 
have quite similar final luminosities, hence we consider the results of these to be indistinguishable 
(at least at the level of accuracy possible for any individual SN progenitor). The KEPLER models
are less luminous at their endpoints by between $0.1-0.2$\,dex which typically translates into a difference
in mass of 1-2\msun. 
In \cite{2009MNRAS.395.1409S} and other work that employed one set of stellar
models, the uncertainty in the theoretical models was taken into account by using luminosities at the end of the He burning. This was somewhat artificial and 
this review uses an alternative approach of taking the end points of three sets
of evolutionary model grids. 

The estimated initial masses of the progenitors are listed in Tables\,\ref{tab:SNII} and \ref{tab:II-lims}. 
The masses for the three lowest luminosity progenitors (SN2003gd, SN2005cs and SN2009md) 
are uncertain in the KEPLER models, as this code does not yet have published endpoints for stars in the 
8-10\msun\ regime. This is a problematic regime in that 2nd dredge up may occur, pushing the 
progenitors to higher luminosities than their more massive counterparts as discussed in 
\cite{2007MNRAS.376L..52E,2002ApJ...565.1089S} and references there in. The quoted, but uncertain estimates from 
employing the KEPLER code are simply the extrapolated differences from the STARS estimates. 

As in \cite{2009MNRAS.395.1409S}, one can assume that the progenitors come from a mass function
of some slope and take  $\alpha = -2.35$ as a reliable standard. A maximum likelihood calculation 
can produce an estimate of the most likely lower mass and upper mass of the distribution, assuming 
the masses follow a Salpeter function. For the masses estimated with the STARS (and rotating Geneva) 
models, the values determined (using the same IDL routine as employed in Smartt et al. 2009) are 
a minimum mass for the distribution of $m_{\rm min} = 9.5^{+0.5}_{-2}$ and a maximum mass of 
$m_{\rm max} = 16.5^{+2.5}_{-2.5}$ where the errors are the 95\% confidence limits (see Figure\,\ref{fig:maxlike}). 
If we employ the KEPLER models, then the values are 
$m_{\rm min} = 10^{+0.5}_{-1.5}$ and a maximum mass of 
$m_{\rm max} = 18.5^{+3}_{-4}$ (again with 95\% confidence limits). These results are illustrated further in 
Figure\,\ref{fig:imf} where the masses are plotted with a Salpeter IMF (cumulative frequency function). 
The plots show that the mass distributions are comfortably reproduced with a standard IMF between the 
lower and upper mass limits from the maximum likelihood calculations, but they need to be truncated at the 
higher mass. If one allows the mass function to vary up to say 30\msun, then the mass distribution cannot 
be reproduced. This is the same basic result as shown in Fig\,\ref{fig:hrd} -  the population of progenitors
is missing the high mass end of the distribution, but this time the IMF is quantitatively considered. 
The maximum likelihood calculation is visualised  in this cumulative frequency plot - given an IMF slope, 
the line fit should go through the error bars of the detections and not conflict with any of the upper limits. 

The lower mass limit to produce a core-collapse SN was estimated in  \cite{2009MNRAS.395.1409S} 
to be $m_{\rm min} = 8.5^{+1.0}_{-1.5}$ from the same method and the sample to that point. 
 \cite{2009ARA&A..47...63S} reviewed the limits from the maximum masses of white dwarf
progenitors, suggesting a convergence at  $m_{\rm min} = 8\pm1$. The two values estimated
here slightly higher: the value from the STARS models is 
$m_{\rm min} = 9.5^{+0.5}_{-2}$ (integer mass models evolved through C-burning down to that 
mass have been calculated) which is not significantly different to that estimated previously given
the errors. The value from the KEPLER models is higher again, at $m_{\rm min} = 10^{+0.5}_{-1.5}$. 
However low mass models (7-10\msun) are not available from KEPLER and the 
values in this luminosity range were estimated  assuming 
the same differential in luminosity between KEPLER and STARS models exists between 
7-10\msun as at 11\msun\ (Figure\,\ref{fig:endpoints}). This is uncertain and the 
lower mass from KEPLER should not be treated as a quantitative estimate : 
$m_{\rm min}$  is critically dependent on the mass estimates for the three lowest luminosity progenitors
and if these are adjusted  down by $\sim$1\msun then the value of $m_{\rm min} = 9.5$ would be 
reproduced. Some further quantitative modelling of stars in this interesting mass range is 
required to reproduce the stellar luminosities and produce either a Fe-core collapse or 
O-Mg-Ne core that collapses through electron capture. Despite this uncertainty at the lower 
end, the existence of a high mass upper limit for type II SNe appears be secure. 
The value is model dependent of course, but the basic result is that type II progenitors are statistically  
lacking above a \logl$\simeq5.1$\,dex. The final model luminosities at this 
value are stars with $m_{\rm max} = 16.5$ (19\msun\ at 95\% confidence) for the STARS and 
Geneva models  and $m_{\rm max} = 18.5$ (21.5\msun\ at 95\% confidence) for the KEPLER models.

While \cite{2009MNRAS.395.1409S} first discussed this as the``red supergiant problem", the lack of detected high mass progenitors is now a 
broader issue for all SN types. This broader issue of a lack of high mass progenitors generally was discussed in 
\cite{2008ApJ...684.1336K}, who took all historical and literature limits to that date.  
The fact that there are now three type IIb SNe in our sample with progenitor detections (Table\,\ref{tab:IIb}) and 
these also have luminosities less than $\sim$5.1\,dex illustrates that the missing high mass star
problem is relevant for all type II SNe (type II-P, II-L, IIb). We address the issue of type IIn progenitors
below. 

Alternative methods of probing the mass function of SN progenitor stars have been 
recently advanced by \cite{2014ApJ...791..105W} and  \cite{2014ApJ...795..170J}. 
These use similar methods of quantifying the stellar population around historical SNe
in galaxies closer than 8\,Mpc and around SN remnants in M31 and M33. They use 
high quality Hubble Space Telescope imaging and careful stellar photometric measurements
to determine the luminosity and masses of stars within the immediate vicinity of 
SNe (typically within 50\,pc).   \cite{2014ApJ...795..170J}  find that 
there is a lack of high mass stars close to the  SN remnants in M31 and M33, suggesting
either the  highest mass stars do not produce SNe, or that 
SNR surveys are biased against finding objects in the very youngest ($<10$\,Myr old) 
starforming regions. \cite{2014ApJ...791..105W}  also suggest that their results 
are compatible with progenitors all coming from masses $M<20$\msun\ although 
the uncertainties do not rule out the possibility of no upper-mass cutoff.

\subsection{Possible explanations}
\label{sec:explain}

The reasons for these missing high mass progenitors are discussed as follows 

\subsubsection{Dust formation and circumstellar extinction}
As discussed in Section\,\ref{sec:IIsum} the extinction toward the progenitors is often estimated from the 
extinction toward the SN itself, or the nearby stellar population. The former  estimates may not
be directly applicable since the circumstellar dust around the progenitor stars can be
destroyed in explosions - as in the case of SN2012aw and SN2008S. 

\cite{2012MNRAS.419.2054W} calculated the dust that could be produced in red supergiant winds 
and the extra extinction that this would produce. The idea is well motivated and valid, 
but \cite{2012ApJ...759...20K}
showed that treating CSM extinction with a slab of ISM material is not physically consistent. 
As shown in  \cite{2012ApJ...759...20K}, the progenitor of SN2012aw was thought to be 
quite a high mass star but correct treatment of radiative transfer in a spherical dust shell 
reduces the progenitor luminosity limit while comfortably fitting the optical, NIR and MIR
detections and limits. The major concern for this sample is that the objects with 
limits only are effectively unconstrained - one could propose that these are all 
high mass objects enshrouded in dust and having no detection at any waveband 
does not allow meaningful constraints. However there is another important constraint 
which comes from x-ray and radio observations of type II SNe. 

\cite{2006ApJ...641.1029C} considered a sample of six type II-P SNe with x-ray 
and radio observations. They estimated the mass-loss densities in the progenitor
star's stellar wind  from the 
thermal x-ray and radio synchrotron flux which originates when the fast 
moving SN ejecta interact with a pre-existing, lower velocity stellar wind
from the progenitor star. The results for these six SNe (which are in the
compilation presented here, either in Table\,\ref{tab:SNII} or \ref{tab:II-lims}), 
were consistent with the mass-loss expected for progenitor red supergiants in the 
mass range  $\sim8-20$\msun.
\cite{2014MNRAS.440.1917D} has now  
compiled all the available literature x-ray detections for type II-P
SNe and used the \cite{2003LNP...598..171C} treatment of the free-free x-ray luminosity from the shocks 
produced  by the SN ejecta and the stellar winds. Red supergiant stars have 
observed mass-loss rates that span a wide range of values  from $<10^{-6}$\msun\,yr$^{-1}$
up to $\sim10^{-4}$\msun\,yr$^{-1}$
\citep{2011A&A...526A.156M}.
The observed  x-ray luminosity is  
$L_{\rm x-ray} \propto (\dot{M}/v)^2$  (where $\dot{M}$ is the stellar mass loss rate and $v$ is
the stellar wind velocity; Chevalier \& Fransson 2003). With a wind velocity of around $\sim$10\kms, the x-ray 
luminosities imply mass-loss rates of less than $\dot{M} < 10^{-5}$\msun\,yr$^{-1}$. 
The mass-loss rates of red supergiants are correlated with stellar luminosities and 
\cite{2014MNRAS.440.1917D} then used the 
\cite{2011A&A...526A.156M} 
relations  ($\dot{M} = 4.7\times10^{-6}(L/10^{5})^{1.7}$) to illustrate that type 
II-P SNe do not have high enough x-ray luminosities to have had progenitors
significantly greater than \logl$\simeq5.2$\,dex or 19\msun. 
Similar arguments were used in \cite{2012ApJ...759...20K} specifically for the
case of SN2012aw to show that the steady wind of 
 $\dot{M} \lesssim10^{-5.5}$  to $10^{-5}$\msun\,yr$^{-1}$ was consistent with 
the x-ray and radio fluxes measured. The existence of obscuring CSM dust 
cannot exist without CSM gas, in the form of stellar winds, with typical 
dust masses around $\sim0.5$\% that of the gas masses \citep[from the typical opacities in][]{2012ApJ...759...20K}. 
Therefore an obscured progenitor should be brighter in x-ray and radio than the typical 
luminosities observed to date.   \cite{2014MNRAS.440.1917D}  shows that 
the IIb and IIL SNe have typically higher x-ray luminosities than the II-P,  with 
values around $\sim10^{39.5}$erg\,s$^{-1}$
\citep[also see][]{2012MNRAS.419.1515D}
.  This is consistent with their higher luminosity
progenitors, in the range of $\logl \simeq 5.1-5.2$\,dex. As remarked upon by 
 \cite{2014MNRAS.440.1917D} the general agreement between the two independent 
methods is remarkably consistent.  While dust obscuration should still be considered carefully
for any progenitor and limit, it appears the sample as a whole is not significantly biased and 
there are no major discrepancies between the optical-NIR luminosity and mass estimates
of progenitors and the SN x-ray and radio fluxes. 

It is interesting to note that it is not just steady wind mass-loss that has been proposed
to produce dense CSM shells, and potentially extra extinction, around progenitors. 
\cite{2014ApJ...780...96S}
have argued that after core neon burning, internal gravity waves can transport a super-Eddington
energy flux out into the stellar envelope and cause a mass ejection of around 1\msun\ of material. 
They suggest that this could occur preferentially in progenitors with $M_{\rm ZAMS}\sim$20\msun\ 
within a few months to a decade of core-collapse, with the most intense mass-loss  
occurring closer to core-collapse. This timescale is quite similar to the typical time differences between 
SN explosions and dates of progenitor detections. The dates of the observations of progenitors, or limits (for
the SNe in Tables 1, 2 and 3) have a range of 2 months to 17 years before the SN discovery dates
(which one can assume is close to the explosion date to within $\sim$weeks at worst). The median 
and standard deviation of the time of data being taken before SN discovery are 
65 and 54 months. 
It is possible that these ejection episodes could create dust shells, however the gas masses
would be inconsistent with the x-ray and radio fluxes, as discussed for the steady mass-loss
scenarios. There is no obvious correlation between probability of discovery and date of data taken,
and some of the discoveries have images taken within 5-9 months before explosion (e.g. 2003gd, 2005cs, 2008bk).

\subsubsection{Biased luminosity estimates and evolution to the blue}
An argument could be made  that the observed limit that we see for red supergiant progenitors 
of  \logl$\simeq 5.1$\,dex is not significantly lower, than the 
highest luminosity red supergiants of  \logl$\simeq 5.45$\,dex
\citep{2009AJ....137.4744L} and that uncertainties in the extinction and 
luminosity estimates from small numbers of broad band fluxes are high enough to 
argue that this is not a discrepancy at all. Rather, this indicates broad and  basic agreement \citep[see][for a discussion on careful model atmosphere techniques]{2013ApJ...767....3D}.  However  the increase in statistics now available,  and the 
statistical and quantitative deficit of high luminosity progenitors as calculated in Section\,\ref{sec:massfunc}
would suggest that this is not a satisfactory answer. 

More broadly, the question then is - if there is no real ``red supergiant problem", then where
Hertzsprung Russell Diagram (HRD)
do these $M_{\rm ZAMS} \gtrsim 18$\msun\ stars evolve to before the
point of core-collapse and what SN do they produce ?  They must end up with an 
iron core and, within the bounds of current theory, they must collapse. 
It does not appear that they 
can all produce WR stars and Ibc SNe for the reasons discussed in Section\,\ref{sec:Ibc}. 
The search for progenitors is not biased against any particular SN type, and  the one 
II-L progenitor (SN2009kr), and the three IIb progenitors do not have luminosities higher than
the observed limit of 5.1\,dex. If  high mass stars ($M>18$\msun) evolve to produce 
visible supernovae explosions, there is no obvious reason we should not detect 
both the supernovae and their progenitors, unless the SNe themselves
are intrinsically faint and have evaded either detection or recognition. 

The type IIn SNe have been suggested as a way out. High mass stars, with high mass-loss rates
produce high enough CSM gas masses that the SNe are observed to be IIn types. 
This is reasonable and indeed corroborated by two detections of likely high mass
progenitors for IIn SNe \citep{2009Natur.458..865G,2011ApJ...732...63S}, but we have not systematically found enough of them to account
for the missing mass range. 
\cite{2011MNRAS.412.1522S} argue for quite a high relative rate of IIn SNe in the LOSS survey and that these
would naturally have high mass progenitor stars. It seems almost certain that SNe 
IIn do come from high mass progenitors (see Section\,\ref{sec:IIn}). Some have detected progenitors (as
in SN2009ip and SN2005gl) and they tend to have
the largest x-ray and radio fluxes that require mass-loss rates which can not be realistically produced
by low to moderate mass stars \cite[][and references therein]{2014MNRAS.440.1917D}. 
However the high relative rate estimated in \cite{2011MNRAS.412.1522S} does not appear
to hold at lower distances 
\citep[see][for a discussion on this topic]{2013MNRAS.436..774E}. There is no reason why we should not detect them
and no reason why they should not appear in Figure\,\ref{fig:hrd} other than there 
are not enough high mass stars producing visible IIn SNe. 

Other work has also hinted that there could be some systematic bias underlying the mass estimates 
from the direct detection of progenitors. Hydrodynamic modelling of type II-P SN lightcurves and velocity measurements has suggested a significant discrepancy between the SN ejected masses and the stellar masses of the progenitors.  The detailed modelling of  \cite{2008A&A...491..507U} and \cite{2009A&A...506..829U} for example 
has estimated ejecta masses for 5 type II SNe (1987A, 1999em, 2003Z, 2005cs 
and 2004et) which would imply initial stellar masses a factor of $\sim$2
higher than those estimated from the progenitor luminosity and stellar evolutionary
tracks. \cite{2009A&A...506..829U} speculate that the hydrodynamics may be effected 
by asymmetry in the explosion  and the lack of treatment of the multi-dimensional effects of   Rayleigh-Taylor mixing between the helium core and the hydrogen envelope. However \cite{2014MNRAS.439.2873S} used a different code
\citep[that of][]{2011ApJ...741...41P} to find better agreement between the 
direct progenitor masses and hydrodynamic masses estimates. 
A major goal for the SN community is to reconcile these independent and 
complimentary estimates as they fundamentally link the progenitor, explosion
mechanism and SN observables. One would hope uniform agreement could 
be found that each method can inform the other of potential systematic 
errors.

\begin{figure}
\begin{center}
\includegraphics[width=7cm]{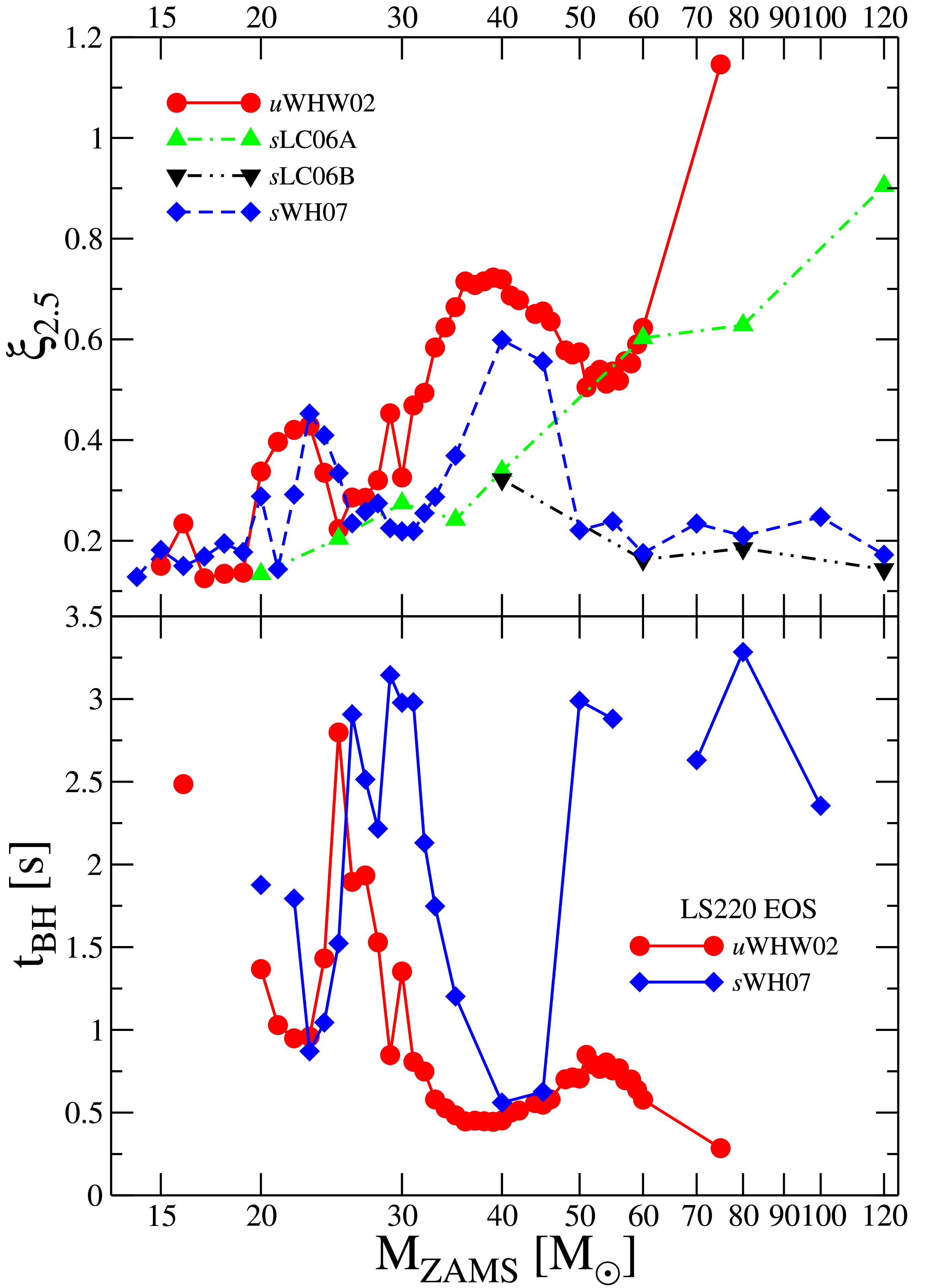}
\caption{From \cite{2011ApJ...730...70O}. The upper plot shows 
the compactness parameter $\xi_{2.55}$ of stellar cores as a function of ZAMS mass for different stellar evolutionary models.  The lower plot shows the time to black hole formation, assuming no explosion has occurred. Reproduced from 
Figure 9 in {\em ``Black Hole Formation in Failing Core-Collapse Supernovae”}, by O’Connor \& Ott 2011, ApJ, 730, 70. 
 }
\label{fig:compactness}
\end{center}
\end{figure}

\begin{figure*}
\begin{center}
\includegraphics[width=16cm]{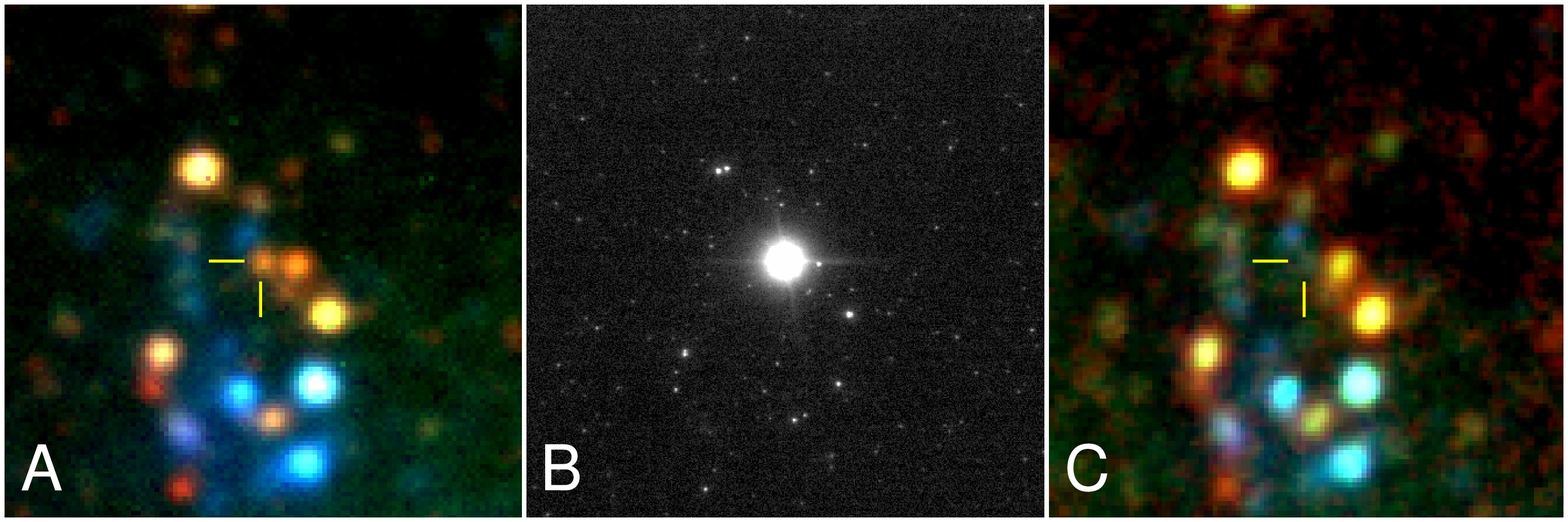}
\includegraphics[width=16cm]{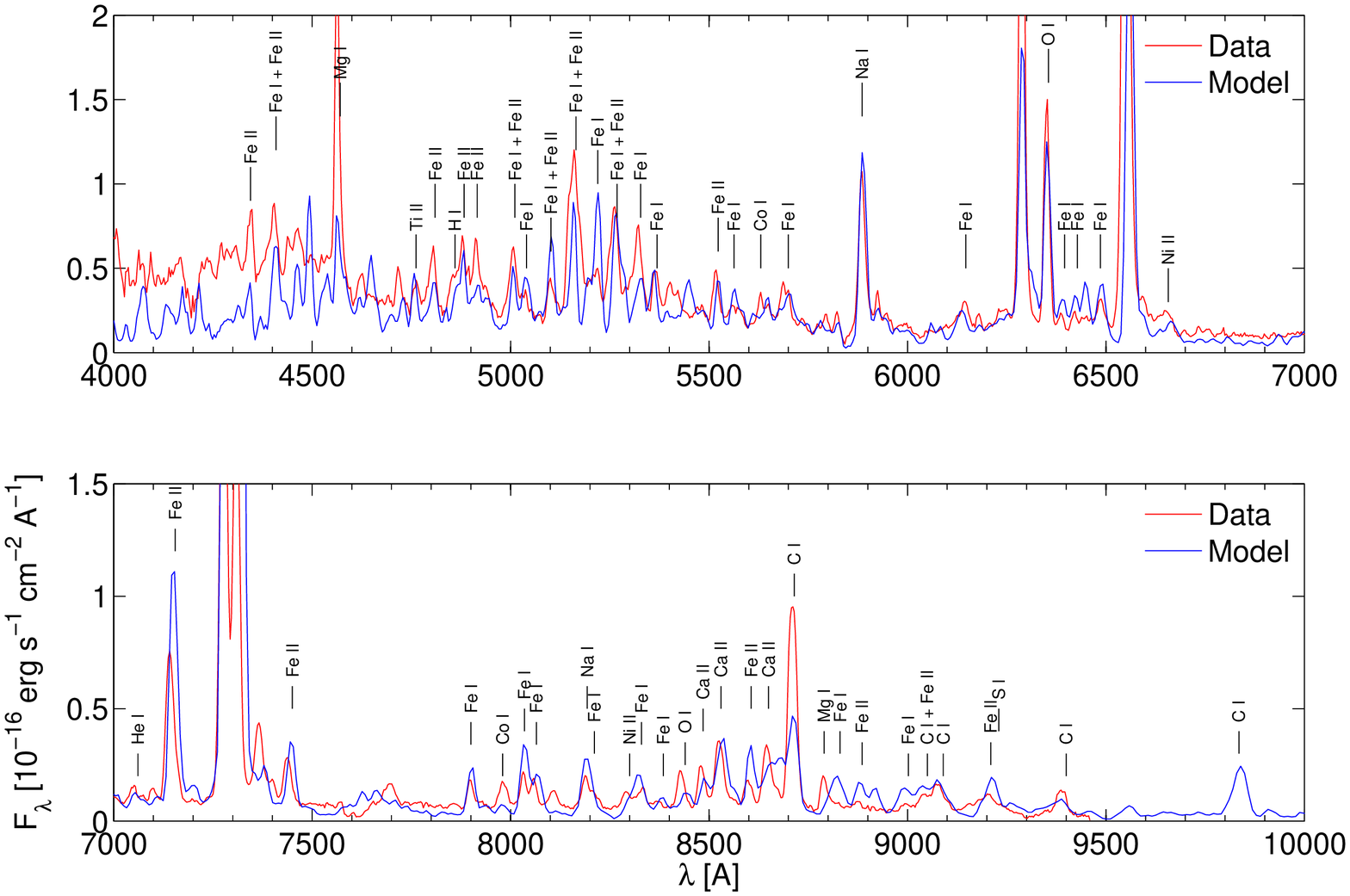}
\caption{{\bf Upper:} The visually striking illustration of the disappearance of the red supergiant progenitor of SN2008bk, 
from \cite{2010arXiv1011.5494M}. Panel A shows the VLT colour image of the progenitor (marked). Panel B shows the VLT NACO image of SN2008bk and the surrounding population at high resolution. Panel C shows an NTT colour image at 
approximately 940 days after explosion illustrating the disappearance of the red source. The quantitative mass estimates of the progenitor are in 
\cite{2014MNRAS.438.1577M}.  {\bf Middle and Lower:} A spectrum of the faint blue source seen  at 547 days after explosion from \cite{2012MNRAS.420.3451M}, showing the nebular phase emission lines from the SN ejecta, with a model from \cite{2012A&A...546A..28J}. 
The model is from an exploded 12\msun\ star, which is in good agreement with the updated progenitor mass estimate in \cite{2014MNRAS.438.1577M} and in the summary table here (see Table\,\ref{tab:SNII}. 
Reproduced from Figure 13, in
{\em ``Constraining the physical properties of Type II-Plateau supernovae using nebular phase spectra''}, K. Maguire et al, 2012, Monthly Notices of the Royal Astronomical Society, 420, 3451 
 }
\label{fig:08bk}
\end{center}
\end{figure*}

\subsection{Black hole formation and failed supernovae}
\label{sec:bh}
On the basis of available data, the above two explanations are  unlikely. 
Therefore the reason for the lack of high mass progenitors may be simply that they do not exist. 
In other words the SN population that we observe does not, on the whole, come from 
stars with initial masses greater than $M_{\rm ZAMS}\sim$18\msun. 
That these high mass stars 
will  evolve to carbon-oxygen cores and subsequently to an Fe core composition
and collapse is inevitable. However if black holes are formed, either directly or by fall back, 
and the neutrino energy deposition is not sufficient to drive the shock through the 
steeper density profiles of these progenitors a failed SN is possible. It is plausible to 
postulate that the reason for the missing high mass progenitors is that they
produce failed SNe that have, so far gone undetected (or unrecognised) in all 
of our efforts to find and quantify transients in the local Universe. From a theoretical 
standpoint this is not surprising at all, the issue of explodability of
the most massive stars has  been a question since the 1980's and 1990s. 
 \cite{1999ApJ...522..413F} explored
black hole formation through direct collapse and fall back and already suggested 
then that stars above 20\msun\  would produce black holes, possibly without a SN explosion. 

The density profiles in stellar cores is a strong function of stellar mass, and recent 
work has revisited the idea that very massive stars may not produce detectable SNe. As discussed in the  introduction in
\citep{2014ApJ...783...10S}
the density gradient in more massive stars is relatively shallow, which will likely
lead to higher accretion rates onto the compact object created at the point of 
collapse. It would also lead to higher ``ram pressure" which would need to be
overcome by the shock (through the neutrino energy deposition mechanism) to 
launch a successful explosion through the star. 
The difficulty in getting successful simulated explosions has reinvigorated the study of 
the explodability of stars and in particular, three independent analyses by 
\cite{2011ApJ...730...70O}, 
\cite{2012ApJ...757...69U}
\cite{2014ApJ...783...10S}
and 
have used the compactness parameter
(defined in O'Connor \& Ott)
which quantifies the core radius within which 
a certain amount of mass is confined in the final progenitor model $\xi= M/R(M)$ where 
$M$ is in solar masses and $R(M)$ is in units of 1000\,km. Defining $\xi_{2.5}$ for 
a value of 2.5\msun\ has been proposed.  All three of these studies suggest that 
the compactness parameter is not a monotonically increasing and simple function of 
initial stellar mass. It will depend critically on mass-loss rate, internal mixing mechanisms
in the stellar model not to mention metallicity and rotation (and very likely mass-transfer in 
binaries will have similar effects to mass-loss). Figure\,\ref{fig:compactness} is from 
\cite{2011ApJ...730...70O}
showing the compactness $\xi_{2.5}$  at bounce, as a function of $M_{\rm ZAMS}$.
The three studies all suggest that at critical values of $\xi_{2.5} > 0.2 - 0.3$, 
successful explosions are difficult to achieve. The figure illustrates that the 
compactness is dependent on the mass-loss recipes used in the stellar models, 
and that below $M_{\rm ZAMS} < 20$\msun, the stars are relatively easy to explode. 
However above that threshold, it becomes significantly more difficult to 
achieve explosions. There are what have been termed ``islands of explodability'' 
such that at some masses, the combination of mass-loss and final core 
structure may make it easier or hard to explode. What is constant in all these
three studies is that above $M_{\rm ZAMS}  >20$\msun, one would expect to 
have failed explosions and that there will be a tendency for stars not to produce
visible SNe.  It may be that there is not a one to one relation and there is no 
simple, single numerical mass threshold at which neutron star or black hole formation 
occurs. Overall though, model stars above this threshold have a tendency to produce
failed SNe at the point of core-collapse  whereas stars below it generally proceed
to explosion. In the same spirit,  \cite{2014MNRAS.445L..99H} have carried out an extensive series of simulations with more than 300 models and argue that a critical value of $\xi_{2.5}\sim0.2$ 
is the likely divide between successful and failed explosions. This 
would result in stars in the initial mass range $\sim16-30$\msun\ not producing 
canonical SNe. 

The observational constraints presented in this review can reasonably be explained, 
and in a quantitative manner, with this model framework. The search for progenitors has
been more difficult than was first thought, with the dearth of high mass and high 
luminosity progenitors an inescapable fact. Arguments can be made concerning
dust, bias, and evolution to very faint WR stars but these have been addressed above.
 The reasonable agreement between the 
explosion theories based on the  core compactness 
and the progenitor constraints mean they should be considered
to be physically linked.  As first noted by 
\cite{2008ApJ...684.1336K}
the prediction of this scenario is that many massive stars (a fraction as high as 
$f = 20 - 30$\%)  end their lives by disappearing without a visible supernovae. 
As the dynamical timescale of a star of red supergiant dimensions is of order a year, 
these failed SNe may give rise to faint, long lived red transients as proposed
by \cite{2013ApJ...769..109L}. However 
\cite{2013ApJ...768L..14P} proposed that a short timescale optical 
transient (lasting 3-10 days) with a faint luminosity of $\sim10^{40}-10^{41}$erg\,s$^{-1}$ could be produced due to the breakout of a shock induced in the 
stellar envelope from neutrino losses in the core. 
\cite{2014arXiv1411.1761G}
have just presented the first results from their novel and ongoing survey for failed stars and this has turned
up one plausible candidate that requires further investigation. One could reasonably
argue that black hole forming cores and failed explosions are not controversial ideas,  rather they are solid theoretical predictions that have been  proposed for more than 15 years  \citep[e.g.][]{1999ApJ...522..413F}. \cite{2014ApJ...785...28K} 
finds that the observed black hole mass function in the Galaxy can be modelled by 
assuming that black hole formation (and hence successful supernova explosion) 
is determined by the core compactness parameter. All of this adds impetus to the 
searches of \cite{2014arXiv1411.1761G}
and surveys for transients in the local Universe to identify the signatures
that have been suggested by, for example, 
 \cite{2013ApJ...769..109L} and  \cite{2013ApJ...768L..14P}. 

A final note on this topic of the explosion mechanism  is the potential of recent work to compare the ejecta
velocities  in successful SNe to the progenitor mass estimates by
\cite{2013MNRAS.436.3224P}. There appears to be a relation between 
the observed ejecta velocities, plateau duration and progenitor mass 
which could be explained if the energy deposition (the kinetic energy 
observed in the ejected envelope) is proportional to the initial mass of the progenitor cubed ($E \propto M^{3}$). If this analysis holds then it could be telling us something 
about the efficiency of the deposition process in a neutrino revived explosion.
As noted by \cite{2013MNRAS.436.3224P}, the paucity of objects with low 
energy and large mass could support the idea of a population of failed SNe. 
This interesting correlation is worth pursuing further both from the observational 
gathering of larger samples and theoretical investigation of the explosion mechanisms. 

\begin{figure}
\begin{center}
\includegraphics[width=7cm]{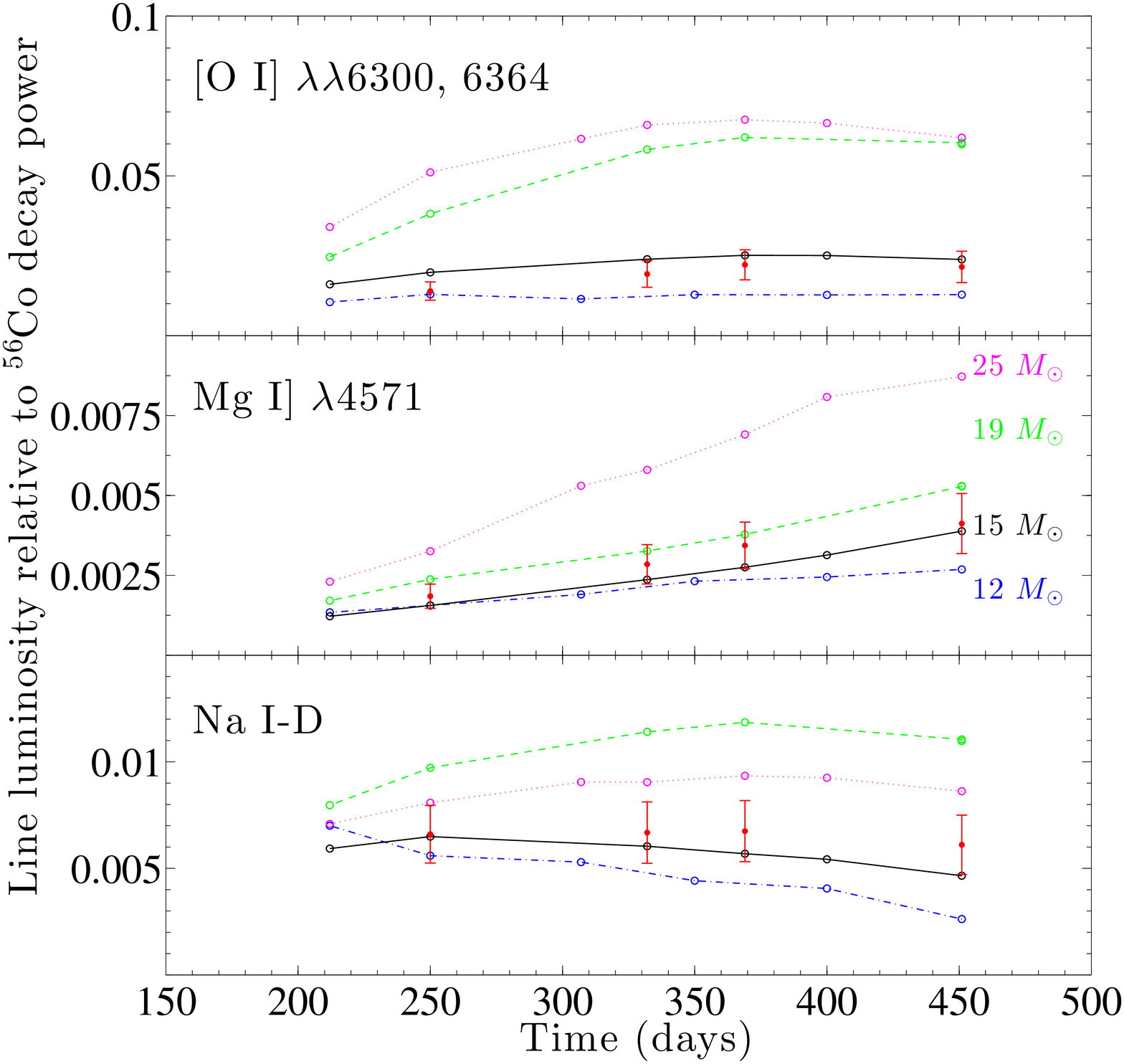}
\includegraphics[width=7cm]{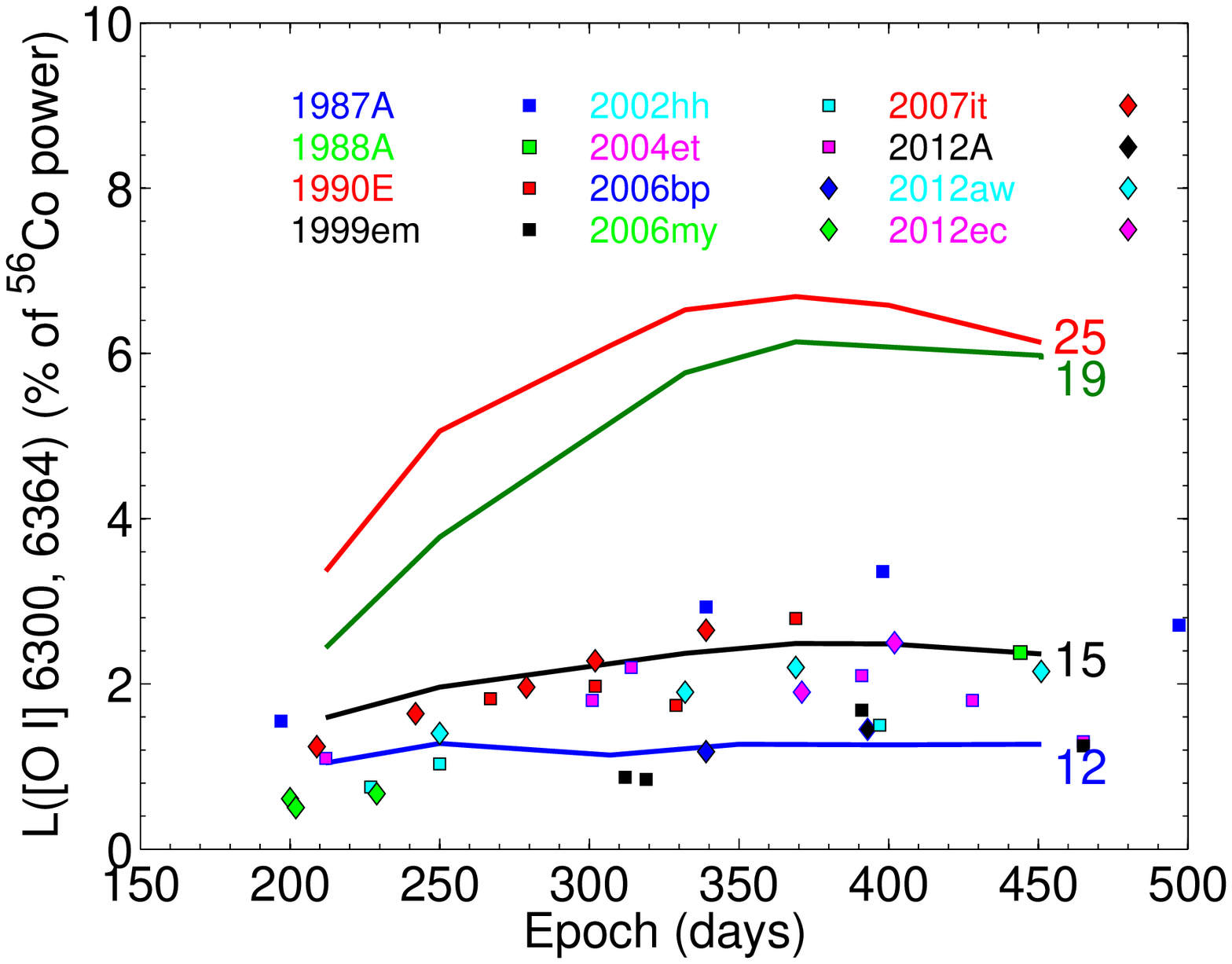}
\caption{{\bf Upper Panel:} This shows the 
observed line luminosities for [O\,{\sc i}], Mg\,{\sc i}] and Na\,{\sc i}
for SN2012aw compared to the nebular models for 12 - 25\msun\ stars
\citep{2014MNRAS.439.3694J}. 
The line strengths support the mass estimate from direct detection of 
the progenitor star of around $M_{\rm ZAMS}=$15\msun\ and are much weaker 
than is predicted expect for exploding high mass progenitor models
of 20-25\msun. 
Reproduced from Figure 4, in
{\em ``The nebular spectra of SN 2012aw and constraints on stellar nucleosynthesis from oxygen emission lines''}, A. Jerkstrand et al, 2014, Monthly Notices of the Royal Astronomical Society, 439, 3649. 
{\bf Lower Panel:} 
A compilation of the [O\,{\sc i}] line 
luminosities compared to the predictions of
exploding models of 12, 15, 19 and 25\msun. 
This again illustrates that type II SNe (predominantly II-P) do not have the  observational signatures expected from higher mass 19-25\msun progenitors. 
Reproduced from Figure 9, in
{\em ``Supersolar Ni/Fe production in the Type IIP SN 2012ec''}, 
\cite{2015MNRAS.448.2482J}
Monthly Notices of the Royal Astronomical Society, 2015, 448, 2482
 }
\label{fig:lineflux}
\end{center}
\end{figure}
\subsection{Nebular spectra and nucleosynthesis}
\label{sec:nucleo}

A potentially powerful  method of testing stellar evolutionary modelling and explosive nucleosynthesis is 
by observing, and physically modelling SNe when they are in their nebular phase. As the ejecta expand 
and become optically thin, the inner regions of the progenitor star become visible and the elements 
synthesised within the He and CO cores during stellar evolution become visible.
Nebular spectra allow the diagnosis of nucleosynthesis
products. This becomes particularly interesting for SNe
with detected progenitors, as the independent estimate
of the main sequence mass from nebular modelling can
test the prediction from the progenitor luminosity. 
An evolutionary model should reproduce the pre-explosion 
stellar luminosity and the oxygen mass (for example) produced during helium and carbon burning. 
\citep[as in][]{1998ApJ...497..431K}. 
Figure\,\ref{fig:08bk} visually illustrates the diagnostic power of combining these two methods. 
The progenitor star is detected, seen to disappear and the faint blue flux visible at the 
star's position can be studied spectroscopically to determine element masses synthesised in the 
progenitor (e.g. the [O\,{\sc i}] doublet at 6300,6364 is a useful diagnostic of the neutral oxygen mass). 
The key to this  of course is having a reliable and sophisticated radiative
transfer model that can reproduce the observed spectra and estimate element masses in 
the ejecta from the absolute and relative strengths of the emission lines. 
One of the leading codes for this has been developed by 
\cite{2011PhDT........90J,2011A&A...530A..45J,2012A&A...546A..28J}. 
There are others such as 
\cite{2010MNRAS.408..827D} and \cite{2010MNRAS.408...87M}
which have been successfully applied. 
The Jerkstrand et al. code, developed in Stockholm,  calculates the physical conditions in the 
expanding nebula and includes non-thermal heating, ionization, and excitation
from the gamma-ray and positron energy deposition of $^{56}$Ni/$^{56}$Co
(and $^{57}$Co and $^{44}$Ti at later phases), and computes thermal and
statistical equilibrium with the latest atomic data.
It  treats multi- line radiative transfer with a Monte Carlo technique and is 
being actively applied to those SNe with 
progenitor detections and limits \citep[e.g.][]{2014MNRAS.439.3694J,2015A&A...573A..12J}. 
These models can successfully reproduce the observed optical, near-IR and (when available) mid-IR 
spectra of type II SNe at late phases \cite[see the results for SN2004et in][]{2012A&A...546A..28J,2012MNRAS.420.3451M}

In particular, the application to SN2012aw allows a self-consistent comparison of the 
progenitor luminosity and the oxygen produced. In a progenitor model, the stellar luminosity 
is determined by the He core luminosity. The He core size and burning rate 
dictates the oxygen production. Thus a progenitor model which has been artificially exploded 
should be able to reproduce the oxygen line strengths and the final luminosity consistently. 
The physical sophistication of the \cite{2011A&A...530A..45J} code now allows this to be done to 
high accuracy.  For SN2012aw the nebular phase spectra are in good agreement with the ejecta from a KEPLER 
model of $M_{\rm ZAMS}=15$ \msun\ progenitor star \citep{2014MNRAS.439.3694J}
and the forbidden neutral oxygen lines can be used to constrain the 
mass of oxygen to be less than  $< 1$ \msun (see Figure\,\ref{fig:lineflux}). 
 This provides a physically consistent match to the 
progenitor luminosity of a KEPLER model of the same mass (14\msun\, as in Table\,\ref{tab:SNII} and illustrated in
Figure\,\ref{fig:endpoints}). 

The [O\,{\sc i}] line strengths (and also the Mg\,{\sc i} and Na\,{\sc i} fluxes) are strongly dependent on progenitor mass. The conclusion 
of  \cite{2014MNRAS.439.3694J}  is that to date, there is no convincing example of a type II-P SN which displays the 
nucleosynthetic products expected from a $M_{\rm ZAMS} > 20$ \msun\ progenitor in its nebular spectra (see Fig\,.\ref{fig:lineflux}).
  \cite{2015A&A...573A..12J}  further modelled the nebular
spectra of the IIb SNe 1993J, 2008ax and 2011dh and constrained the initial mass of the 
progenitors to be $M_{\rm ZAMS} = 12-16$\msun\ from the line strengths of the oxygen lines. 
Together with progenitor detections, constraints on the mass-loss rates, these nebular spectra and 
the radiative transfer analysis there is a consistent picture that there are no type II SNe in the local
Universe which come from high mass progenitors. While the nebular spectral analysis provides the strong 
progenitor constraints useful for this review it is becoming a powerful way to probe the enrichment of
SNe and to allow quantitative checks on what role various SNe play in galactic chemical evolution models.

\subsection{Nucleosynthesis and chemical evolution}
\label{sec:chemevol}
If all stars above  a certain main-sequence mass do form black holes then they are not likely to eject the elements
synthesised during stellar evolution. The important implications of this were recently studied by 
\cite{2013ApJ...769...99B} who found that the solar abundance pattern (from oxygen to strontium, at 
$A\simeq$90) are fit well if there is no cut-off assumed. This  calculation had all stars up to 120\msun\ 
produce successful supernovae and produce nucleosynthetic yields according to 
\cite{2007PhR...442..269W}. If the limiting main sequence mass that produces a black hole and
no supernova (and therefore no enrichment) is reduced to $M_{\rm BH}=25$\msun\ then the 
solar elemental abundance pattern is still comfortably reproduced. However there are unpalatable 
consequences if this mass limit is as low as $M_{\rm BH}=18$\msun, for which there is 
some direct evidence as discussed earlier in this review. For this limit to be compatible
with the observed abundance pattern, \cite{2013ApJ...769...99B}  show that the mass-loss rates
should be lower than currently assumed (to keep the carbon abundance under control), 
and the reaction rate for $^{22}{\rm Ne(}\alpha,n{\rm )Mg}^{25}$ needs to be near the experimental 
maximum rate in order to reproduce the $s-$process isotopes. In addition the historical 
supernova rate needs to be 3 times higher than that when no cut-off is assumed. The 
non-monotonic distribution of compactness discussed in Section\,\ref{sec:bh} which results 
in some higher mass stars exploding in the islands of explodability would mitigate these
problems. However that would require significant numbers of successful explosions above
about 22\msun. One way out is that  if supernovae from the most massive stars ($M>20$\msun)
are successful in enrichment but
are more often enshrouded in dust and invisible to current surveys. The discussion
in Section\,\ref{sec:explain} disfavours this scenario. Searches for SNe in very high starformation rate
galaxies such as dusty luminous infra-red galaxies suggest that more than about 80\% of
expected core-collapse SNe are missed in optical and near infra-red searches \citep{2012ApJ...756..111M}. For normal galaxies, the missing fraction is more likely to be 10-15\% 
\citep[again estimated in][]{2012ApJ...756..111M}
and these missing SNe would 
need to be from the most massive stars to explain the deficit of high mass progenitors and ease the 
tension with the elemental production budget. Further work  on searches for SNe specifically in the 
radio and mid-IR searches would be required to tie this down securely.

\section{CONCLUSION}
\label{sec:concl}
Drawing together the observational results by the community over the period
1999-2013, together with the theoretical concepts on exploding stars, the following conclusions are proposed
\begin{itemize}
\item There  are 45 supernovae with either detected progenitors or upper limits in the volume limited sample and either none (or one, if SN2009ip is included) has a 
progenitor above the luminosity limit of \logl$\simeq5.1$ (or an equivalent 
initial mass of $M\simeq18$\msun). A Salpeter IMF would suggest there should be 13 objects and the probability of finding 0 or 1 is $p = 3\times10^{-5}$. 
\item  The various possible biases that have been suggested to explain this deficit were explored in this review including circumstellar dust, selection bias and luminosity 
analysis errors. It does not seem likely that they can explain the deficit. 
\item One possible explanation is that most, or nearly all, stars above 
$M\gtrsim18$\msun\ evolve into WR stars that are hot and faint at the point of 
core-collapse when they produce Ibc SNe.  However this ignores the growing evidence that the Ibc SN population are produced mostly by binary systems with masses in 
the range $8 - 20$\msun. 
\item Explosion modelling by several theoretical groups actually predict that stars 
above the mass limit of  $M\simeq18$\msun\  have a tendency to produce failed SNe and form black holes. Above this limit there are islands of explodability that can 
produce neutron stars and successful explosions. There is no monotonic 
relationship between the core ``compactness" parameter (which will dictate the
likelihood of a successful explosion) and initial mass. There is 
reasonable agreement between the 
observed limits for the luminosity of progenitors and the theory of explosions. 
\item The nebular spectra of core-collapse SNe which have identified progenitors is a new and powerful way to probe the progenitor mass range and nucleosynthesis in 
massive stars. The results to date support the idea that the progenitors discovered to 
date are in the $8-17$\msun\ range and illustrate that the supernovae from 
high mass stars which  are thought to produce the bulk of cosmic oxygen
have not yet been found. 
\end{itemize}

\begin{acknowledgements}
 I thank  Morgan Fraser, John Eldridge, Anders Jerkstrand for their initial 
comments and  my  collaborators over many years in this area, including
Justyn Maund, Seppo Mattila, Mark Crockett and Maggie Hendry.  I also thank 
Dan Milisavljevic,
Rob Fesen, 
Dan Patnaude,
Jos\'{e} Groh,
Stan Woosley, 
Avishay Gal-Yam, 
Joe Anderson, 
Dovi Poznanski, 
Nolan Walborn, 
Edward Young
for comments on the initial  version and discussion. 
It is a pleasure to acknowledge funding from the European Research Council under the European Union's Seventh Framework Programme (FP7/2007-2013)/ERC Grant agreement n$^{\rm o}$ [291222] and STFC grants ST/I001123/1 and ST/L000709/1. I thank the organisers 
and participants of  the 2014 CAASTRO Annual Scientific Conference ``Supernovae in the Local Universe : celebrating 10,000 days of SN1987A'' for an immensely enjoyable meeting.
\end{acknowledgements}


\begin{thebibliography}{178}
\expandafter\ifx\csname natexlab\endcsname\relax\def\natexlab#1{#1}\fi

\bibitem[{{Aldering} {et~al.}(1994){Aldering}, {Humphreys}, \&
  {Richmond}}]{1994AJ....107..662A}
{Aldering}, G., {Humphreys}, R.~M., \& {Richmond}, M. 1994, \aj, 107, 662

\bibitem[{{Anderson} {et~al.}(2012){Anderson}, {Habergham}, {James}, \&
  {Hamuy}}]{2012MNRAS.424.1372A}
{Anderson}, J.~P., {Habergham}, S.~M., {James}, P.~A., \& {Hamuy}, M. 2012,
  \mnras, 424, 1372

\bibitem[{{Anderson} \& {James}(2008)}]{2008MNRAS.390.1527A}
{Anderson}, J.~P. \& {James}, P.~A. 2008, \mnras, 390, 1527

\bibitem[{{Benvenuto} {et~al.}(2013){Benvenuto}, {Bersten}, \&
  {Nomoto}}]{2013ApJ...762...74B}
{Benvenuto}, O.~G., {Bersten}, M.~C., \& {Nomoto}, K. 2013, \apj, 762, 74

\bibitem[{{Bersten} {et~al.}(2014){Bersten}, {Benvenuto}, {Folatelli},
  {Nomoto}, {Kuncarayakti}, {Srivastav}, {Anupama}, {Quimby}, \&
  {Sahu}}]{2014AJ....148...68B}
{Bersten}, M.~C., {Benvenuto}, O.~G., {Folatelli}, G., {Nomoto}, K.,
  {Kuncarayakti}, H., {Srivastav}, S., {Anupama}, G.~C., {Quimby}, R., \&
  {Sahu}, D.~K. 2014, \aj, 148, 68

\bibitem[{{Bose} {et~al.}(2013){Bose}, {Kumar}, {Sutaria}, {Kumar}, {Roy},
  {Bhatt}, {Pandey}, {Chandola}, {Sagar}, {Misra}, \&
  {Chakraborti}}]{2013MNRAS.433.1871B}
{Bose}, S., {Kumar}, B., {Sutaria}, F., {Kumar}, B., {Roy}, R., {Bhatt}, V.~K.,
  {Pandey}, S.~B., {Chandola}, H.~C., {Sagar}, R., {Misra}, K., \&
  {Chakraborti}, S. 2013, \mnras, 433, 1871

\bibitem[{{Botticella} {et~al.}(2009){Botticella}, {Pastorello}, {Smartt},
  {Meikle}, {Benetti}, {Kotak}, {Cappellaro}, {Crockett}, {Mattila}, {Sereno},
  {Patat}, {Tsvetkov}, {van Loon}, {Abraham}, {Agnoletto}, {Arbour}, {Benn},
  {di Rico}, {Elias-Rosa}, {Gorshanov}, {Harutyunyan}, {Hunter}, {Lorenzi},
  {Keenan}, {Maguire}, {Mendez}, {Mobberley}, {Navasardyan}, {Ries},
  {Stanishev}, {Taubenberger}, {Trundle}, {Turatto}, \&
  {Volkov}}]{2009MNRAS.398.1041B}
{Botticella}, M.~T., {Pastorello}, A., {Smartt}, S.~J., {Meikle}, W.~P.~S.,
  {Benetti}, S., {Kotak}, R., {Cappellaro}, E., {Crockett}, R.~M., {Mattila},
  S., {Sereno}, M., {Patat}, F., {Tsvetkov}, D., {van Loon}, J.~T., {Abraham},
  D., {Agnoletto}, I., {Arbour}, R., {Benn}, C., {di Rico}, G., {Elias-Rosa},
  N., {Gorshanov}, D.~L., {Harutyunyan}, A., {Hunter}, D., {Lorenzi}, V.,
  {Keenan}, F.~P., {Maguire}, K., {Mendez}, J., {Mobberley}, M., {Navasardyan},
  H., {Ries}, C., {Stanishev}, V., {Taubenberger}, S., {Trundle}, C.,
  {Turatto}, M., \& {Volkov}, I.~M. 2009, \mnras, 398, 1041

\bibitem[{{Botticella} {et~al.}(2012){Botticella}, {Smartt}, {Kennicutt},
  {Cappellaro}, {Sereno}, \& {Lee}}]{2012A&A...537A.132B}
{Botticella}, M.~T., {Smartt}, S.~J., {Kennicutt}, R.~C., {Cappellaro}, E.,
  {Sereno}, M., \& {Lee}, J.~C. 2012, \aap, 537, 132

\bibitem[{{Brown} \& {Woosley}(2013)}]{2013ApJ...769...99B}
{Brown}, J.~M. \& {Woosley}, S.~E. 2013, \apj, 769, 99

\bibitem[{{Cao} {et~al.}(2013){Cao}, {Kasliwal}, {Arcavi}, {Horesh}, {Hancock},
  {Valenti}, {Cenko}, {Kulkarni}, {Gal-Yam}, {Gorbikov}, {Ofek}, {Sand},
  {Yaron}, {Graham}, {Silverman}, {Wheeler}, {Marion}, {Walker}, {Mazzali},
  {Howell}, {Li}, {Kong}, {Bloom}, {Nugent}, {Surace}, {Masci}, {Carpenter},
  {Degenaar}, \& {Gelino}}]{2013ApJ...775L...7C}
{Cao}, Y., {Kasliwal}, M.~M., {Arcavi}, I., {Horesh}, A., {Hancock}, P.,
  {Valenti}, S., {Cenko}, S.~B., {Kulkarni}, S.~R., {Gal-Yam}, A., {Gorbikov},
  E., {Ofek}, E.~O., {Sand}, D., {Yaron}, O., {Graham}, M., {Silverman}, J.~M.,
  {Wheeler}, J.~C., {Marion}, G.~H., {Walker}, E.~S., {Mazzali}, P., {Howell},
  D.~A., {Li}, K.~L., {Kong}, A.~K.~H., {Bloom}, J.~S., {Nugent}, P.~E.,
  {Surace}, J., {Masci}, F., {Carpenter}, J., {Degenaar}, N., \& {Gelino},
  C.~R. 2013, \apjl, 775, L7

\bibitem[{{Chevalier} \& {Fransson}(2003)}]{2003LNP...598..171C}
{Chevalier}, R.~A. \& {Fransson}, C. 2003, in Lecture Notes in Physics, Berlin
  Springer Verlag, Vol. 598, Supernovae and Gamma-Ray Bursters, ed.
  K.~{Weiler}, 171--194

\bibitem[{{Chevalier} {et~al.}(2006){Chevalier}, {Fransson}, \&
  {Nymark}}]{2006ApJ...641.1029C}
{Chevalier}, R.~A., {Fransson}, C., \& {Nymark}, T.~K. 2006, \apj, 641, 1029

\bibitem[{{Crockett} {et~al.}(2008){Crockett}, {Eldridge}, {Smartt},
  {Pastorello}, {Gal-Yam}, {Fox}, {Leonard}, {Kasliwal}, {Mattila}, {Maund},
  {Stephens}, \& {Danziger}}]{2008MNRAS.391L...5C}
{Crockett}, R.~M., {Eldridge}, J.~J., {Smartt}, S.~J., {Pastorello}, A.,
  {Gal-Yam}, A., {Fox}, D.~B., {Leonard}, D.~C., {Kasliwal}, M.~M., {Mattila},
  S., {Maund}, J.~R., {Stephens}, A.~W., \& {Danziger}, I.~J. 2008, \mnras,
  391, L5

\bibitem[{{Crockett} {et~al.}(2011){Crockett}, {Smartt}, {Pastorello},
  {Eldridge}, {Stephens}, {Maund}, \& {Mattila}}]{2011MNRAS.410.2767C}
{Crockett}, R.~M., {Smartt}, S.~J., {Pastorello}, A., {Eldridge}, J.~J.,
  {Stephens}, A.~W., {Maund}, J.~R., \& {Mattila}, S. 2011, \mnras, 410, 2767

\bibitem[{{Crowther}(2007)}]{2007ARA&A..45..177C}
{Crowther}, P.~A. 2007, \araa, 45, 177

\bibitem[{{Crowther}(2013)}]{2013MNRAS.428.1927C}
---. 2013, \mnras, 428, 1927

\bibitem[{{Dall'Ora} {et~al.}(2014){Dall'Ora}, {Botticella}, {Pumo},
  {Zampieri}, {Tomasella}, {Pignata}, {Bayless}, {Pritchard}, {Taubenberger},
  {Kotak}, {Inserra}, {Della Valle}, {Cappellaro}, {Benetti}, {Benitez},
  {Bufano}, {Elias-Rosa}, {Fraser}, {Haislip}, {Harutyunyan}, {Howell},
  {Hsiao}, {Iijima}, {Kankare}, {Kuin}, {Maund}, {Morales-Garoffolo},
  {Morrell}, {Munari}, {Ochner}, {Pastorello}, {Patat}, {Phillips}, {Reichart},
  {Roming}, {Siviero}, {Smartt}, {Sollerman}, {Taddia}, {Valenti}, \&
  {Wright}}]{2014ApJ...787..139D}
{Dall'Ora}, M., {Botticella}, M.~T., {Pumo}, M.~L., {Zampieri}, L.,
  {Tomasella}, L., {Pignata}, G., {Bayless}, A.~J., {Pritchard}, T.~A.,
  {Taubenberger}, S., {Kotak}, R., {Inserra}, C., {Della Valle}, M.,
  {Cappellaro}, E., {Benetti}, S., {Benitez}, S., {Bufano}, F., {Elias-Rosa},
  N., {Fraser}, M., {Haislip}, J.~B., {Harutyunyan}, A., {Howell}, D.~A.,
  {Hsiao}, E.~Y., {Iijima}, T., {Kankare}, E., {Kuin}, P., {Maund}, J.~R.,
  {Morales-Garoffolo}, A., {Morrell}, N., {Munari}, U., {Ochner}, P.,
  {Pastorello}, A., {Patat}, F., {Phillips}, M.~M., {Reichart}, D., {Roming},
  P.~W.~A., {Siviero}, A., {Smartt}, S.~J., {Sollerman}, J., {Taddia}, F.,
  {Valenti}, S., \& {Wright}, D. 2014, \apj, 787, 139

\bibitem[{{Davies} {et~al.}(2013){Davies}, {Kudritzki}, {Plez}, {Trager},
  {Lan{\c c}on}, {Gazak}, {Bergemann}, {Evans}, \&
  {Chiavassa}}]{2013ApJ...767....3D}
{Davies}, B., {Kudritzki}, R.-P., {Plez}, B., {Trager}, S., {Lan{\c c}on}, A.,
  {Gazak}, Z., {Bergemann}, M., {Evans}, C., \& {Chiavassa}, A. 2013, \apj,
  767, 3

\bibitem[{{Dessart} {et~al.}(2010){Dessart}, {Livne}, \&
  {Waldman}}]{2010MNRAS.408..827D}
{Dessart}, L., {Livne}, E., \& {Waldman}, R. 2010, \mnras, 408, 827

\bibitem[{{Drout} {et~al.}(2011){Drout}, {Soderberg}, {Gal-Yam}, {Cenko},
  {Fox}, {Leonard}, {Sand}, {Moon}, {Arcavi}, \& {Green}}]{2011ApJ...741...97D}
{Drout}, M.~R., {Soderberg}, A.~M., {Gal-Yam}, A., {Cenko}, S.~B., {Fox},
  D.~B., {Leonard}, D.~C., {Sand}, D.~J., {Moon}, D.-S., {Arcavi}, I., \&
  {Green}, Y. 2011, \apj, 741, 97

\bibitem[{{Dwarkadas}(2014)}]{2014MNRAS.440.1917D}
{Dwarkadas}, V.~V. 2014, \mnras, 440, 1917

\bibitem[{{Dwarkadas} \& {Gruszko}(2012)}]{2012MNRAS.419.1515D}
{Dwarkadas}, V.~V. \& {Gruszko}, J. 2012, \mnras, 419, 1515

\bibitem[{{Eldridge} {et~al.}(2015){Eldridge}, {Fraser}, {Maund}, \&
  {Smartt}}]{2015MNRAS.446.2689E}
{Eldridge}, J.~J., {Fraser}, M., {Maund}, J.~R., \& {Smartt}, S.~J. 2015,
  \mnras, 446, 2689

\bibitem[{{Eldridge} {et~al.}(2013){Eldridge}, {Fraser}, {Smartt}, {Maund}, \&
  {Crockett}}]{2013MNRAS.436..774E}
{Eldridge}, J.~J., {Fraser}, M., {Smartt}, S.~J., {Maund}, J.~R., \&
  {Crockett}, R.~M. 2013, \mnras, 436, 774

\bibitem[{{Eldridge} {et~al.}(2008){Eldridge}, {Izzard}, \&
  {Tout}}]{2008MNRAS.384.1109E}
{Eldridge}, J.~J., {Izzard}, R.~G., \& {Tout}, C.~A. 2008, \mnras, 384, 1109

\bibitem[{{Eldridge} {et~al.}(2007){Eldridge}, {Mattila}, \&
  {Smartt}}]{2007MNRAS.376L..52E}
{Eldridge}, J.~J., {Mattila}, S., \& {Smartt}, S.~J. 2007, \mnras, 376, L52

\bibitem[{{Eldridge} \& {Tout}(2004)}]{2004MNRAS.353...87E}
{Eldridge}, J.~J. \& {Tout}, C.~A. 2004, \mnras, 353, 87

\bibitem[{{Elias-Rosa} {et~al.}(2013){Elias-Rosa}, {Pastorello}, {Maund},
  {Tak{\'a}ts}, {Fraser}, {Smartt}, {Benetti}, {Pignata}, {Sand}, \&
  {Valenti}}]{2013MNRAS.436L.109E}
{Elias-Rosa}, N., {Pastorello}, A., {Maund}, J.~R., {Tak{\'a}ts}, K., {Fraser},
  M., {Smartt}, S.~J., {Benetti}, S., {Pignata}, G., {Sand}, D., \& {Valenti},
  S. 2013, \mnras, 436, L109

\bibitem[{{Elias-Rosa} {et~al.}(2010){Elias-Rosa}, {Van Dyk}, {Li}, {Miller},
  {Silverman}, {Ganeshalingam}, {Boden}, {Kasliwal}, {Vink{\'o}}, {Cuillandre},
  {Filippenko}, {Steele}, {Bloom}, {Griffith}, {Kleiser}, \&
  {Foley}}]{2010ApJ...714L.254E}
{Elias-Rosa}, N., {Van Dyk}, S.~D., {Li}, W., {Miller}, A.~A., {Silverman},
  J.~M., {Ganeshalingam}, M., {Boden}, A.~F., {Kasliwal}, M.~M., {Vink{\'o}},
  J., {Cuillandre}, J.-C., {Filippenko}, A.~V., {Steele}, T.~N., {Bloom},
  J.~S., {Griffith}, C.~V., {Kleiser}, I.~K.~W., \& {Foley}, R.~J. 2010, \apjl,
  714, L254

\bibitem[{{Elias-Rosa} {et~al.}(2009){Elias-Rosa}, {Van Dyk}, {Li}, {Morrell},
  {Gonzalez}, {Hamuy}, {Filippenko}, {Cuillandre}, {Foley}, \&
  {Smith}}]{2009ApJ...706.1174E}
{Elias-Rosa}, N., {Van Dyk}, S.~D., {Li}, W., {Morrell}, N., {Gonzalez}, S.,
  {Hamuy}, M., {Filippenko}, A.~V., {Cuillandre}, J.-C., {Foley}, R.~J., \&
  {Smith}, N. 2009, \apj, 706, 1174

\bibitem[{{Elias-Rosa} {et~al.}(2011){Elias-Rosa}, {Van Dyk}, {Li},
  {Silverman}, {Foley}, {Ganeshalingam}, {Mauerhan}, {Kankare}, {Jha},
  {Filippenko}, {Beckman}, {Berger}, {Cuillandre}, \&
  {Smith}}]{2011ApJ...742....6E}
{Elias-Rosa}, N., {Van Dyk}, S.~D., {Li}, W., {Silverman}, J.~M., {Foley},
  R.~J., {Ganeshalingam}, M., {Mauerhan}, J.~C., {Kankare}, E., {Jha}, S.,
  {Filippenko}, A.~V., {Beckman}, J.~E., {Berger}, E., {Cuillandre}, J.-C., \&
  {Smith}, N. 2011, \apj, 742, 6

\bibitem[{{Eriksen} {et~al.}(2009){Eriksen}, {Arnett}, {McCarthy}, \&
  {Young}}]{2009ApJ...697...29E}
{Eriksen}, K.~A., {Arnett}, D., {McCarthy}, D.~W., \& {Young}, P. 2009, \apj,
  697, 29

\bibitem[{{Filippenko}(1997)}]{1997ARA&A..35..309F}
{Filippenko}, A.~V. 1997, \araa, 35, 309

\bibitem[{{Filippenko} {et~al.}(1993){Filippenko}, {Matheson}, \&
  {Ho}}]{1993ApJ...415L.103F}
{Filippenko}, A.~V., {Matheson}, T., \& {Ho}, L.~C. 1993, \apjl, 415, L103

\bibitem[{{Folatelli} {et~al.}(2014){Folatelli}, {Bersten}, {Benvenuto}, {Van
  Dyk}, {Kuncarayakti}, {Maeda}, {Nozawa}, {Nomoto}, {Hamuy}, \&
  {Quimby}}]{2014ApJ...793L..22F}
{Folatelli}, G., {Bersten}, M.~C., {Benvenuto}, O.~G., {Van Dyk}, S.~D.,
  {Kuncarayakti}, H., {Maeda}, K., {Nozawa}, T., {Nomoto}, K., {Hamuy}, M., \&
  {Quimby}, R.~M. 2014, \apjl, 793, L22

\bibitem[{{Foley} {et~al.}(2011){Foley}, {Berger}, {Fox}, {Levesque},
  {Challis}, {Ivans}, {Rhoads}, \& {Soderberg}}]{2011ApJ...732...32F}
{Foley}, R.~J., {Berger}, E., {Fox}, O., {Levesque}, E.~M., {Challis}, P.~J.,
  {Ivans}, I.~I., {Rhoads}, J.~E., \& {Soderberg}, A.~M. 2011, \apj, 732, 32

\bibitem[{{Fox} {et~al.}(2014){Fox}, {Azalee Bostroem}, {Van Dyk},
  {Filippenko}, {Fransson}, {Matheson}, {Cenko}, {Chandra}, {Dwarkadas}, {Li},
  {Parker}, \& {Smith}}]{2014ApJ...790...17F}
{Fox}, O.~D., {Azalee Bostroem}, K., {Van Dyk}, S.~D., {Filippenko}, A.~V.,
  {Fransson}, C., {Matheson}, T., {Cenko}, S.~B., {Chandra}, P., {Dwarkadas},
  V., {Li}, W., {Parker}, A.~H., \& {Smith}, N. 2014, \apj, 790, 17

\bibitem[{{Fraser}(2011)}]{2011PhD_MFT}
{Fraser}, M. 2011, PhD thesis, Queen's University of Belfast, (2011)

\bibitem[{{Fraser} {et~al.}(2011){Fraser}, {Ergon}, {Eldridge}, {Valenti},
  {Pastorello}, {Sollerman}, {Smartt}, {Agnoletto}, {Arcavi}, {Benetti},
  {Botticella}, {Bufano}, {Campillay}, {Crockett}, {Gal-Yam}, {Kankare},
  {Leloudas}, {Maguire}, {Mattila}, {Maund}, {Salgado}, {Stephens},
  {Taubenberger}, \& {Turatto}}]{2011MNRAS.417.1417F}
{Fraser}, M., {Ergon}, M., {Eldridge}, J.~J., {Valenti}, S., {Pastorello}, A.,
  {Sollerman}, J., {Smartt}, S.~J., {Agnoletto}, I., {Arcavi}, I., {Benetti},
  S., {Botticella}, M.-T., {Bufano}, F., {Campillay}, A., {Crockett}, R.~M.,
  {Gal-Yam}, A., {Kankare}, E., {Leloudas}, G., {Maguire}, K., {Mattila}, S.,
  {Maund}, J.~R., {Salgado}, F., {Stephens}, A., {Taubenberger}, S., \&
  {Turatto}, M. 2011, \mnras, 417, 1417

\bibitem[{{Fraser} {et~al.}(2013{\natexlab{a}}){Fraser}, {Inserra},
  {Jerkstrand}, {Kotak}, {Pignata}, {Benetti}, {Botticella}, {Bufano},
  {Childress}, {Mattila}, {Pastorello}, {Smartt}, {Turatto}, {Yuan},
  {Anderson}, {Bayliss}, {Bauer}, {Chen}, {F{\"o}rster Bur{\'o}n}, {Gal-Yam},
  {Haislip}, {Knapic}, {Le Guillou}, {Marchi}, {Mazzali}, {Molinaro}, {Moore},
  {Reichart}, {Smareglia}, {Smith}, {Sternberg}, {Sullivan}, {Tak{\'a}ts},
  {Tucker}, {Valenti}, {Yaron}, {Young}, \& {Zhou}}]{2013MNRAS.433.1312F}
{Fraser}, M., {Inserra}, C., {Jerkstrand}, A., {Kotak}, R., {Pignata}, G.,
  {Benetti}, S., {Botticella}, M.-T., {Bufano}, F., {Childress}, M., {Mattila},
  S., {Pastorello}, A., {Smartt}, S.~J., {Turatto}, M., {Yuan}, F., {Anderson},
  J.~P., {Bayliss}, D.~D.~R., {Bauer}, F.~E., {Chen}, T.-W., {F{\"o}rster
  Bur{\'o}n}, F., {Gal-Yam}, A., {Haislip}, J.~B., {Knapic}, C., {Le Guillou},
  L., {Marchi}, S., {Mazzali}, P., {Molinaro}, M., {Moore}, J.~P., {Reichart},
  D., {Smareglia}, R., {Smith}, K.~W., {Sternberg}, A., {Sullivan}, M.,
  {Tak{\'a}ts}, K., {Tucker}, B.~E., {Valenti}, S., {Yaron}, O., {Young},
  D.~R., \& {Zhou}, G. 2013{\natexlab{a}}, \mnras, 433, 1312

\bibitem[{{Fraser} {et~al.}(2013{\natexlab{b}}){Fraser}, {Magee}, {Kotak},
  {Smartt}, {Smith}, {Polshaw}, {Drake}, {Boles}, {Lee}, {Burgett}, {Chambers},
  {Draper}, {Flewelling}, {Hodapp}, {Kaiser}, {Kudritzki}, {Magnier}, {Price},
  {Tonry}, {Wainscoat}, \& {Waters}}]{2013ApJ...779L...8F}
{Fraser}, M., {Magee}, M., {Kotak}, R., {Smartt}, S.~J., {Smith}, K.~W.,
  {Polshaw}, J., {Drake}, A.~J., {Boles}, T., {Lee}, C.-H., {Burgett}, W.~S.,
  {Chambers}, K.~C., {Draper}, P.~W., {Flewelling}, H., {Hodapp}, K.~W.,
  {Kaiser}, N., {Kudritzki}, R.-P., {Magnier}, E.~A., {Price}, P.~A., {Tonry},
  J.~L., {Wainscoat}, R.~J., \& {Waters}, C. 2013{\natexlab{b}}, \apjl, 779, L8

\bibitem[{{Fraser} {et~al.}(2012){Fraser}, {Maund}, {Smartt}, {Botticella},
  {Dall'Ora}, {Inserra}, {Tomasella}, {Benetti}, {Ciroi}, {Eldridge}, {Ergon},
  {Kotak}, {Mattila}, {Ochner}, {Pastorello}, {Reilly}, {Sollerman},
  {Stephens}, {Taddia}, \& {Valenti}}]{2012ApJ...759L..13F}
{Fraser}, M., {Maund}, J.~R., {Smartt}, S.~J., {Botticella}, M.-T., {Dall'Ora},
  M., {Inserra}, C., {Tomasella}, L., {Benetti}, S., {Ciroi}, S., {Eldridge},
  J.~J., {Ergon}, M., {Kotak}, R., {Mattila}, S., {Ochner}, P., {Pastorello},
  A., {Reilly}, E., {Sollerman}, J., {Stephens}, A., {Taddia}, F., \&
  {Valenti}, S. 2012, \apjl, 759, L13

\bibitem[{{Fraser} {et~al.}(2014){Fraser}, {Maund}, {Smartt}, {Kotak},
  {Lawrence}, {Bruce}, {Valenti}, {Yuan}, {Benetti}, {Chen}, {Gal-Yam},
  {Inserra}, \& {Young}}]{2014MNRAS.439L..56F}
{Fraser}, M., {Maund}, J.~R., {Smartt}, S.~J., {Kotak}, R., {Lawrence}, A.,
  {Bruce}, A., {Valenti}, S., {Yuan}, F., {Benetti}, S., {Chen}, T.-W.,
  {Gal-Yam}, A., {Inserra}, C., \& {Young}, D.~R. 2014, \mnras, 439, L56

\bibitem[{{Fraser} {et~al.}(2010){Fraser}, {Tak{\'a}ts}, {Pastorello},
  {Smartt}, {Mattila}, {Botticella}, {Valenti}, {Ergon}, {Sollerman}, {Arcavi},
  {Benetti}, {Bufano}, {Crockett}, {Danziger}, {Gal-Yam}, {Maund},
  {Taubenberger}, \& {Turatto}}]{2010ApJ...714L.280F}
{Fraser}, M., {Tak{\'a}ts}, K., {Pastorello}, A., {Smartt}, S.~J., {Mattila},
  S., {Botticella}, M.-T., {Valenti}, S., {Ergon}, M., {Sollerman}, J.,
  {Arcavi}, I., {Benetti}, S., {Bufano}, F., {Crockett}, R.~M., {Danziger},
  I.~J., {Gal-Yam}, A., {Maund}, J.~R., {Taubenberger}, S., \& {Turatto}, M.
  2010, \apjl, 714, L280

\bibitem[{{Fremling} {et~al.}(2014){Fremling}, {Sollerman}, {Taddia}, {Ergon},
  {Valenti}, {Arcavi}, {Ben-Ami}, {Cao}, {Cenko}, {Filippenko}, {Gal-Yam}, \&
  {Howell}}]{2014A&A...565A.114F}
{Fremling}, C., {Sollerman}, J., {Taddia}, F., {Ergon}, M., {Valenti}, S.,
  {Arcavi}, I., {Ben-Ami}, S., {Cao}, Y., {Cenko}, S.~B., {Filippenko}, A.~V.,
  {Gal-Yam}, A., \& {Howell}, D.~A. 2014, \aap, 565, 114

\bibitem[{{Fryer}(1999)}]{1999ApJ...522..413F}
{Fryer}, C.~L. 1999, \apj, 522, 413

\bibitem[{{Gal-Yam} {et~al.}(2014){Gal-Yam}, {Arcavi}, {Ofek}, {Ben-Ami},
  {Cenko}, {Kasliwal}, {Cao}, {Yaron}, {Tal}, {Silverman}, {Horesh}, {De Cia},
  {Taddia}, {Sollerman}, {Perley}, {Vreeswijk}, {Kulkarni}, {Nugent},
  {Filippenko}, \& {Wheeler}}]{2014Natur.509..471G}
{Gal-Yam}, A., {Arcavi}, I., {Ofek}, E.~O., {Ben-Ami}, S., {Cenko}, S.~B.,
  {Kasliwal}, M.~M., {Cao}, Y., {Yaron}, O., {Tal}, D., {Silverman}, J.~M.,
  {Horesh}, A., {De Cia}, A., {Taddia}, F., {Sollerman}, J., {Perley}, D.,
  {Vreeswijk}, P.~M., {Kulkarni}, S.~R., {Nugent}, P.~E., {Filippenko}, A.~V.,
  \& {Wheeler}, J.~C. 2014, \nat, 509, 471

\bibitem[{{Gal-Yam} \& {Leonard}(2009)}]{2009Natur.458..865G}
{Gal-Yam}, A. \& {Leonard}, D.~C. 2009, \nat, 458, 865

\bibitem[{{Gal-Yam} {et~al.}(2007){Gal-Yam}, {Leonard}, {Fox}, {Cenko},
  {Soderberg}, {Moon}, {Sand}, {Li}, {Filippenko}, {Aldering}, \&
  {Copin}}]{2007ApJ...656..372G}
{Gal-Yam}, A., {Leonard}, D.~C., {Fox}, D.~B., {Cenko}, S.~B., {Soderberg},
  A.~M., {Moon}, D.-S., {Sand}, D.~J., {Li}, W., {Filippenko}, A.~V.,
  {Aldering}, G., \& {Copin}, Y. 2007, \apj, 656, 372

\bibitem[{{Gerke} {et~al.}(2014){Gerke}, {Kochanek}, \&
  {Stanek}}]{2014arXiv1411.1761G}
{Gerke}, J.~R., {Kochanek}, C.~S., \& {Stanek}, K.~Z. 2014, MNRAS, in press, arXiv:1411.1761

\bibitem[{{Graham} {et~al.}(2014){Graham}, {Sand}, {Valenti}, {Howell},
  {Parrent}, {Halford}, {Zaritsky}, {Bianco}, {Rest}, \&
  {Dilday}}]{2014ApJ...787..163G}
{Graham}, M.~L., {Sand}, D.~J., {Valenti}, S., {Howell}, D.~A., {Parrent}, J.,
  {Halford}, M., {Zaritsky}, D., {Bianco}, F., {Rest}, A., \& {Dilday}, B.
  2014, \apj, 787, 163

\bibitem[{{Graur} \& {Maoz}(2012)}]{2012ATel.4535....1G}
{Graur}, O. \& {Maoz}, D. 2012, The Astronomer's Telegram, 4535, 1

\bibitem[{{Groh}(2014)}]{2014A&A...572L..11G}
{Groh}, J.~H. 2014, \aap, 572, L11

\bibitem[{{Groh} {et~al.}(2013{\natexlab{a}}){Groh}, {Georgy}, \&
  {Ekstr{\"o}m}}]{2013A&A...558L...1G}
{Groh}, J.~H., {Georgy}, C., \& {Ekstr{\"o}m}, S. 2013{\natexlab{a}}, \aap,
  558, L1

\bibitem[{{Groh} {et~al.}(2013{\natexlab{b}}){Groh}, {Meynet}, \&
  {Ekstr{\"o}m}}]{2013A&A...550L...7G}
{Groh}, J.~H., {Meynet}, G., \& {Ekstr{\"o}m}, S. 2013{\natexlab{b}}, \aap,
  550, L7

\bibitem[{{Groh} {et~al.}(2013{\natexlab{c}}){Groh}, {Meynet}, {Georgy}, \&
  {Ekstr{\"o}m}}]{2013A&A...558A.131G}
{Groh}, J.~H., {Meynet}, G., {Georgy}, C., \& {Ekstr{\"o}m}, S.
  2013{\natexlab{c}}, \aap, 558, A131

\bibitem[{{Hendry} {et~al.}(2006){Hendry}, {Smartt}, {Crockett}, {Maund},
  {Gal-Yam}, {Moon}, {Cenko}, {Fox}, {Kudritzki}, {Benn}, \&
  {{\O}stensen}}]{2006MNRAS.369.1303H}
{Hendry}, M.~A., {Smartt}, S.~J., {Crockett}, R.~M., {Maund}, J.~R., {Gal-Yam},
  A., {Moon}, D.-S., {Cenko}, S.~B., {Fox}, D.~W., {Kudritzki}, R.~P., {Benn},
  C.~R., \& {{\O}stensen}, R. 2006, \mnras, 369, 1303

\bibitem[{{Hirschi} {et~al.}(2004){Hirschi}, {Meynet}, \&
  {Maeder}}]{2004A&A...425..649H}
{Hirschi}, R., {Meynet}, G., \& {Maeder}, A. 2004, \aap, 425, 649

\bibitem[{{Horiuchi} {et~al.}(2011){Horiuchi}, {Beacom}, {Kochanek}, {Prieto},
  {Stanek}, \& {Thompson}}]{2011ApJ...738..154H}
{Horiuchi}, S., {Beacom}, J.~F., {Kochanek}, C.~S., {Prieto}, J.~L., {Stanek},
  K.~Z., \& {Thompson}, T.~A. 2011, \apj, 738, 154

\bibitem[{{Horiuchi} {et~al.}(2014){Horiuchi}, {Nakamura}, {Takiwaki},
  {Kotake}, \& {Tanaka}}]{2014MNRAS.445L..99H}
{Horiuchi}, S., {Nakamura}, K., {Takiwaki}, T., {Kotake}, K., \& {Tanaka}, M.
  2014, \mnras, 445, L99

\bibitem[{{Hwang} \& {Laming}(2009)}]{2009ApJ...703..883H}
{Hwang}, U. \& {Laming}, J.~M. 2009, \apj, 703, 883

\bibitem[{{Hwang} \& {Laming}(2012)}]{2012ApJ...746..130H}
---. 2012, \apj, 746, 130

\bibitem[{{Ivezic} \& {Elitzur}(1997)}]{1997MNRAS.287..799I}
{Ivezic}, Z. \& {Elitzur}, M. 1997, \mnras, 287, 799

\bibitem[{{Jennings} {et~al.}(2014){Jennings}, {Williams}, {Murphy},
  {Dalcanton}, {Gilbert}, {Dolphin}, {Weisz}, \&
  {Fouesneau}}]{2014ApJ...795..170J}
{Jennings}, Z.~G., {Williams}, B.~F., {Murphy}, J.~W., {Dalcanton}, J.~J.,
  {Gilbert}, K.~M., {Dolphin}, A.~E., {Weisz}, D.~R., \& {Fouesneau}, M. 2014,
  \apj, 795, 170

\bibitem[{{Jerkstrand}(2011)}]{2011PhDT........90J}
{Jerkstrand}, A. 2011, PhD thesis, PhD Thesis, University of Stockholm, Faculty
  of Science, Department of Astronomy (2011).~Advisor: Claes Fransson

\bibitem[{{Jerkstrand} {et~al.}(2015{\natexlab{a}}){Jerkstrand}, {Ergon},
  {Smartt}, {Fransson}, {Sollerman}, {Taubenberger}, {Bersten}, \&
  {Spyromilio}}]{2015A&A...573A..12J}
{Jerkstrand}, A., {Ergon}, M., {Smartt}, S.~J., {Fransson}, C., {Sollerman},
  J., {Taubenberger}, S., {Bersten}, M., \& {Spyromilio}, J.
  2015{\natexlab{a}}, \aap, 573, 12

\bibitem[{{Jerkstrand} {et~al.}(2011){Jerkstrand}, {Fransson}, \&
  {Kozma}}]{2011A&A...530A..45J}
{Jerkstrand}, A., {Fransson}, C., \& {Kozma}, C. 2011, \aap, 530, 45

\bibitem[{{Jerkstrand} {et~al.}(2012){Jerkstrand}, {Fransson}, {Maguire},
  {Smartt}, {Ergon}, \& {Spyromilio}}]{2012A&A...546A..28J}
{Jerkstrand}, A., {Fransson}, C., {Maguire}, K., {Smartt}, S., {Ergon}, M., \&
  {Spyromilio}, J. 2012, \aap, 546, 28

\bibitem[{{Jerkstrand} {et~al.}(2014){Jerkstrand}, {Smartt}, {Fraser},
  {Fransson}, {Sollerman}, {Taddia}, \& {Kotak}}]{2014MNRAS.439.3694J}
{Jerkstrand}, A., {Smartt}, S.~J., {Fraser}, M., {Fransson}, C., {Sollerman},
  J., {Taddia}, F., \& {Kotak}, R. 2014, \mnras, 439, 3694

\bibitem[{{Jerkstrand} {et~al.}(2015{\natexlab{b}}){Jerkstrand}, {Smartt},
  {Sollerman}, {Inserra}, {Fraser}, {Spyromilio}, {Fransson}, {Chen},
  {Barbarino}, {Dall'Ora}, {Botticella}, {Della Valle}, {Gal-Yam}, {Valenti},
  {Maguire}, {Mazzali}, \& {Tomasella}}]{2015MNRAS.448.2482J}
{Jerkstrand}, A., {Smartt}, S.~J., {Sollerman}, J., {Inserra}, C., {Fraser},
  M., {Spyromilio}, J., {Fransson}, C., {Chen}, T.-W., {Barbarino}, C.,
  {Dall'Ora}, M., {Botticella}, M.~T., {Della Valle}, M., {Gal-Yam}, A.,
  {Valenti}, S., {Maguire}, K., {Mazzali}, P., \& {Tomasella}, L.
  2015{\natexlab{b}}, \mnras, 448, 2482

\bibitem[{{Kennicutt} {et~al.}(2003){Kennicutt}, {Armus}, {Bendo}, {Calzetti},
  {Dale}, {Draine}, {Engelbracht}, {Gordon}, {Grauer}, {Helou}, {Hollenbach},
  {Jarrett}, {Kewley}, {Leitherer}, {Li}, {Malhotra}, {Regan}, {Rieke},
  {Rieke}, {Roussel}, {Smith}, {Thornley}, \& {Walter}}]{2003PASP..115..928K}
{Kennicutt}, Jr., R.~C., {Armus}, L., {Bendo}, G., {Calzetti}, D., {Dale},
  D.~A., {Draine}, B.~T., {Engelbracht}, C.~W., {Gordon}, K.~D., {Grauer},
  A.~D., {Helou}, G., {Hollenbach}, D.~J., {Jarrett}, T.~H., {Kewley}, L.~J.,
  {Leitherer}, C., {Li}, A., {Malhotra}, S., {Regan}, M.~W., {Rieke}, G.~H.,
  {Rieke}, M.~J., {Roussel}, H., {Smith}, J.-D.~T., {Thornley}, M.~D., \&
  {Walter}, F. 2003, \pasp, 115, 928

\bibitem[{{Kochanek}(2014)}]{2014ApJ...785...28K}
{Kochanek}, C.~S. 2014, \apj, 785, 28

\bibitem[{{Kochanek} {et~al.}(2008){Kochanek}, {Beacom}, {Kistler}, {Prieto},
  {Stanek}, {Thompson}, \& {Y{\"u}ksel}}]{2008ApJ...684.1336K}
{Kochanek}, C.~S., {Beacom}, J.~F., {Kistler}, M.~D., {Prieto}, J.~L.,
  {Stanek}, K.~Z., {Thompson}, T.~A., \& {Y{\"u}ksel}, H. 2008, \apj, 684, 1336

\bibitem[{{Kochanek} {et~al.}(2012){Kochanek}, {Khan}, \&
  {Dai}}]{2012ApJ...759...20K}
{Kochanek}, C.~S., {Khan}, R., \& {Dai}, X. 2012, \apj, 759, 20

\bibitem[{{Kotak} \& {Vink}(2006)}]{2006A&A...460L...5K}
{Kotak}, R. \& {Vink}, J.~S. 2006, \aap, 460, L5

\bibitem[{{Kozma} \& {Fransson}(1998)}]{1998ApJ...497..431K}
{Kozma}, C. \& {Fransson}, C. 1998, \apj, 497, 431

\bibitem[{{Krause} {et~al.}(2008){Krause}, {Tanaka}, {Usuda}, {Hattori},
  {Goto}, {Birkmann}, \& {Nomoto}}]{2008Natur.456..617K}
{Krause}, O., {Tanaka}, M., {Usuda}, T., {Hattori}, T., {Goto}, M., {Birkmann},
  S., \& {Nomoto}, K. 2008, \nat, 456, 617

\bibitem[{{Kudritzki} \& {Puls}(2000)}]{2000ARA&A..38..613K}
{Kudritzki}, R.-P. \& {Puls}, J. 2000, \araa, 38, 613

\bibitem[{{Langer}(2012)}]{2012ARA&A..50..107L}
{Langer}, N. 2012, \araa, 50, 107

\bibitem[{{Leaman} {et~al.}(2011){Leaman}, {Li}, {Chornock}, \&
  {Filippenko}}]{2011MNRAS.412.1419L}
{Leaman}, J., {Li}, W., {Chornock}, R., \& {Filippenko}, A.~V. 2011, \mnras,
  412, 1419

\bibitem[{{Levesque} {et~al.}(2006){Levesque}, {Massey}, {Olsen}, {Plez},
  {Meynet}, \& {Maeder}}]{2006ApJ...645.1102L}
{Levesque}, E.~M., {Massey}, P., {Olsen}, K.~A.~G., {Plez}, B., {Meynet}, G.,
  \& {Maeder}, A. 2006, \apj, 645, 1102

\bibitem[{{Levesque} {et~al.}(2009){Levesque}, {Massey}, {Plez}, \&
  {Olsen}}]{2009AJ....137.4744L}
{Levesque}, E.~M., {Massey}, P., {Plez}, B., \& {Olsen}, K.~A.~G. 2009, \aj,
  137, 4744

\bibitem[{{Li} {et~al.}(2009){Li}, {Cenko}, \&
  {Filippenko}}]{2009CBET.1656....1L}
{Li}, W., {Cenko}, S.~B., \& {Filippenko}, A.~V. 2009, Central Bureau
  Electronic Telegrams, 1656, 1

\bibitem[{{Li} {et~al.}(2011){Li}, {Leaman}, {Chornock}, {Filippenko},
  {Poznanski}, {Ganeshalingam}, {Wang}, {Modjaz}, {Jha}, {Foley}, \&
  {Smith}}]{2011MNRAS.412.1441L}
{Li}, W., {Leaman}, J., {Chornock}, R., {Filippenko}, A.~V., {Poznanski}, D.,
  {Ganeshalingam}, M., {Wang}, X., {Modjaz}, M., {Jha}, S., {Foley}, R.~J., \&
  {Smith}, N. 2011, \mnras, 412, 1441

\bibitem[{{Li} {et~al.}(2005){Li}, {Van Dyk}, {Filippenko}, \&
  {Cuillandre}}]{2005PASP..117..121L}
{Li}, W., {Van Dyk}, S.~D., {Filippenko}, A.~V., \& {Cuillandre}, J.-C. 2005,
  \pasp, 117, 121

\bibitem[{{Li} {et~al.}(2006){Li}, {Van Dyk}, {Filippenko}, {Cuillandre},
  {Jha}, {Bloom}, {Riess}, \& {Livio}}]{2006ApJ...641.1060L}
{Li}, W., {Van Dyk}, S.~D., {Filippenko}, A.~V., {Cuillandre}, J.-C., {Jha},
  S., {Bloom}, J.~S., {Riess}, A.~G., \& {Livio}, M. 2006, \apj, 641, 1060

\bibitem[{{Li} {et~al.}(2007){Li}, {Wang}, {Van Dyk}, {Cuillandre}, {Foley}, \&
  {Filippenko}}]{2007ApJ...661.1013L}
{Li}, W., {Wang}, X., {Van Dyk}, S.~D., {Cuillandre}, J.-C., {Foley}, R.~J., \&
  {Filippenko}, A.~V. 2007, \apj, 661, 1013

\bibitem[{{Lovegrove} \& {Woosley}(2013)}]{2013ApJ...769..109L}
{Lovegrove}, E. \& {Woosley}, S.~E. 2013, \apj, 769, 109

\bibitem[{{Lyman} {et~al.}(2014){Lyman}, {Bersier}, {James}, {Mazzali},
  {Eldridge}, {Fraser}, \& {Pian}}]{2014arXiv1406.3667L}
{Lyman}, J., {Bersier}, D., {James}, P., {Mazzali}, P., {Eldridge}, J.,
  {Fraser}, M., \& {Pian}, E. 2014, MNRAS, submitted, arXiv:1406.3667

\bibitem[{{Mackey} {et~al.}(2014){Mackey}, {Mohamed}, {Gvaramadze}, {Kotak},
  {Langer}, {Meyer}, {Moriya}, \& {Neilson}}]{2014Natur.512..282M}
{Mackey}, J., {Mohamed}, S., {Gvaramadze}, V.~V., {Kotak}, R., {Langer}, N.,
  {Meyer}, D.~M.-A., {Moriya}, T.~J., \& {Neilson}, H.~R. 2014, \nat, 512, 282

\bibitem[{{Maguire} {et~al.}(2012){Maguire}, {Jerkstrand}, {Smartt},
  {Fransson}, {Pastorello}, {Benetti}, {Valenti}, {Bufano}, \&
  {Leloudas}}]{2012MNRAS.420.3451M}
{Maguire}, K., {Jerkstrand}, A., {Smartt}, S.~J., {Fransson}, C., {Pastorello},
  A., {Benetti}, S., {Valenti}, S., {Bufano}, F., \& {Leloudas}, G. 2012,
  \mnras, 420, 3451

\bibitem[{{Ma{\'{\i}}z-Apell{\'a}niz}
  {et~al.}(2004){Ma{\'{\i}}z-Apell{\'a}niz}, {Bond}, {Siegel}, {Lipkin},
  {Maoz}, {Ofek}, \& {Poznanski}}]{2004ApJ...615L.113M}
{Ma{\'{\i}}z-Apell{\'a}niz}, J., {Bond}, H.~E., {Siegel}, M.~H., {Lipkin}, Y.,
  {Maoz}, D., {Ofek}, E.~O., \& {Poznanski}, D. 2004, \apjl, 615, L113

\bibitem[{{Margutti} {et~al.}(2014){Margutti}, {Milisavljevic}, {Soderberg},
  {Chornock}, {Zauderer}, {Murase}, {Guidorzi}, {Sanders}, {Kuin}, {Fransson},
  {Levesque}, {Chandra}, {Berger}, {Bianco}, {Brown}, {Challis},
  {Chatzopoulos}, {Cheung}, {Choi}, {Chomiuk}, {Chugai}, {Contreras}, {Drout},
  {Fesen}, {Foley}, {Fong}, {Friedman}, {Gall}, {Gehrels}, {Hjorth}, {Hsiao},
  {Kirshner}, {Im}, {Leloudas}, {Lunnan}, {Marion}, {Martin}, {Morrell},
  {Neugent}, {Omodei}, {Phillips}, {Rest}, {Silverman}, {Strader},
  {Stritzinger}, {Szalai}, {Utterback}, {Vinko}, {Wheeler}, {Arnett},
  {Campana}, {Chevalier}, {Ginsburg}, {Kamble}, {Roming}, {Pritchard}, \&
  {Stringfellow}}]{2014ApJ...780...21M}
{Margutti}, R., {Milisavljevic}, D., {Soderberg}, A.~M., {Chornock}, R.,
  {Zauderer}, B.~A., {Murase}, K., {Guidorzi}, C., {Sanders}, N.~E., {Kuin},
  P., {Fransson}, C., {Levesque}, E.~M., {Chandra}, P., {Berger}, E., {Bianco},
  F.~B., {Brown}, P.~J., {Challis}, P., {Chatzopoulos}, E., {Cheung}, C.~C.,
  {Choi}, C., {Chomiuk}, L., {Chugai}, N., {Contreras}, C., {Drout}, M.~R.,
  {Fesen}, R., {Foley}, R.~J., {Fong}, W., {Friedman}, A.~S., {Gall}, C.,
  {Gehrels}, N., {Hjorth}, J., {Hsiao}, E., {Kirshner}, R., {Im}, M.,
  {Leloudas}, G., {Lunnan}, R., {Marion}, G.~H., {Martin}, J., {Morrell}, N.,
  {Neugent}, K.~F., {Omodei}, N., {Phillips}, M.~M., {Rest}, A., {Silverman},
  J.~M., {Strader}, J., {Stritzinger}, M.~D., {Szalai}, T., {Utterback}, N.~B.,
  {Vinko}, J., {Wheeler}, J.~C., {Arnett}, D., {Campana}, S., {Chevalier}, R.,
  {Ginsburg}, A., {Kamble}, A., {Roming}, P.~W.~A., {Pritchard}, T., \&
  {Stringfellow}, G. 2014, \apj, 780, 21

\bibitem[{{Massey} {et~al.}(1995){Massey}, {Lang}, {Degioia-Eastwood}, \&
  {Garmany}}]{1995ApJ...438..188M}
{Massey}, P., {Lang}, C.~C., {Degioia-Eastwood}, K., \& {Garmany}, C.~D. 1995,
  \apj, 438, 188

\bibitem[{{Matheson} {et~al.}(2000){Matheson}, {Filippenko}, {Ho}, {Barth}, \&
  {Leonard}}]{2000AJ....120.1499M}
{Matheson}, T., {Filippenko}, A.~V., {Ho}, L.~C., {Barth}, A.~J., \& {Leonard},
  D.~C. 2000, \aj, 120, 1499

\bibitem[{{Mattila} {et~al.}(2012){Mattila}, {Dahlen}, {Efstathiou}, {Kankare},
  {Melinder}, {Alonso-Herrero}, {P{\'e}rez-Torres}, {Ryder},
  {V{\"a}is{\"a}nen}, \& {{\"O}stlin}}]{2012ApJ...756..111M}
{Mattila}, S., {Dahlen}, T., {Efstathiou}, A., {Kankare}, E., {Melinder}, J.,
  {Alonso-Herrero}, A., {P{\'e}rez-Torres}, M.~{\'A}., {Ryder}, S.,
  {V{\"a}is{\"a}nen}, P., \& {{\"O}stlin}, G. 2012, \apj, 756, 111

\bibitem[{{Mattila} {et~al.}(2013){Mattila}, {Fraser}, {Smartt}, {Meikle},
  {Romero-Ca{\~n}izales}, {Crockett}, \& {Stephens}}]{2013MNRAS.431.2050M}
{Mattila}, S., {Fraser}, M., {Smartt}, S.~J., {Meikle}, W.~P.~S.,
  {Romero-Ca{\~n}izales}, C., {Crockett}, R.~M., \& {Stephens}, A. 2013,
  \mnras, 431, 2050

\bibitem[{{Mattila} {et~al.}(2010){Mattila}, {Smartt}, {Maund}, {Benetti}, \&
  {Ergon}}]{2010arXiv1011.5494M}
{Mattila}, S., {Smartt}, S., {Maund}, J., {Benetti}, S., \& {Ergon}, M. 2010,
  ArXiv:1011.5494

\bibitem[{{Mattila} {et~al.}(2008){Mattila}, {Smartt}, {Eldridge}, {Maund},
  {Crockett}, \& {Danziger}}]{2008ApJ...688L..91M}
{Mattila}, S., {Smartt}, S.~J., {Eldridge}, J.~J., {Maund}, J.~R., {Crockett},
  R.~M., \& {Danziger}, I.~J. 2008, \apjl, 688, L91

\bibitem[{{Mauerhan} {et~al.}(2013){Mauerhan}, {Smith}, {Filippenko},
  {Blanchard}, {Blanchard}, {Casper}, {Cenko}, {Clubb}, {Cohen}, {Fuller},
  {Li}, \& {Silverman}}]{2013MNRAS.430.1801M}
{Mauerhan}, J.~C., {Smith}, N., {Filippenko}, A.~V., {Blanchard}, K.~B.,
  {Blanchard}, P.~K., {Casper}, C.~F.~E., {Cenko}, S.~B., {Clubb}, K.~I.,
  {Cohen}, D.~P., {Fuller}, K.~L., {Li}, G.~Z., \& {Silverman}, J.~M. 2013,
  \mnras, 430, 1801

\bibitem[{{Maund} {et~al.}(2011){Maund}, {Fraser}, {Ergon}, {Pastorello},
  {Smartt}, {Sollerman}, {Benetti}, {Botticella}, {Bufano}, {Danziger},
  {Kotak}, {Magill}, {Stephens}, \& {Valenti}}]{2011ApJ...739L..37M}
{Maund}, J.~R., {Fraser}, M., {Ergon}, M., {Pastorello}, A., {Smartt}, S.~J.,
  {Sollerman}, J., {Benetti}, S., {Botticella}, M.-T., {Bufano}, F.,
  {Danziger}, I.~J., {Kotak}, R., {Magill}, L., {Stephens}, A.~W., \&
  {Valenti}, S. 2011, \apjl, 739, L37

\bibitem[{{Maund} {et~al.}(2013){Maund}, {Fraser}, {Smartt}, {Botticella},
  {Barbarino}, {Childress}, {Gal-Yam}, {Inserra}, {Pignata}, {Reichart},
  {Schmidt}, {Sollerman}, {Taddia}, {Tomasella}, {Valenti}, \&
  {Yaron}}]{2013MNRAS.431L.102M}
{Maund}, J.~R., {Fraser}, M., {Smartt}, S.~J., {Botticella}, M.~T.,
  {Barbarino}, C., {Childress}, M., {Gal-Yam}, A., {Inserra}, C., {Pignata},
  G., {Reichart}, D., {Schmidt}, B., {Sollerman}, J., {Taddia}, F.,
  {Tomasella}, L., {Valenti}, S., \& {Yaron}, O. 2013, \mnras, 431, L102

\bibitem[{{Maund} {et~al.}(2014{\natexlab{a}}){Maund}, {Mattila},
  {Ramirez-Ruiz}, \& {Eldridge}}]{2014MNRAS.438.1577M}
{Maund}, J.~R., {Mattila}, S., {Ramirez-Ruiz}, E., \& {Eldridge}, J.~J.
  2014{\natexlab{a}}, \mnras, 438, 1577

\bibitem[{{Maund} {et~al.}(2014{\natexlab{b}}){Maund}, {Reilly}, \&
  {Mattila}}]{2014MNRAS.438..938M}
{Maund}, J.~R., {Reilly}, E., \& {Mattila}, S. 2014{\natexlab{b}}, \mnras, 438,
  938

\bibitem[{{Maund} \& {Smartt}(2005)}]{2005MNRAS.360..288M}
{Maund}, J.~R. \& {Smartt}, S.~J. 2005, \mnras, 360, 288

\bibitem[{{Maund} \& {Smartt}(2009)}]{2009Sci...324..486M}
---. 2009, Science, 324, 486

\bibitem[{{Maund} {et~al.}(2005){Maund}, {Smartt}, \&
  {Danziger}}]{2005MNRAS.364L..33M}
{Maund}, J.~R., {Smartt}, S.~J., \& {Danziger}, I.~J. 2005, \mnras, 364, L33

\bibitem[{{Maund} {et~al.}(2004){Maund}, {Smartt}, {Kudritzki},
  {Podsiadlowski}, \& {Gilmore}}]{2004Natur.427..129M}
{Maund}, J.~R., {Smartt}, S.~J., {Kudritzki}, R.~P., {Podsiadlowski}, P., \&
  {Gilmore}, G.~F. 2004, \nat, 427, 129

\bibitem[{{Mauron} \& {Josselin}(2011)}]{2011A&A...526A.156M}
{Mauron}, N. \& {Josselin}, E. 2011, \aap, 526, 156

\bibitem[{{Maza} {et~al.}(2009){Maza}, {Hamuy}, {Antezana}, {Gonzalez},
  {Lopez}, {Silva}, {Folatelli}, {Iturra}, {Cartier}, {Forster}, {Marchi},
  {Rojas}, {Pignata}, {Conuel}, {Reichart}, {Ivarsen}, {Haislip}, {Crain},
  {Foster}, {Nysewander}, \& {Lacluyze}}]{2009CBET.1928....1M}
{Maza}, J., {Hamuy}, M., {Antezana}, R., {Gonzalez}, L., {Lopez}, P., {Silva},
  S., {Folatelli}, G., {Iturra}, D., {Cartier}, R., {Forster}, F., {Marchi},
  S., {Rojas}, A., {Pignata}, G., {Conuel}, B., {Reichart}, D., {Ivarsen}, K.,
  {Haislip}, J., {Crain}, A., {Foster}, D., {Nysewander}, M., \& {Lacluyze}, A.
  2009, Central Bureau Electronic Telegrams, 1928, 1

\bibitem[{{Mazzali} {et~al.}(2010){Mazzali}, {Maurer}, {Valenti}, {Kotak}, \&
  {Hunter}}]{2010MNRAS.408...87M}
{Mazzali}, P.~A., {Maurer}, I., {Valenti}, S., {Kotak}, R., \& {Hunter}, D.
  2010, \mnras, 408, 87

\bibitem[{{Milisavljevic} \& {Fesen}(2015)}]{2015Sci...347..526M}
{Milisavljevic}, D. \& {Fesen}, R.~A. 2015, Science, 347, 526

\bibitem[{{Nakano} \& {Itagaki}(2004)}]{2004CBET...74....1N}
{Nakano}, S. \& {Itagaki}, K. 2004, Central Bureau Electronic Telegrams, 74, 1

\bibitem[{{Nomoto} {et~al.}(1993){Nomoto}, {Suzuki}, {Shigeyama}, {Kumagai},
  {Yamaoka}, \& {Saio}}]{1993Natur.364..507N}
{Nomoto}, K., {Suzuki}, T., {Shigeyama}, T., {Kumagai}, S., {Yamaoka}, H., \&
  {Saio}, H. 1993, \nat, 364, 507

\bibitem[{{Nomoto} {et~al.}(1995){Nomoto}, {Iwamoto}, \&
  {Suzuki}}]{1995PhR...256..173N}
{Nomoto}, K.~I., {Iwamoto}, K., \& {Suzuki}, T. 1995, \physrep, 256, 173

\bibitem[{{O'Connor} \& {Ott}(2011)}]{2011ApJ...730...70O}
{O'Connor}, E. \& {Ott}, C.~D. 2011, \apj, 730, 70

\bibitem[{{Ofek} {et~al.}(2014){Ofek}, {Sullivan}, {Shaviv}, {Steinbok},
  {Arcavi}, {Gal-Yam}, {Tal}, {Kulkarni}, {Nugent}, {Ben-Ami}, {Kasliwal},
  {Cenko}, {Laher}, {Surace}, {Bloom}, {Filippenko}, {Silverman}, \&
  {Yaron}}]{2014ApJ...789..104O}
{Ofek}, E.~O., {Sullivan}, M., {Shaviv}, N.~J., {Steinbok}, A., {Arcavi}, I.,
  {Gal-Yam}, A., {Tal}, D., {Kulkarni}, S.~R., {Nugent}, P.~E., {Ben-Ami}, S.,
  {Kasliwal}, M.~M., {Cenko}, S.~B., {Laher}, R., {Surace}, J., {Bloom}, J.~S.,
  {Filippenko}, A.~V., {Silverman}, J.~M., \& {Yaron}, O. 2014, \apj, 789, 104

\bibitem[{{Paczy{\'n}ski}(1967)}]{1967AcA....17..355P}
{Paczy{\'n}ski}, B. 1967, \actaa, 17, 355

\bibitem[{{Pastorello} {et~al.}(2013){Pastorello}, {Cappellaro}, {Inserra},
  {Smartt}, {Pignata}, {Benetti}, {Valenti}, {Fraser}, {Tak{\'a}ts}, {Benitez},
  {Botticella}, {Brimacombe}, {Bufano}, {Cellier-Holzem}, {Costado}, {Cupani},
  {Curtis}, {Elias-Rosa}, {Ergon}, {Fynbo}, {Hambsch}, {Hamuy}, {Harutyunyan},
  {Ivarson}, {Kankare}, {Martin}, {Kotak}, {LaCluyze}, {Maguire}, {Mattila},
  {Maza}, {McCrum}, {Miluzio}, {Norgaard-Nielsen}, {Nysewander}, {Ochner},
  {Pan}, {Pumo}, {Reichart}, {Tan}, {Taubenberger}, {Tomasella}, {Turatto}, \&
  {Wright}}]{2013ApJ...767....1P}
{Pastorello}, A., {Cappellaro}, E., {Inserra}, C., {Smartt}, S.~J., {Pignata},
  G., {Benetti}, S., {Valenti}, S., {Fraser}, M., {Tak{\'a}ts}, K., {Benitez},
  S., {Botticella}, M.~T., {Brimacombe}, J., {Bufano}, F., {Cellier-Holzem},
  F., {Costado}, M.~T., {Cupani}, G., {Curtis}, I., {Elias-Rosa}, N., {Ergon},
  M., {Fynbo}, J.~P.~U., {Hambsch}, F.-J., {Hamuy}, M., {Harutyunyan}, A.,
  {Ivarson}, K.~M., {Kankare}, E., {Martin}, J.~C., {Kotak}, R., {LaCluyze},
  A.~P., {Maguire}, K., {Mattila}, S., {Maza}, J., {McCrum}, M., {Miluzio}, M.,
  {Norgaard-Nielsen}, H.~U., {Nysewander}, M.~C., {Ochner}, P., {Pan}, Y.-C.,
  {Pumo}, M.~L., {Reichart}, D.~E., {Tan}, T.~G., {Taubenberger}, S.,
  {Tomasella}, L., {Turatto}, M., \& {Wright}, D. 2013, \apj, 767, 1

\bibitem[{{Pastorello} {et~al.}(2007){Pastorello}, {Smartt}, {Mattila},
  {Eldridge}, {Young}, {Itagaki}, {Yamaoka}, {Navasardyan}, {Valenti}, {Patat},
  {Agnoletto}, {Augusteijn}, {Benetti}, {Cappellaro}, {Boles}, {Bonnet-Bidaud},
  {Botticella}, {Bufano}, {Cao}, {Deng}, {Dennefeld}, {Elias-Rosa},
  {Harutyunyan}, {Keenan}, {Iijima}, {Lorenzi}, {Mazzali}, {Meng}, {Nakano},
  {Nielsen}, {Smoker}, {Stanishev}, {Turatto}, {Xu}, \&
  {Zampieri}}]{2007Natur.447..829P}
{Pastorello}, A., {Smartt}, S.~J., {Mattila}, S., {Eldridge}, J.~J., {Young},
  D., {Itagaki}, K., {Yamaoka}, H., {Navasardyan}, H., {Valenti}, S., {Patat},
  F., {Agnoletto}, I., {Augusteijn}, T., {Benetti}, S., {Cappellaro}, E.,
  {Boles}, T., {Bonnet-Bidaud}, J.-M., {Botticella}, M.~T., {Bufano}, F.,
  {Cao}, C., {Deng}, J., {Dennefeld}, M., {Elias-Rosa}, N., {Harutyunyan}, A.,
  {Keenan}, F.~P., {Iijima}, T., {Lorenzi}, V., {Mazzali}, P.~A., {Meng}, X.,
  {Nakano}, S., {Nielsen}, T.~B., {Smoker}, J.~V., {Stanishev}, V., {Turatto},
  M., {Xu}, D., \& {Zampieri}, L. 2007, \nat, 447, 829

\bibitem[{{Patnaude} \& {Fesen}(2009)}]{2009ApJ...697..535P}
{Patnaude}, D.~J. \& {Fesen}, R.~A. 2009, \apj, 697, 535

\bibitem[{{Piro}(2013)}]{2013ApJ...768L..14P}
{Piro}, A.~L. 2013, \apjl, 768, L14

\bibitem[{{Podsiadlowski} {et~al.}(1993){Podsiadlowski}, {Hsu}, {Joss}, \&
  {Ross}}]{1993Natur.364..509P}
{Podsiadlowski}, P., {Hsu}, J.~J.~L., {Joss}, P.~C., \& {Ross}, R.~R. 1993,
  \nat, 364, 509

\bibitem[{{Podsiadlowski} {et~al.}(1992){Podsiadlowski}, {Joss}, \&
  {Hsu}}]{1992ApJ...391..246P}
{Podsiadlowski}, P., {Joss}, P.~C., \& {Hsu}, J.~J.~L. 1992, \apj, 391, 246

\bibitem[{{Poznanski}(2013)}]{2013MNRAS.436.3224P}
{Poznanski}, D. 2013, \mnras, 436, 3224

\bibitem[{{Poznanski} {et~al.}(2011){Poznanski}, {Ganeshalingam}, {Silverman},
  \& {Filippenko}}]{2011MNRAS.415L..81P}
{Poznanski}, D., {Ganeshalingam}, M., {Silverman}, J.~M., \& {Filippenko},
  A.~V. 2011, \mnras, 415, L81

\bibitem[{{Poznanski} {et~al.}(2012){Poznanski}, {Prochaska}, \&
  {Bloom}}]{2012MNRAS.426.1465P}
{Poznanski}, D., {Prochaska}, J.~X., \& {Bloom}, J.~S. 2012, \mnras, 426, 1465

\bibitem[{{Prieto} {et~al.}(2013){Prieto}, {Brimacombe}, {Drake}, \&
  {Howerton}}]{2013ApJ...763L..27P}
{Prieto}, J.~L., {Brimacombe}, J., {Drake}, A.~J., \& {Howerton}, S. 2013,
  \apjl, 763, L27

\bibitem[{{Prieto} {et~al.}(2008){Prieto}, {Kistler}, {Thompson}, {Y{\"u}ksel},
  {Kochanek}, {Stanek}, {Beacom}, {Martini}, {Pasquali}, \&
  {Bechtold}}]{2008ApJ...681L...9P}
{Prieto}, J.~L., {Kistler}, M.~D., {Thompson}, T.~A., {Y{\"u}ksel}, H.,
  {Kochanek}, C.~S., {Stanek}, K.~Z., {Beacom}, J.~F., {Martini}, P.,
  {Pasquali}, A., \& {Bechtold}, J. 2008, \apjl, 681, L9

\bibitem[{{Pumo} \& {Zampieri}(2011)}]{2011ApJ...741...41P}
{Pumo}, M.~L. \& {Zampieri}, L. 2011, \apj, 741, 41

\bibitem[{{Rest} {et~al.}(2008){Rest}, {Welch}, {Suntzeff}, {Oaster},
  {Lanning}, {Olsen}, {Smith}, {Becker}, {Bergmann}, {Challis}, {Clocchiatti},
  {Cook}, {Damke}, {Garg}, {Huber}, {Matheson}, {Minniti}, {Prieto}, \&
  {Wood-Vasey}}]{2008ApJ...681L..81R}
{Rest}, A., {Welch}, D.~L., {Suntzeff}, N.~B., {Oaster}, L., {Lanning}, H.,
  {Olsen}, K., {Smith}, R.~C., {Becker}, A.~C., {Bergmann}, M., {Challis}, P.,
  {Clocchiatti}, A., {Cook}, K.~H., {Damke}, G., {Garg}, A., {Huber}, M.~E.,
  {Matheson}, T., {Minniti}, D., {Prieto}, J.~L., \& {Wood-Vasey}, W.~M. 2008,
  \apjl, 681, L81

\bibitem[{{Sana} {et~al.}(2012){Sana}, {de Mink}, {de Koter}, {Langer},
  {Evans}, {Gieles}, {Gosset}, {Izzard}, {Le Bouquin}, \&
  {Schneider}}]{2012Sci...337..444S}
{Sana}, H., {de Mink}, S.~E., {de Koter}, A., {Langer}, N., {Evans}, C.~J.,
  {Gieles}, M., {Gosset}, E., {Izzard}, R.~G., {Le Bouquin}, J.-B., \&
  {Schneider}, F.~R.~N. 2012, Science, 337, 444

\bibitem[{{Shiode} \& {Quataert}(2014)}]{2014ApJ...780...96S}
{Shiode}, J.~H. \& {Quataert}, E. 2014, \apj, 780, 96

\bibitem[{{Shivvers} {et~al.}(2014){Shivvers}, {Mauerhan}, {Leonard},
  {Filippenko}, \& {Fox}}]{2014arXiv1408.1404S}
{Shivvers}, I., {Mauerhan}, J.~C., {Leonard}, D.~C., {Filippenko}, A.~V., \&
  {Fox}, O.~D. 2014, ArXiv:1408.1404

\bibitem[{{Singer} {et~al.}(2004){Singer}, {Pugh}, \&
  {Li}}]{2004IAUC.8297....2S}
{Singer}, D., {Pugh}, H., \& {Li}, W. 2004, \iauc, 8297, 2

\bibitem[{{Smartt}(2009)}]{2009ARA&A..47...63S}
{Smartt}, S.~J. 2009, \araa, 47, 63

\bibitem[{{Smartt} {et~al.}(2009){Smartt}, {Eldridge}, {Crockett}, \&
  {Maund}}]{2009MNRAS.395.1409S}
{Smartt}, S.~J., {Eldridge}, J.~J., {Crockett}, R.~M., \& {Maund}, J.~R. 2009,
  \mnras, 395, 1409

\bibitem[{{Smartt} {et~al.}(2002){Smartt}, {Gilmore}, {Tout}, \&
  {Hodgkin}}]{2002ApJ...565.1089S}
{Smartt}, S.~J., {Gilmore}, G.~F., {Tout}, C.~A., \& {Hodgkin}, S.~T. 2002,
  \apj, 565, 1089

\bibitem[{{Smartt} {et~al.}(2001){Smartt}, {Gilmore}, {Trentham}, {Tout}, \&
  {Frayn}}]{2001ApJ...556L..29S}
{Smartt}, S.~J., {Gilmore}, G.~F., {Trentham}, N., {Tout}, C.~A., \& {Frayn},
  C.~M. 2001, \apjl, 556, L29

\bibitem[{{Smartt} {et~al.}(2003){Smartt}, {Maund}, {Gilmore}, {Tout},
  {Kilkenny}, \& {Benetti}}]{2003MNRAS.343..735S}
{Smartt}, S.~J., {Maund}, J.~R., {Gilmore}, G.~F., {Tout}, C.~A., {Kilkenny},
  D., \& {Benetti}, S. 2003, \mnras, 343, 735

\bibitem[{{Smartt} {et~al.}(2004){Smartt}, {Maund}, {Hendry}, {Tout},
  {Gilmore}, {Mattila}, \& {Benn}}]{2004Sci...303..499S}
{Smartt}, S.~J., {Maund}, J.~R., {Hendry}, M.~A., {Tout}, C.~A., {Gilmore},
  G.~F., {Mattila}, S., \& {Benn}, C.~R. 2004, Science, 303, 499

\bibitem[{{Smith} \& {Arnett}(2014)}]{2014ApJ...785...82S}
{Smith}, N. \& {Arnett}, W.~D. 2014, \apj, 785, 82

\bibitem[{{Smith} {et~al.}(2011{\natexlab{a}}){Smith}, {Li}, {Filippenko}, \&
  {Chornock}}]{2011MNRAS.412.1522S}
{Smith}, N., {Li}, W., {Filippenko}, A.~V., \& {Chornock}, R.
  2011{\natexlab{a}}, \mnras, 412, 1522

\bibitem[{{Smith} {et~al.}(2011{\natexlab{b}}){Smith}, {Li}, {Miller},
  {Silverman}, {Filippenko}, {Cuillandre}, {Cooper}, {Matheson}, \& {Van
  Dyk}}]{2011ApJ...732...63S}
{Smith}, N., {Li}, W., {Miller}, A.~A., {Silverman}, J.~M., {Filippenko},
  A.~V., {Cuillandre}, J.-C., {Cooper}, M.~C., {Matheson}, T., \& {Van Dyk},
  S.~D. 2011{\natexlab{b}}, \apj, 732, 63

\bibitem[{{Smith} {et~al.}(2015){Smith}, {Mauerhan}, {Cenko}, {Kasliwal},
  {Silverman}, {Filippenko}, {Gal-Yam}, {Clubb}, {Graham}, {Leonard}, {Horst},
  {Williams}, {Andrews}, {Kulkarni}, {Nugent}, {Sullivan}, {Maguire}, {Xu}, \&
  {Ben-Ami}}]{2015MNRAS.449.1876S}
{Smith}, N., {Mauerhan}, J.~C., {Cenko}, S.~B., {Kasliwal}, M.~M., {Silverman},
  J.~M., {Filippenko}, A.~V., {Gal-Yam}, A., {Clubb}, K.~I., {Graham}, M.~L.,
  {Leonard}, D.~C., {Horst}, J.~C., {Williams}, G.~G., {Andrews}, J.~E.,
  {Kulkarni}, S.~R., {Nugent}, P., {Sullivan}, M., {Maguire}, K., {Xu}, D., \&
  {Ben-Ami}, S. 2015, \mnras, 449, 1876

\bibitem[{{Smith} {et~al.}(2014){Smith}, {Mauerhan}, \&
  {Prieto}}]{2014MNRAS.438.1191S}
{Smith}, N., {Mauerhan}, J.~C., \& {Prieto}, J.~L. 2014, \mnras, 438, 1191

\bibitem[{{Smith} {et~al.}(2010){Smith}, {Miller}, {Li}, {Filippenko},
  {Silverman}, {Howard}, {Nugent}, {Marcy}, {Bloom}, {Ghez}, {Lu}, {Yelda},
  {Bernstein}, \& {Colucci}}]{2010AJ....139.1451S}
{Smith}, N., {Miller}, A., {Li}, W., {Filippenko}, A.~V., {Silverman}, J.~M.,
  {Howard}, A.~W., {Nugent}, P., {Marcy}, G.~W., {Bloom}, J.~S., {Ghez}, A.~M.,
  {Lu}, J., {Yelda}, S., {Bernstein}, R.~A., \& {Colucci}, J.~E. 2010, \aj,
  139, 1451

\bibitem[{{Smith} \& {Owocki}(2006)}]{2006ApJ...645L..45S}
{Smith}, N. \& {Owocki}, S.~P. 2006, \apjl, 645, L45

\bibitem[{{Soderberg} {et~al.}(2012){Soderberg}, {Margutti}, {Zauderer},
  {Krauss}, {Katz}, {Chomiuk}, {Dittmann}, {Nakar}, {Sakamoto}, {Kawai},
  {Hurley}, {Barthelmy}, {Toizumi}, {Morii}, {Chevalier}, {Gurwell},
  {Petitpas}, {Rupen}, {Alexander}, {Levesque}, {Fransson}, {Brunthaler},
  {Bietenholz}, {Chugai}, {Grindlay}, {Copete}, {Connaughton}, {Briggs},
  {Meegan}, {von Kienlin}, {Zhang}, {Rau}, {Golenetskii}, {Mazets}, \&
  {Cline}}]{2012ApJ...752...78S}
{Soderberg}, A.~M., {Margutti}, R., {Zauderer}, B.~A., {Krauss}, M., {Katz},
  B., {Chomiuk}, L., {Dittmann}, J.~A., {Nakar}, E., {Sakamoto}, T., {Kawai},
  N., {Hurley}, K., {Barthelmy}, S., {Toizumi}, T., {Morii}, M., {Chevalier},
  R.~A., {Gurwell}, M., {Petitpas}, G., {Rupen}, M., {Alexander}, K.~D.,
  {Levesque}, E.~M., {Fransson}, C., {Brunthaler}, A., {Bietenholz}, M.~F.,
  {Chugai}, N., {Grindlay}, J., {Copete}, A., {Connaughton}, V., {Briggs}, M.,
  {Meegan}, C., {von Kienlin}, A., {Zhang}, X., {Rau}, A., {Golenetskii}, S.,
  {Mazets}, E., \& {Cline}, T. 2012, \apj, 752, 78

\bibitem[{{Spiro} {et~al.}(2014){Spiro}, {Pastorello}, {Pumo}, {Zampieri},
  {Turatto}, {Smartt}, {Benetti}, {Cappellaro}, {Valenti}, {Agnoletto},
  {Altavilla}, {Aoki}, {Brocato}, {Corsini}, {Di Cianno}, {Elias-Rosa},
  {Hamuy}, {Enya}, {Fiaschi}, {Folatelli}, {Desidera}, {Harutyunyan}, {Howell},
  {Kawka}, {Kobayashi}, {Leibundgut}, {Minezaki}, {Navasardyan}, {Nomoto},
  {Mattila}, {Pietrinferni}, {Pignata}, {Raimondo}, {Salvo}, {Schmidt},
  {Sollerman}, {Spyromilio}, {Taubenberger}, {Valentini}, {Vennes}, \&
  {Yoshii}}]{2014MNRAS.439.2873S}
{Spiro}, S., {Pastorello}, A., {Pumo}, M.~L., {Zampieri}, L., {Turatto}, M.,
  {Smartt}, S.~J., {Benetti}, S., {Cappellaro}, E., {Valenti}, S., {Agnoletto},
  I., {Altavilla}, G., {Aoki}, T., {Brocato}, E., {Corsini}, E.~M., {Di
  Cianno}, A., {Elias-Rosa}, N., {Hamuy}, M., {Enya}, K., {Fiaschi}, M.,
  {Folatelli}, G., {Desidera}, S., {Harutyunyan}, A., {Howell}, D.~A., {Kawka},
  A., {Kobayashi}, Y., {Leibundgut}, B., {Minezaki}, T., {Navasardyan}, H.,
  {Nomoto}, K., {Mattila}, S., {Pietrinferni}, A., {Pignata}, G., {Raimondo},
  G., {Salvo}, M., {Schmidt}, B.~P., {Sollerman}, J., {Spyromilio}, J.,
  {Taubenberger}, S., {Valentini}, G., {Vennes}, S., \& {Yoshii}, Y. 2014,
  \mnras, 439, 2873

\bibitem[{{Sukhbold} \& {Woosley}(2014)}]{2014ApJ...783...10S}
{Sukhbold}, T. \& {Woosley}, S.~E. 2014, \apj, 783, 10

\bibitem[{{Szczygie{\l}} {et~al.}(2012){Szczygie{\l}}, {Gerke}, {Kochanek}, \&
  {Stanek}}]{2012ApJ...747...23S}
{Szczygie{\l}}, D.~M., {Gerke}, J.~R., {Kochanek}, C.~S., \& {Stanek}, K.~Z.
  2012, \apj, 747, 23

\bibitem[{{Takats} {et~al.}(2015){Takats}, {Pignata}, {Pumo}, {Paillas},
  {Zampieri}, {Elias-Rosa}, {Benetti}, {Bufano}, {Cappellaro}, {Ergon},
  {Fraser}, {Hamuy}, {Inserra}, {Kankare}, {Smartt}, {Stritzinger}, {Van Dyk},
  {Haislip}, {LaCluyze}, {Moore}, \& {Reichart}}]{2015arXiv150402404T}
{Takats}, K., {Pignata}, G., {Pumo}, M.~L., {Paillas}, E., {Zampieri}, L.,
  {Elias-Rosa}, N., {Benetti}, S., {Bufano}, F., {Cappellaro}, E., {Ergon}, M.,
  {Fraser}, M., {Hamuy}, M., {Inserra}, C., {Kankare}, E., {Smartt}, S.~J.,
  {Stritzinger}, M.~D., {Van Dyk}, S.~D., {Haislip}, J.~B., {LaCluyze}, A.~P.,
  {Moore}, J.~P., \& {Reichart}, D. 2015, ArXiv e-prints

\bibitem[{{Tak{\'a}ts} {et~al.}(2014){Tak{\'a}ts}, {Pumo}, {Elias-Rosa},
  {Pastorello}, {Pignata}, {Paillas}, {Zampieri}, {Anderson}, {Vink{\'o}},
  {Benetti}, {Botticella}, {Bufano}, {Campillay}, {Cartier}, {Ergon},
  {Folatelli}, {Foley}, {F{\"o}rster}, {Hamuy}, {Hentunen}, {Kankare},
  {Leloudas}, {Morrell}, {Nissinen}, {Phillips}, {Smartt}, {Stritzinger},
  {Taubenberger}, {Valenti}, {Van Dyk}, {Haislip}, {LaCluyze}, {Moore}, \&
  {Reichart}}]{2014MNRAS.438..368T}
{Tak{\'a}ts}, K., {Pumo}, M.~L., {Elias-Rosa}, N., {Pastorello}, A., {Pignata},
  G., {Paillas}, E., {Zampieri}, L., {Anderson}, J.~P., {Vink{\'o}}, J.,
  {Benetti}, S., {Botticella}, M.-T., {Bufano}, F., {Campillay}, A., {Cartier},
  R., {Ergon}, M., {Folatelli}, G., {Foley}, R.~J., {F{\"o}rster}, F., {Hamuy},
  M., {Hentunen}, V.-P., {Kankare}, E., {Leloudas}, G., {Morrell}, N.,
  {Nissinen}, M., {Phillips}, M.~M., {Smartt}, S.~J., {Stritzinger}, M.,
  {Taubenberger}, S., {Valenti}, S., {Van Dyk}, S.~D., {Haislip}, J.~B.,
  {LaCluyze}, A.~P., {Moore}, J.~P., \& {Reichart}, D. 2014, \mnras, 438, 368

\bibitem[{{Tomasella} {et~al.}(2013){Tomasella}, {Cappellaro}, {Fraser},
  {Pumo}, {Pastorello}, {Pignata}, {Benetti}, {Bufano}, {Dennefeld},
  {Harutyunyan}, {Iijima}, {Jerkstrand}, {Kankare}, {Kotak}, {Magill},
  {Nascimbeni}, {Ochner}, {Siviero}, {Smartt}, {Sollerman}, {Stanishev},
  {Taddia}, {Taubenberger}, {Turatto}, {Valenti}, {Wright}, \&
  {Zampieri}}]{2013MNRAS.434.1636T}
{Tomasella}, L., {Cappellaro}, E., {Fraser}, M., {Pumo}, M.~L., {Pastorello},
  A., {Pignata}, G., {Benetti}, S., {Bufano}, F., {Dennefeld}, M.,
  {Harutyunyan}, A., {Iijima}, T., {Jerkstrand}, A., {Kankare}, E., {Kotak},
  R., {Magill}, L., {Nascimbeni}, V., {Ochner}, P., {Siviero}, A., {Smartt},
  S., {Sollerman}, J., {Stanishev}, V., {Taddia}, F., {Taubenberger}, S.,
  {Turatto}, M., {Valenti}, S., {Wright}, D.~E., \& {Zampieri}, L. 2013,
  \mnras, 434, 1636

\bibitem[{{Trundle} {et~al.}(2008){Trundle}, {Kotak}, {Vink}, \&
  {Meikle}}]{2008A&A...483L..47T}
{Trundle}, C., {Kotak}, R., {Vink}, J.~S., \& {Meikle}, W.~P.~S. 2008, \aap,
  483, L47

\bibitem[{{Ugliano} {et~al.}(2012){Ugliano}, {Janka}, {Marek}, \&
  {Arcones}}]{2012ApJ...757...69U}
{Ugliano}, M., {Janka}, H.-T., {Marek}, A., \& {Arcones}, A. 2012, \apj, 757,
  69

\bibitem[{{Utrobin} \& {Chugai}(2008)}]{2008A&A...491..507U}
{Utrobin}, V.~P. \& {Chugai}, N.~N. 2008, \aap, 491, 507

\bibitem[{{Utrobin} \& {Chugai}(2009)}]{2009A&A...506..829U}
---. 2009, \aap, 506, 829

\bibitem[{{Valenti} {et~al.}(2008){Valenti}, {Benetti}, {Cappellaro}, {Patat},
  {Mazzali}, {Turatto}, {Hurley}, {Maeda}, {Gal-Yam}, {Foley}, {Filippenko},
  {Pastorello}, {Challis}, {Frontera}, {Harutyunyan}, {Iye}, {Kawabata},
  {Kirshner}, {Li}, {Lipkin}, {Matheson}, {Nomoto}, {Ofek}, {Ohyama}, {Pian},
  {Poznanski}, {Salvo}, {Sauer}, {Schmidt}, {Soderberg}, \&
  {Zampieri}}]{2008MNRAS.383.1485V}
{Valenti}, S., {Benetti}, S., {Cappellaro}, E., {Patat}, F., {Mazzali}, P.,
  {Turatto}, M., {Hurley}, K., {Maeda}, K., {Gal-Yam}, A., {Foley}, R.~J.,
  {Filippenko}, A.~V., {Pastorello}, A., {Challis}, P., {Frontera}, F.,
  {Harutyunyan}, A., {Iye}, M., {Kawabata}, K., {Kirshner}, R.~P., {Li}, W.,
  {Lipkin}, Y.~M., {Matheson}, T., {Nomoto}, K., {Ofek}, E.~O., {Ohyama}, Y.,
  {Pian}, E., {Poznanski}, D., {Salvo}, M., {Sauer}, D.~N., {Schmidt}, B.~P.,
  {Soderberg}, A., \& {Zampieri}, L. 2008, \mnras, 383, 1485

\bibitem[{{Van Dyk} {et~al.}(2012{\natexlab{a}}){Van Dyk}, {Cenko},
  {Poznanski}, {Arcavi}, {Gal-Yam}, {Filippenko}, {Silverio}, {Stockton},
  {Cuillandre}, {Marcy}, {Howard}, \& {Isaacson}}]{2012ApJ...756..131V}
{Van Dyk}, S.~D., {Cenko}, S.~B., {Poznanski}, D., {Arcavi}, I., {Gal-Yam}, A.,
  {Filippenko}, A.~V., {Silverio}, K., {Stockton}, A., {Cuillandre}, J.-C.,
  {Marcy}, G.~W., {Howard}, A.~W., \& {Isaacson}, H. 2012{\natexlab{a}}, \apj,
  756, 131

\bibitem[{{Van Dyk} {et~al.}(2012{\natexlab{b}}){Van Dyk}, {Davidge},
  {Elias-Rosa}, {Taubenberger}, {Li}, {Levesque}, {Howerton}, {Pignata},
  {Morrell}, {Hamuy}, \& {Filippenko}}]{2012AJ....143...19V}
{Van Dyk}, S.~D., {Davidge}, T.~J., {Elias-Rosa}, N., {Taubenberger}, S., {Li},
  W., {Levesque}, E.~M., {Howerton}, S., {Pignata}, G., {Morrell}, N., {Hamuy},
  M., \& {Filippenko}, A.~V. 2012{\natexlab{b}}, \aj, 143, 19

\bibitem[{{Van Dyk} {et~al.}(2002){Van Dyk}, {Garnavich}, {Filippenko},
  {H{\"o}flich}, {Kirshner}, {Kurucz}, \& {Challis}}]{2002PASP..114.1322V}
{Van Dyk}, S.~D., {Garnavich}, P.~M., {Filippenko}, A.~V., {H{\"o}flich}, P.,
  {Kirshner}, R.~P., {Kurucz}, R.~L., \& {Challis}, P. 2002, \pasp, 114, 1322

\bibitem[{{Van Dyk} {et~al.}(2011){Van Dyk}, {Li}, {Cenko}, {Kasliwal},
  {Horesh}, {Ofek}, {Kraus}, {Silverman}, {Arcavi}, {Filippenko}, {Gal-Yam},
  {Quimby}, {Kulkarni}, {Yaron}, \& {Polishook}}]{2011ApJ...741L..28V}
{Van Dyk}, S.~D., {Li}, W., {Cenko}, S.~B., {Kasliwal}, M.~M., {Horesh}, A.,
  {Ofek}, E.~O., {Kraus}, A.~L., {Silverman}, J.~M., {Arcavi}, I.,
  {Filippenko}, A.~V., {Gal-Yam}, A., {Quimby}, R.~M., {Kulkarni}, S.~R.,
  {Yaron}, O., \& {Polishook}, D. 2011, \apjl, 741, L28

\bibitem[{{Van Dyk} {et~al.}(2003{\natexlab{a}}){Van Dyk}, {Li}, \&
  {Filippenko}}]{2003PASP..115....1V}
{Van Dyk}, S.~D., {Li}, W., \& {Filippenko}, A.~V. 2003{\natexlab{a}}, \pasp,
  115, 1

\bibitem[{{Van Dyk} {et~al.}(2003{\natexlab{b}}){Van Dyk}, {Li}, \&
  {Filippenko}}]{2003PASP..115.1289V}
---. 2003{\natexlab{b}}, \pasp, 115, 1289

\bibitem[{{Van Dyk} {et~al.}(1999){Van Dyk}, {Peng}, {Barth}, \&
  {Filippenko}}]{1999AJ....118.2331V}
{Van Dyk}, S.~D., {Peng}, C.~Y., {Barth}, A.~J., \& {Filippenko}, A.~V. 1999,
  \aj, 118, 2331

\bibitem[{{Van Dyk} {et~al.}(2013){Van Dyk}, {Zheng}, {Clubb}, {Filippenko},
  {Cenko}, {Smith}, {Fox}, {Kelly}, {Shivvers}, \&
  {Ganeshalingam}}]{2013ApJ...772L..32V}
{Van Dyk}, S.~D., {Zheng}, W., {Clubb}, K.~I., {Filippenko}, A.~V., {Cenko},
  S.~B., {Smith}, N., {Fox}, O.~D., {Kelly}, P.~L., {Shivvers}, I., \&
  {Ganeshalingam}, M. 2013, \apjl, 772, L32

\bibitem[{{Van Dyk} {et~al.}(2014){Van Dyk}, {Zheng}, {Fox}, {Cenko}, {Clubb},
  {Filippenko}, {Foley}, {Miller}, {Smith}, {Kelly}, {Lee}, {Ben-Ami}, \&
  {Gal-Yam}}]{2014AJ....147...37V}
{Van Dyk}, S.~D., {Zheng}, W., {Fox}, O.~D., {Cenko}, S.~B., {Clubb}, K.~I.,
  {Filippenko}, A.~V., {Foley}, R.~J., {Miller}, A.~A., {Smith}, N., {Kelly},
  P.~L., {Lee}, W.~H., {Ben-Ami}, S., \& {Gal-Yam}, A. 2014, \aj, 147, 37

\bibitem[{{Vanbeveren} {et~al.}(1998){Vanbeveren}, {De Loore}, \& {Van
  Rensbergen}}]{1998A&ARv...9...63V}
{Vanbeveren}, D., {De Loore}, C., \& {Van Rensbergen}, W. 1998, \aapr, 9, 63

\bibitem[{{Vink} \& {de Koter}(2005)}]{2005A&A...442..587V}
{Vink}, J.~S. \& {de Koter}, A. 2005, \aap, 442, 587

\bibitem[{{Walmswell} \& {Eldridge}(2012)}]{2012MNRAS.419.2054W}
{Walmswell}, J.~J. \& {Eldridge}, J.~J. 2012, \mnras, 419, 2054

\bibitem[{{Williams} {et~al.}(2014){Williams}, {Peterson}, {Murphy}, {Gilbert},
  {Dalcanton}, {Dolphin}, \& {Jennings}}]{2014ApJ...791..105W}
{Williams}, B.~F., {Peterson}, S., {Murphy}, J., {Gilbert}, K., {Dalcanton},
  J.~J., {Dolphin}, A.~E., \& {Jennings}, Z.~G. 2014, \apj, 791, 105

\bibitem[{{Woosley} {et~al.}(1994){Woosley}, {Eastman}, {Weaver}, \&
  {Pinto}}]{1994ApJ...429..300W}
{Woosley}, S.~E., {Eastman}, R.~G., {Weaver}, T.~A., \& {Pinto}, P.~A. 1994,
  \apj, 429, 300

\bibitem[{{Woosley} \& {Heger}(2007)}]{2007PhR...442..269W}
{Woosley}, S.~E. \& {Heger}, A. 2007, \physrep, 442, 269

\bibitem[{{Yoon} {et~al.}(2012){Yoon}, {Gr{\"a}fener}, {Vink}, {Kozyreva}, \&
  {Izzard}}]{2012A&A...544L..11Y}
{Yoon}, S.-C., {Gr{\"a}fener}, G., {Vink}, J.~S., {Kozyreva}, A., \& {Izzard},
  R.~G. 2012, \aap, 544, L11

\bibitem[{{Yoon} \& {Langer}(2005)}]{2005A&A...443..643Y}
{Yoon}, S.-C. \& {Langer}, N. 2005, \aap, 443, 643

\bibitem[{{Yoon} {et~al.}(2010){Yoon}, {Woosley}, \&
  {Langer}}]{2010ApJ...725..940Y}
{Yoon}, S.-C., {Woosley}, S.~E., \& {Langer}, N. 2010, \apj, 725, 940

\end{thebibliography}


\end{document}